\documentclass[10pt,preprint]{aastex}
\usepackage{epsf,amssymb,amsmath}
\def\deg{\ifmmode {^\circ}\else {$^\circ$}\fi}
\def\degree{\ifmmode {^\circ}\else {$^\circ$}\fi}
\def\mum{\ifmmode {\rm \,\mu {\rm m}}\else $\rm \,\mu {\rm m}$\fi}
\def\arcsec{\ifmmode ^{\prime \prime}\else $^{\prime \prime}$\fi}

\def\inch{\ifmmode ^{\prime \prime}\else $^{\prime \prime}$\fi}
\def\gs{\ifmmode {{\rm g~s^{-1}}}\else ${\rm g~s^{-1}}$\fi}
\def\msunyr{\ifmmode {M_{\odot}~{\rm yr^{-1}}}\else $M_{\odot}~{\rm yr^{-1}}$\fi}
\def\msun{\ifmmode {M_{\odot}}\else $M_{\odot}$\fi}
\def\rsun{\ifmmode {R_{\odot}}\else $R_{\odot}$\fi}
\def\lsun{\ifmmode {L_{\odot}}\else $L_{\odot}$\fi}
\def\mstar{\ifmmode {M_{\star}}\else $M_{\star}$\fi}
\def\rstar{\ifmmode {R_{\star}}\else $R_{\star}$\fi}
\def\tstar{\ifmmode {T_{\star}}\else $T_{\star}$\fi}
\def\lstar{\ifmmode {L_{\star}}\else $L_{\star}$\fi}
\def\mwd{\ifmmode {M_{wd}}\else $M_{wd}$\fi}
\def\rwd{\ifmmode {R_{wd}}\else $R_{wd}$\fi}
\def\twd{\ifmmode {T_{wd}}\else $T_{wd}$\fi}
\def\lwd{\ifmmode {L_{wd}}\else $L_{wd}$\fi}
\def\md{\ifmmode {M_d}\else $M_d$\fi}
\def\ld{\ifmmode {L_d}\else $L_d$\fi}
\def\ad{\ifmmode A_d\else $A_d$\fi}
\def\ldlwd{\ifmmode L_d / L_{wd}\else $L_d / L_{wd}$\fi}
\def\ldlstar{\ifmmode L_d / L_\star\else $L_d / L_{\star}$\fi}
\def\rearth{\ifmmode {\rm R_{\oplus}}\else $\rm R_{\oplus}$\fi}
\def\mearth{\ifmmode {\rm M_{\oplus}}\else $\rm M_{\oplus}$\fi}
\def\qc{\ifmmode Q_c\else $Q_c$\fi}
\def\qdstar{\ifmmode Q_D^\star\else $Q_D^\star$\fi}
\def\rt{\ifmmode r_t\else $r_t$\fi}
\def\vc{\ifmmode v_c\else $v_c$\fi}
\def\vsqd{\ifmmode v^2 / Q_D^\star\else $v^2 / Q_D^\star$\fi}
\def\kms{\ifmmode {\rm km~s^{-1}}\else $\rm km~s^{-1}$\fi}
\def\ms{\ifmmode {\rm m~s^{-1}}\else $\rm m~s^{-1}$\fi}
\def\vrel{\ifmmode v_{rel}\else $v_{rel}$\fi}
\def\mdot{\ifmmode \dot{M}\else $\dot{M}$\fi}
\def\mdotz{\ifmmode \dot{M}_0\else $\dot{M}_0$\fi}
\def\mesc{\ifmmode m_{esc}\else $m_{esc}$\fi}
\def\rmin{\ifmmode r_{min}\else $r_{min}$\fi}
\def\rmax{\ifmmode r_{max}\else $r_{max}$\fi}
\def\xmax{\ifmmode x_{max}\else $x_{max}$\fi}
\def\mmin{\ifmmode m_{min}\else $m_{min}$\fi}
\def\mmax{\ifmmode m_{max}\else $m_{max}$\fi}
\def\rmind{\ifmmode r_{min,d}\else $r_{min,d}$\fi}
\def\rmaxd{\ifmmode r_{max,d}\else $r_{max,d}$\fi}
\def\mmaxd{\ifmmode m_{max,d}\else $m_{max,d}$\fi}
\def\vrad{\ifmmode v_{rad}\else $v_{rad}$\fi}
\def\qz{\ifmmode q_{0}\else $q_{0}$\fi}
\def\qi{\ifmmode q_{i}\else $q_{i}$\fi}
\def\ql{\ifmmode q_{l}\else $q_{l}$\fi}
\def\qs{\ifmmode q_{s}\else $q_{s}$\fi}
\def\vhill{\ifmmode v_H\else $r_H$\fi}
\def\rhill{\ifmmode r_H\else $r_H$\fi}
\def\Rhill{\ifmmode R_H\else $R_H$\fi}
\def\rbrk{\ifmmode r_{brk}\else $r_{brk}$\fi}
\def\rdamp{\ifmmode r_{damp}\else $r_{damp}$\fi}
\def\rin{\ifmmode r_{in}\else $r_{in}$\fi}
\def\rout{\ifmmode r_{out}\else $r_{out}$\fi}
\def\tin{\ifmmode t_{in}\else $t_{in}$\fi}
\def\tout{\ifmmode t_{out}\else $t_{out}$\fi}
\def\ain{\ifmmode a_{in}\else $a_{in}$\fi}
\def\aout{\ifmmode a_{out}\else $a_{out}$\fi}
\def\r0{\ifmmode r_{0}\else $r_{0}$\fi}
\def\R0{\ifmmode R_{0}\else $R_{0}$\fi}
\def\m0{\ifmmode m_{0}\else $m_{0}$\fi}
\def\mone{\ifmmode m_{1}\else $m_{1}$\fi}
\def\mtwo{\ifmmode m_{2}\else $m_{2}$\fi}
\def\atwo{\ifmmode a_{2}\else $a_{2}$\fi}
\def\etwo{\ifmmode e_{2}\else $e_{2}$\fi}
\def\mf{\ifmmode m_{f}\else $m_{f}$\fi}
\def\af{\ifmmode a_{f}\else $a_{f}$\fi}
\def\ef{\ifmmode e_{f}\else $e_{f}$\fi}
\def\M0{\ifmmode M_{0}\else $M_{0}$\fi}
\def\amax{\ifmmode a_{max}\else $a_{max}$\fi}
\def\a0{\ifmmode a_{0}\else $a_{0}$\fi}
\def\e0{\ifmmode e_{0}\else $e_{0}$\fi}
\def\v0{\ifmmode v_{0}\else $v_{0}$\fi}
\def\xm{\ifmmode x_{m}\else $x_{m}$\fi}
\def\sigz{\ifmmode \Sigma_0\else $\Sigma_0$\fi}
\def\ergg{\ifmmode {\rm erg~g^{-1}}\else ${\rm erg~g^{-1}}$\fi}
\def\ergs{\ifmmode {\rm erg~s^{-1}}\else ${\rm erg~s^{-1}}$\fi}
\def\gyr{\ifmmode {\rm g~yr^{-1}}\else ${\rm g~yr^{-1}}$\fi}
\def\cms{\ifmmode {\rm cm~s^{-1}}\else ${\rm cm~s^{-1}}$\fi}
\def\gcms{\ifmmode {\rm g~cm^{-2}}\else $\rm g~cm^{-2}$\fi}
\def\gcmc{\ifmmode {\rm g~cm^{-3}}\else $\rm g~cm^{-3}$\fi}
\def\atil{\ifmmode {\tilde{a}}\else $\tilde{a}$\fi}
\def\ttil{\ifmmode {\tilde{t}}\else $\tilde{t}$\fi}
\def\sqrttt{\ifmmode {\tilde{t}^{1/2}}\else $\tilde{t}^{1/2}$\fi}

\def\orch{{\it Orchestra}}
\def\nh{{\it New Horizons}}
\def\pc{Pluto--Charon}
\def\mp{\ifmmode m_P\else $m_P$\fi}
\def\mc{\ifmmode m_C\else $m_C$\fi}
\def\mh{\ifmmode m_H\else $m_H$\fi}
\def\mk{\ifmmode m_K\else $m_K$\fi}
\def\ms{\ifmmode m_S\else $m_S$\fi}
\def\mn{\ifmmode m_N\else $m_N$\fi}
\def\rp{\ifmmode r_P\else $r_P$\fi}
\def\rc{\ifmmode r_C\else $r_C$\fi}
\def\apc{\ifmmode a_{PC}\else $a_{PC}$\fi}
\def\mpc{\ifmmode m_{PC}\else $m_{PC}$\fi}
\def\epc{\ifmmode e_{PC}\else $e_{PC}$\fi}

\begin{document}

\title{Craters on Charon: Impactors From a Collisional Cascade Among Trans-Neptunian Objects}
\vskip 7ex
\author{Scott J. Kenyon}
\affil{Smithsonian Astrophysical Observatory,
60 Garden Street, Cambridge, MA 02138}
\email{e-mail: skenyon@cfa.harvard.edu}

\author{Benjamin C. Bromley}
\affil{Department of Physics \& Astronomy, University of Utah,
201 JFB, Salt Lake City, UT 84112}
\email{e-mail: bromley@physics.utah.edu}

\begin{abstract}

We consider whether equilibrium size distributions from collisional cascades
match the frequency of impactors derived from \nh\ crater counts on Charon 
\citep{singer2019}. Using an analytic model and a suite of numerical 
simulations, we demonstrate that collisional cascades generate wavy size 
distributions; the morphology of the waves depends on the binding energy 
of solids \qdstar\ and the collision velocity \vc. For an adopted minimum 
size of solids, \rmin\ = 1~\mum, and collision velocity \vc\ = 1--3~\kms, 
the waves are rather insensitive to the gravitational component of \qdstar. 
If the bulk strength component of \qdstar\ is $Q_s r^{e_s}$ for particles 
with radius $r$, size distributions with small $Q_s$ are much wavier than 
those with large $Q_s$; systems with $e_s \approx -0.4$ have stronger waves 
than systems with $e_s \approx 0$. Detailed comparisons with the \nh\ data 
suggest that a collisional cascade among solids with a bulk strength 
intermediate between weak ice \citep[e.g.,][]{lein2012} and normal ice 
\citep[e.g.,][]{schlicht2013} produces size distributions fairly similar 
to the size distribution of impactors on Charon. If the surface density 
$\Sigma$ of the protosolar nebula varies with semimajor axis $a$ as
$\Sigma \approx 30~{\rm g~cm^{-2}} (a / {\rm 1~au})^{-3/2}$, the time scale 
for a cascade to generate an approximate equilibrium is 100--300~Myr at 
45~au and 10--30~Myr at 25~au.  Although it is necessary to perform more 
complete evolutionary calculations of the Kuiper belt, collisional cascades 
are a viable model for producing the size distribution of solids that 
impacted Charon throughout its history.
\end{abstract}


\section{INTRODUCTION}
\label{sec: intro}

Beyond the orbit of Neptune, the Solar System contains a vast population
of icy objects with radii ranging from $r \sim$ 1000~km \citep[e.g., 
Eris and Pluto;][]{brown2007a,sicardy2011b,stern2018} to $r \lesssim$ 1~km
\citep{schlicht2009,schlicht2012,liu2015}.  Although many of these 
trans-Neptunian objects (TNOs) have roughly circular orbits with semimajor 
axes $a \approx$ 35--50~au, others have highly elliptical orbits, 
$e \gtrsim$ 0.5, with $a \gtrsim$ 100~au \citep[e.g.,][]{gladman2008,
petit2011,petit2017}. Among all of the dynamical classes, the total 
mass in TNOs inferred from ground-based optical surveys is $\lesssim$ 
0.1--0.2~\mearth\ \citep[e.g.,][]{fraser2014,adams2014,alexandersen2016,
lawler2018,pitjeva2018}.

A recent analysis of data from the \nh\ mission creates tension between the 
number of TNOs {\it predicted} from deep surveys with ground-based optical 
telescopes and {\it derived} from direct counts of craters on Charon. Together 
with dynamical estimates based on the current population of Jupiter family 
comets \citep[e.g.,][and references therein]{levison1997,emelyan2004,volk2008,
brasser2015}, the optical surveys require a steep size distribution, where the 
predicted number of 1~km TNOs is $\sim 10^5$ times larger than the observed
number of 100~km TNOs. Normalizing the \nh\ results to the ground-based data 
at 10--20~km, the observed number of craters from 1--10~km TNOs is reasonably 
consistent with a steep size distribution.  However, the number of 0.1--1~km 
TNOs derived from the \nh\ observations is roughly two orders of magnitude 
smaller than expected from an extrapolation of ground-based measurements 
\citep{singer2019}. The deficit of small craters on Charon also disagrees with 
most theoretical (coagulation) models of planet formation, which predict counts more 
similar to those implied by ground-based surveys \citep[e.g.,][]{kb2012,schlicht2013}.

Here, we consider whether the size distribution derived from craters on
Charon is consistent with expectations for a collisional cascade, where
high velocity impacts continually erode the material from 1--100~km objects.
Analytical equilibrium size distributions for collisional cascades share 
features with the \nh\ results; we derive physical parameters for cascades 
that broadly match the Charon cratering record. Numerical simulations with
similar parameters agree rather well with the analytic predictions. 
Comparing the theoretical results with the \nh\ observations, solids with
the bulk strength of weak to normal ice have size distributions that resemble
the data. Because the time scale for generating these size distributions is
only 20--30~Myr (100--300~Myr) at 25~au (45~au), the impactors on Charon are
plausibly derived from a collisional cascade during the formation of the
Solar System.

Although it is possible to construct a model that follows the time evolution of 
the population of TNOs, the impact rate of these TNOs on Pluto-Charon
\citep[e.g.,][]{greenstreet2015,bierhaus2015}, and the dynamical evolution of 
the gas giant planets \citep[e.g.,][]{malhotra1993,malhotra1995,levison2003b,
levison2008}, we focus on a simpler issue. The goal is to understand how the 
features in the TNO size distribution depend on initial conditions, model parameters 
and time.  By eliminating model size distributions that do not match the \nh\ data, 
we limit the space of plausible models to be investigated in a more detailed 
evolutionary calculation. We return to issues involving the impact rate and the
long-term dynamical evolution of the Solar System in \S\ref{sec: disc}.

To set the stage for this study, we begin with an observational 
background (\S\ref{sec: obs}).  After reviewing previous theoretical 
approaches (\S\ref{sec: theor}), we consider whether analytic
(\S\ref{sec: casc-an}) or numerical (\S\ref{sec: casc-num}) collisional 
cascade models can match the \nh\ observations. After discussing the
implications of this analysis (\S\ref{sec: disc}), we conclude with
a brief summary (\S\ref{sec: summary}).

\vskip 6ex
\section{OBSERVATIONAL BACKGROUND}
\label{sec: obs}

The discovery of the Kuiper Belt dramatically changed our understanding
of the extent and dynamical structure of the Solar System.
Data from the first few surveys \citep[e.g.,][]{luu1988,jewitt1993,
jewitt1995,williams1995,irwin1995,jewitt1996,luu1997,luu1998,
trujillo2000,trujillo2001b} detected `Kuiper belt objects' (KBOs)
(i) in the 3:2 resonance with Neptune, 
(ii) in roughly circular orbits just outside Neptune's orbit, and 
(iii) in very elliptical orbits ($e \gtrsim$ 0.5) with perihelia close 
to Neptune's orbit. Subsequent deep imaging programs revealed an 
exquisite dynamical richness among Solar System objects orbiting beyond 
Neptune \citep[e.g.,][]{gladman2002,luu2002,brown2004,gladman2008,petit2011,
bernard2020}.  
Today, there are $\sim$ 3500 TNOs with semimajor axis $a \gtrsim$ 30~au listed 
in the database of the International Astronomical Union's Minor Planet Center.

Dynamical classifications of KBOs and TNOs are based on their current 
orbital elements and the gravitational influence of the four gas giant 
planets \citep[e.g.,][]{levison1996,duncan1997b,gladman2002,dones2004,
morbi2004,elliot2005,delsanti2006,gladman2008,petit2011,khain2020}.  Many 
TNOs are in orbital resonance with Neptune. The 3:2 (Plutinos), 5:3, 7:4, 
and 2:1 (Twotinos) resonances are well-populated \citep[e.g.,][]{chiang2003,
petit2011,li2014a,li2014b,volk2016,li2020}; recent discoveries include TNOs 
in the 21:5 \citep{holman2018}, the 9:2 \citep{bannister2016b}, 
and the 9:1 \citep{volk2018} resonances with Neptune. Various non-resonant 
TNOs have orbits with perihelion distances $q$ close to or inside Neptune's 
semimajor axis, $a_N \approx$ 30~au. The scattering (or scattered) disk 
objects (SDOs) have $a \gg a_N$. The SDOs are distinguished from Centaurs, 
which have $a < a_N$, $q >$ 7.35~au, and a Tisserand parameter, 
\begin{equation}
T_J = \frac{a_J}{a} + 2 \left ( \frac{a}{a_J} (1 - e^2) \right )^{1/2} {\rm cos~\imath} ~ > 3 ~ ,
\end{equation}
where $a_J$ is the semimajor axis of Jupiter, and $\imath$ is the orbital
inclination of the Centaur or SDO. 
On more distant orbits, detached TNOs have $q \gtrsim a_N$ and $e \gtrsim$ 
0.24; inner Oort cloud objects have $a \gtrsim$ 2000~au.

The classical KBOs are non-resonant TNOs on fairly low $e$ orbits. 
\citet{gladman2008} identified three dynamical classes: inner ($a <$ 39.4~au;
orbit interior to the 3:2 resonance with Neptune), main ($a \approx$ 42--48~au), 
and outer ($a >$ 48.4~au and $e <$ 0.24; exterior to the 2:1 resonance with 
Neptune). Main classical objects are often call `cubewanos' after the first 
known KBO, 1992~QB$_1$ \citep{jewitt1993}, and are divided into a `hot' 
component ($\imath >$ 5\deg) and a `cold' component ($\imath <$ 5\deg).
\citet{petit2011} isolated three components of the main classical belt -- 
hot, kernel, and stirred -- with specific ranges in $a$, $e$, and $\imath$
\citep[see also][]{petit2017}. 
More recent observations confirm the kernel and indicate the cold classical
belt extends beyond the 2:1 resonance \citep{bannister2016a,bannister2018}.

Measuring the size distribution(s) of TNOs from deep imaging observations 
requires several steps \citep[e.g.,][and references therein]{jlt1998,luu1998,
trujillo2001,luu2002,bernstein2004,petit2006,fuentes2008,fraser2008,gilhutton2009,
fraser2009a,fuentes2009}.  In the most direct approach, analysis of large 
samples of TNOs with good orbits yields $\Sigma(<H)$ (in units of deg$^{-2}$), 
the cumulative sky surface density brighter than an absolute magnitude 
$H$\footnote{Usually, $H$ is measured in the broadband V, R, or $r$ 
filters; sometimes, the bluer $g$ filter is used.}.  Converting the 
frequency of absolute magnitudes to the frequency of object radii $r$ 
requires knowledge of the geometric albedo $p$. If every TNO is a diffuse
disk reflector, the relation between absolute magnitude, geometric albedo,
and radius is straightforward \citep[e.g.,][]{harris1997}:
\begin{equation}
r \approx \frac{665}{p_{\lambda}^{1/2}} ~ 10^{-0.2 ~ H_{\lambda}} ~ {\rm km}.
\end{equation}
Although $H$ is derived for several wavelengths $\lambda$, the albedo is often
quoted in the V-band, $p_V$. 

Early observations of TNOs suggested $\Sigma$ followed a simple power law
\citep[e.g.,][]{jlt1998,luu1998,chiang1999,sheppard2000,trujillo2001,gladman2001}
\begin{equation}
{\rm log} ~ \Sigma(<H) = \alpha ~ (H - H_0) ~ ,
\end{equation}
where $\alpha$ is the slope and $H_0$ is a reference brightness.
Deeper surveys for objects with $H \lesssim$ 12--13 revealed a break 
(or `knee') in the power-law and a shallower slope to larger $H$ 
\citep[e.g.,][]{bernstein2004,petit2006,fraser2008,fuentes2008,
fraser2009a,fuentes2009}. Although the surface density is often 
continuous across the break, several analyses favor a `divot' model, 
where the surface density for magnitudes just below the break is 
significantly smaller than the surface density above the break 
\citep[e.g.,][]{shankman2013,shankman2016,alexandersen2016,lawler2018}.
Combining these approaches into a single set of expressions, the
cumulative sky surface density is
\begin{equation}
\Sigma(<H) = \Sigma_b ~ \delta_b ~ 10^{\alpha_b~(H~-~H_b)} + \Sigma_f ~ \delta_f ~ ( 10^{\alpha_f~(H~-~H_b)} ~ - ~ 1) ~ ,
\end{equation}
where
$H_b$ is the absolute magnitude at the break,
$\Sigma_b$ ($\Sigma_f$) is the surface density of bright (faint) objects at the break,
$\alpha_b$ ($\alpha_f$) is the power-law slope for bright (faint) objects above (below) 
the break, and
\begin{equation}
\delta_{b,f} = \begin{cases}
                0, 1 & H ~ \le ~ H_b \\
                1, 0 & H ~ \ge ~ H_b \\
                \end{cases}
\end{equation}
To make a stronger connection to several published analyses, we follow 
\citet{shankman2013,shankman2016} and define the contrast
\begin{equation}
c_{bf} = \frac{\Sigma_b}{\Sigma_f} ~ .
\end{equation}
When $c_{bf}$ = 1, the surface density distribution is continuous across 
the break. Models with $c_{bf} > 1$ have a pronounced divot at the break.

Deriving $n(r)$, the differential size distribution of TNOs as a function 
of radius $r$, from fits to the surface density distribution requires an 
expression for the albedo $p_V(r)$.  When $n(r) \propto r^{-q}$ and 
$p_V(r) \propto r^{-\beta}$ \citep[e.g.][]{jlt1998,fraser2008}, there is 
a simple relation between $q$, $\alpha$, and $\beta$:
\begin{equation}
q = 5 ~ \alpha ~ (1 - \beta /2) ~ + ~ 1 ~ .
\label{eq: q-alpha}
\end{equation}
For a cumulative size distribution, $n(<r)$ or $n(>r)$, with power-law index 
$q^{\prime}$, $q^\prime = q - 1$. Measurements of $\alpha$ (from large 
magnitude-limited surveys of TNOs) and $\beta$ (from mid-IR and far-IR 
observations that constrain the reflected and thermal emission of TNOs) 
yield $q$.

Over the past decade, analyses of $\Sigma(H)$ yield a fairly consistent picture
for the location of the break and the slope of the magnitude distribution above
the break \citep[e.g.,][]{petit2011,gladman2012,shankman2013,fraser2014,adams2014,
schwamb2014,alexandersen2016,shankman2016,lawler2018}. For the hot classical KBOs,
the SDOs, and the Plutinos, $\alpha_b \approx$ 0.8--1.0 and $H_b \approx$ 7.7--8.3. 
The bright-end slope for the cold classical KBOs is much steeper, $\alpha_b \approx$
1.2--1.5. \citet{fraser2014} derive a brighter magnitude for the break that has
not been tested by other analyses of the cold KBOs. 

Curiously, the bright-end slope for KBOs in the 2:1 and 5:3 resonances with 
Neptune is similar to that of other dynamically hot objects, $\alpha_b \approx$ 
0.9--1 \citep{adams2014}. However, the $\alpha_b \approx$ 1.3 for KBOs in the
5:2 and 7:4 resonances is closer to the slope derived for the cold classical 
KBOs. With many fewer objects than the other TNO populations, the errors in the 
slopes for the size distributions of objects in the 5:2 and 7:4 resonances are 
significant. Larger samples might establish whether these differences in 
$\alpha_b$ are real.

Independent analyses also agree on the overall population of the different 
classes of TNOs.  The SDOs are the most populous group, with $\sim$ 37,500 
objects brighter than $H$ = 8 \citep{shankman2013,adams2014,shankman2016,
lawler2018}.  There are $\sim$ 20,000 classical KBOs, with $\sim 400$ in 
the inner belt and the rest roughly evenly divided between the main belt 
and the outer belt \citep{petit2011,adams2014}. Within the main classical 
belt, $\sim$ 10\% lie within the kernel; the remainder are roughly evenly 
divided between the hot and stirred populations. 

For many resonances, the relative populations are plagued by small number
statistics \citep[e.g.,][]{adams2014,bannister2016b,volk2018,bannister2018}.
However, statistics for the Plutinos suggest $\sim$ 3000 objects with
$H \le$ 8 \citep[e.g.,][]{gladman2012,adams2014,volk2016,alexandersen2016}.
The 2:1 ($\sim$ 1500) and the 5:2 ($\sim$ 1000) are the next most populous.
The number of TNOs in the 5:2 resonance is larger than predicted by early 
dynamical models \citep[e.g.,][]{chiang2003b,levison2008}; recent analyses 
address this issue \citep{malhotra2018,yu2018}.

Despite the general agreement on $\alpha_b$ and $H_b$, the range of possibilities 
for $\alpha_f$ and $c_{bf}$ is much broader \citep[e.g.,][]{fraser2014,shankman2013,
alexandersen2016,shankman2016,lawler2018}. Results for the cold KBOs suggest 
$\alpha_f \approx$ 0.2 and $c_{bf}$ = 1.  Analyses of data for Plutinos and SDOs 
yield larger slopes, $\alpha_f \approx$ 0.2--0.6, and various contrasts, 
$c_{bf} \approx$ 1--10. As outlined by \citet{shankman2013,shankman2016} for the
SDOs, the population at $H \approx$ 15--20 is constrained by the population of 
Jupiter family comets which are believed to originate among the SDOs \citep[and 
perhaps the Plutinos;][]{levison1997,emelyan2004,emelyan2005,volk2008,disisto2009,
brasser2015}. If the Plutino/SDO size distribution consists only of two power laws,
current data favor models with $\alpha_f \approx$ 0.5--0.6 and $c_{bf} \approx$ 3--6.
However, a three component size distribution enables models with smaller $\alpha_f$
and $c_{bf} \approx$ 1 for $H \approx$ 8--12, providing the slope at $H > 12$ 
becomes steep enough to match the constraints for Jupiter family comets at much
fainter magnitudes. 

Combined with ground-based observations, data from the {\it Herschel} and 
{\it Spitzer} satellites enable direct measurements of the geometric albedo $p_V$ 
and the radii $r$ of TNOs \citep[e.g.,][]{vilenius2012,
vilenius2014,duffard2014,kovalenko2017,vilenius2018}. Despite small samples 
limited to the largest TNOs, the distribution of measured albedo is very 
broad, $p_V \approx$ 0.03--0.80; the albedo correlates with orbital inclination
but not size.  Typical values for the hot KBOs, $p_V \approx$ 0.09, are smaller 
than the albedos for cold KBOs, $p_V \approx$ 0.14. However, the Haumea family 
(and other TNOs like Makamake) have $p_V \gtrsim$ 0.5. 

Converting the brightness of TNOs at the break in the {\it surface density} 
distribution, $H_b \approx$ 7.7--8.3, to a radius for TNOs at the break in 
the {\it size} distribution requires a geometric albedo. For simplicity, we 
assume the albedo is independent of wavelength in the optical.  With a typical 
$p_V \approx$ 0.10, the faintest TNOs 
directly observed in ground-based or space-based surveys 
have a radius $r \approx$ 10~km; the
break occurs at a radius $r_b \approx$ 45--60~km.  Reducing (increasing) $p_V$ 
to 0.05 (0.15) raises (lowers) the break radius to $\sim$ 65--85~km (30--45~km).  
Because the range of albedos is much larger than the uncertainty in $H_b$, the 
uncertainty in $r_b$ is dominated by the uncertainty in the adopted $p_V$.

With $\beta \approx 0$ in the relation $p_V(r) \propto r^{-\beta}$, the slope
$q$ of $n(r)$ depends only on $\alpha$ (eq.~\ref{eq: q-alpha}).  For bright 
objects detected in large, ground-based surveys, $q_b \approx$ 5--6 for hot 
KBOs, Plutinos, and SDOs; $q_b \approx$ 7--9 for cold KBOs. These slopes are 
larger than results from the more direct measurements of smaller samples 
from {\it Herschel} and {\it Spitzer}: $q \approx$ 4.2 for the Haumea family, 
$q \approx$ 3.3 for hot KBOs with $r \approx$ 50--250~km, and
$q \approx$ 6 for cold KBOs with $r \approx$ 80--150~km \citep{vilenius2014,
vilenius2018}. Still, the trends are similar: the slope for hot KBOs (including 
SDOs and Plutinos) is much shallower than for cold KBOs.  Among fainter KBOs, 
the preferred $\alpha_f \approx$ 0.5--0.6 implies $q_f \approx$ 3.5--4 
\citep{shankman2013,alexandersen2016,shankman2016}.  Allowing the smaller 
$\alpha$ = 0.2--0.4 from some studies results in $q_f \approx$ 2--3.

Results from other approaches for deriving $\alpha_f$ provide an interesting
contrast with the analyses of the ground-based optical surveys.  The 
$q_f \approx$ 3.5--4 needed to match the optical data at $H \approx$ 8--11 
{\it and} the postulated source population of Jupiter family comets at 
$H \approx$ 17 is identical to the $q_f \approx$ 3.8--3.9 derived from 
several detections of stellar occultations by 0.25--0.50 km KBOs 
\citep{schlicht2009,schlicht2012,liu2015}.  Lack of detections in another 
occultation survey implies $q \lesssim$ 3.3--3.8 \citep{bianco2010,zhang2013}. 
However, number counts of craters on Pluto and Charon from \nh\ suggest 
a shallower $q_f \approx$ 3 ($\alpha_f \approx$ 0.4) at $r \approx$ 
1--10~km and an even shallower $q_f \approx$ 1.75 ($\alpha_f \approx$ 
0.15) at $r \approx$ 0.1--1~km \citep{singer2019}. 

\begin{figure}[t!]
\begin{center}
\includegraphics[width=4.5in]{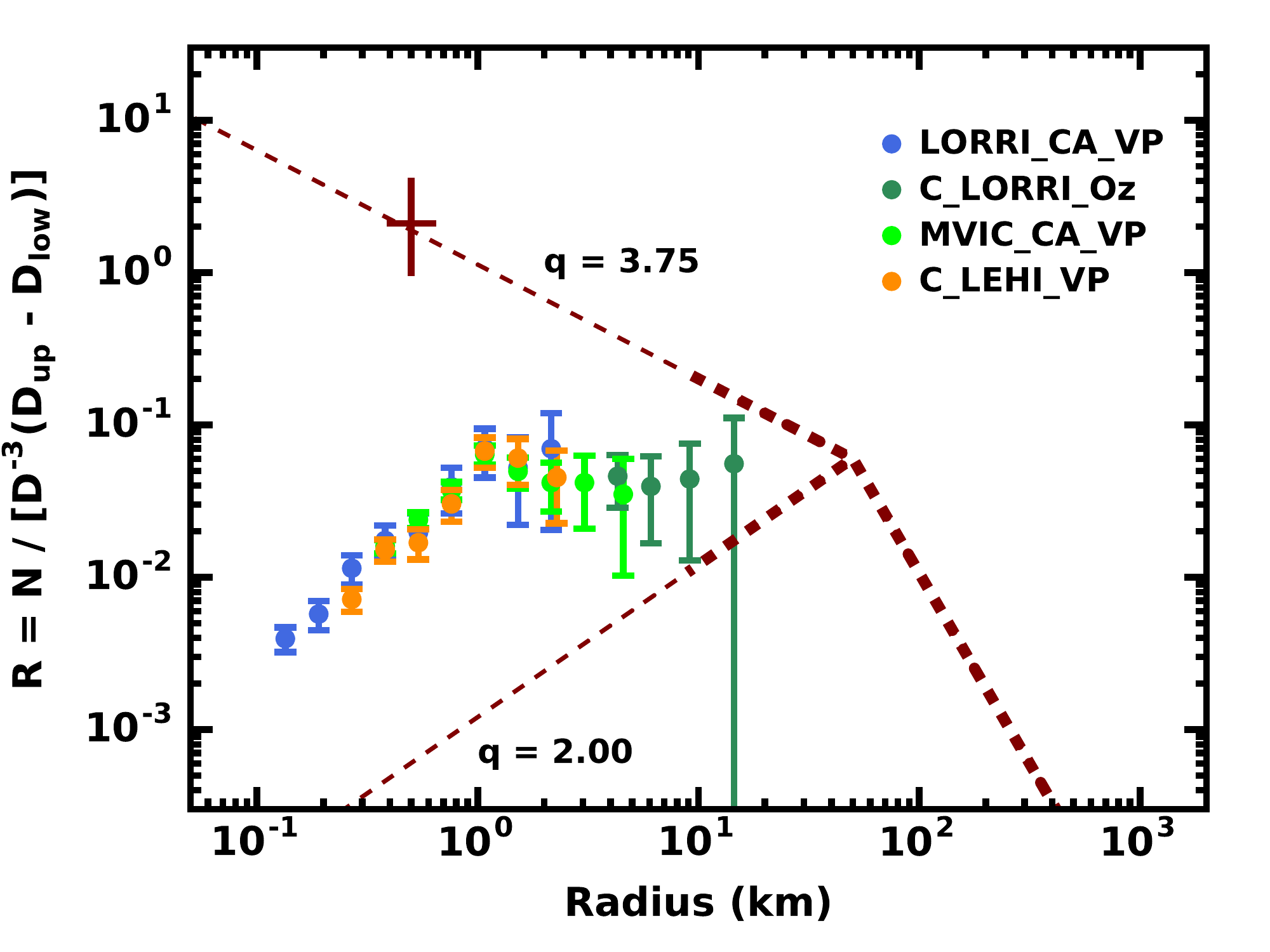}
\vskip -2ex
\caption{\label{fig: obs1}
Comparison of number counts for the radii of impactors derived from 
craters detected on \nh\ images of Charon \citep[filled points with 
error bars;][]{singer2019} with broken power-law size distributions 
derived from ground-based observations of TNOs (dashed lines).  The 
impactor diameter $D$ is $D = D_c / 6.25$, where $D_c$ is the crater 
diameter.  Dashed lines are normalized to yield an $R$ value at 
$r \sim$ 50~km which roughly matches the \nh\ data at $r \sim$ 10--20~km. 
The legend indicates the combination of instrument and geological 
feature for each \nh\ data set \citep{singer2019}.  The slopes of 
the two broken power-laws match observations for SDOs and other TNOs
larger than the `break radius' at $r \approx$ 50~km and span the 
range of possible slopes for smaller objects below the break 
\citep[e.g.,][]{fraser2014,lawler2018}.  Thicker lines cover the 
approximate size range detected by ground-based observations of SDOs; 
thinner lines are extrapolations.  The $+$ along the upper dashed 
line represents the constraints derived from occultation observations.
}
\end{center}
\end{figure}

Fig.~\ref{fig: obs1} illustrates the tension between the ground-based and
the \nh\ measurements of TNO size distributions. In this Figure, the $y$-axis 
plots $R(D) = N / (D^{-3} (D_{up} - D_{low})$ as in \citet{singer2019}; 
$N$ is the number of objects in a mass bin with maximum diameter $D_{up}$, 
minimum diameter $D_{low}$, and central diameter $D$. 
We adopt a simple relation between impactor diameter $D$ and crater
diameter, $D_c$, $D = D_c / 6.25$, which neglects the slightly steeper 
than linear relation between $D$ and $D_c$ and represents a compromise 
among relations for different materials \citep[e.g.,][]{hols1993,
housen2011,singer2013,singer2015,singer2019}.
In Figures derived from
analytic theory or numerical calculations later in this paper, we use the radius
$r = D/2$ and $r = (r_{low} r_{up})^{1/2}$; $R(r) = N / (r^{-3} (r_{up} - r_{low})$ 
and $R(D) = 4 R(r)$. The {\it shape} of the size distribution is then independent 
of $r$ or $D$. To avoid confusion, we use {\it size distribution} to refer to 
$n(r)$, $R(r)$, and $R(D)$; the slope $q$ is always derived from 
$n(r) \propto r^{-q}$.  In systems with $q$ = 3, $R(r)$ is independent of $r$.

To derive an appropriate $R(r)$ from ground-based observations, we consider 
several broken power-laws for TNOs with $H \le$ 12 (indicated by the thick 
dashed lines). To place these power-laws on the same scale as the \nh\ data,
we adopt $R$ = 0.06 at $r$ = 50~km for the TNO power-laws. With this
normalization, the \nh\ measurements at 1--10~km fall well below the $q$ = 
3.5--4.0 needed to match the occultation data (the large '+' along the upper 
dashed line) and to produce enough Jupiter family comets with $H \approx$ 17 
($r \approx$ 1~km). Adopting a different normalization, $R \approx$ 0.4--0.5
at 4--5~km, allows a crude match between the ground-based and \nh\ data at 
1--10~km; however, the \nh\ data at 0.1--1~km still fall well below the level
required to agree with the occultation measurements.

At $r \approx$ 0.1--1~km, the slope from the \nh\ data is at the lower end of 
the allowed ranges derived from cold KBOs and some model fits to observations 
of much larger Plutinos and the SDOs with $H \le$ 12 and $r \gtrsim$ 10--20~km 
for $p_V \approx$ 0.10 \citep{fraser2014,alexandersen2016,shankman2016,lawler2018}. 
Although it is possible to consider a normalization for the TNO power-laws that
places the \nh\ data at 0.1--1~km along the $q$ = 2 branch of the figure, this 
choice leaves the \nh\ data at 1--10~km even farther from the limits derived 
from Jupiter family comets and occultations.

In these examples, the two broken power-laws have 
no divot at the break in the size distribution at 50~km ($c_{bf}$ = 1).
Shifting the $q$ =
2.00 and $q$ = 3.75 power-laws vertically downward by a factor of 10 allows a match 
to the \nh\ data at 1--2~km and roughly agrees with the \nh\ data at 2--10~km. However,
the $c_{bf} \approx$ 10 implied by this shift is larger than the $c_{bf} \approx$ 
3--6 derived in some analyses of ground-based observations for KBOs with
$r \gtrsim$ 10--20 km \citep[e.g.,][]{alexandersen2016,shankman2016,lawler2018}. 
With this shift, there is still a stark disagreement between the \nh\ data at 
0.1--1~km and the slope required to match ground-based observations of 
$r \gtrsim$ 10--20~km TNOs and the frequency of Jupiter family comets with
radii $r \sim$ 1~km.

The results from \nh\ resemble size distributions derived from numerical 
calculations of collisional cascades \citep[e.g.,][]{kb2004c,kb2012}. In
many cascades, 
\begin{equation}
n(r) \propto \begin{cases}
                r^{-q_s} & r < r_s \\
                r^{-q_i} & r_s \le r \le r_l \\
                r^{-q_l} & r > r_l \\
                \end{cases}
\label{eq: nr-kbo}
\end{equation}
where $r_l \gtrsim$ 10--20~km, $r_s \approx$ 0.1--1~km, $q_l \gtrsim$ 3.5, 
$q_s \approx$ 3.5--4.0, and $q_i \approx$ 0--3. A goal of this study is 
to establish whether we can derive the physical characteristics of a
cascade that yields a $n(r)$ which matches the \nh\ results.

One way to restore the balance between the ground-based and \nh\ analyses
is to associate most of the \pc\ impactors with hot KBOs from the classical 
belt and Plutinos and to place the source of Jupiter family comets mostly 
within the SDOs. Indeed, \citet{greenstreet2015} show that four TNO populations
(hot and stirred classical TNOs, classical outer TNOs, and Plutinos) contribute
roughly equally to impactors on Pluto--Charon \citep[see also][]{bierhaus2015}.
TNOs in other dynamical classes should impact Pluto--Charon at much lower rates.
Completed prior to the \nh\ flyby of Pluto--Charon, both studies assumed a variety
of size distributions for small TNOs. Our goal is to explore the origins of the 
features in the size distribution of TNOs. The next section briefly reviews 
analytic and numerical models that attempt to explain these features.

\vskip 6ex
\section{THEORETICAL BACKGROUND}
\label{sec: theor}

Soon after the discovery of 1992~QB$_1$, theorists began to consider 
the formation and long-term collisional evolution of TNOs. Early efforts
examined the growth of the largest KBOs from 1--10~km planetesimals 
\citep[e.g.,][]{stern1995,stern1996,stcol1997a,kl1998,kl1999a,durda2000,
kenyon2002} or the collisional erosion of $n(r)$ and the total mass in TNOs 
\citep[e.g.,][]{davis1997,stcol1997b}. These investigations established that 
(i) the time scale for 100--1000~ km KBOs to grow from a swarm of 1--10~km 
planetesimals within a smooth disk or ring is 20--100~Myr, 
(ii) once large KBOs form, destructive collisions can remove $\gtrsim$ 90\% 
of the initial mass in the Kuiper belt, (iii) the current $n(r)$ for KBOs 
has had a roughly constant shape for $\sim$ 4 Gyr, and (iv) small TNOs with 
$r \lesssim$ 50--100~km are fragments of high velocity impacts.  Initial fits 
of the models to observations of the size distribution at large sizes were 
encouraging \citep[e..g.,][]{kl1999b}.

Since these pioneering studies, several analytic treatments have focused 
on the physical origin of the break in the power-law size distribution at 
$r \approx$ 30--100~km \citep[e.g.][]{kb2004c,pan2005,fraser2009b}. We 
follow previous work and compare $Q_b$ the binding energy of two colliding 
TNOs with their center-of-mass collision energy per unit mass
\begin{equation}
\qc\ = 0.5 \left ( \frac{m_1 m_2 v^2}{m_{12}^2} \right ) ~ = \frac{q v^2}{2~(1 + q)^2} ,
\label{eq: qc}
\end{equation}
where $v$ is the collision velocity, $m_1$ and $m_2 = \theta m_1 ~ (\theta \le 1)$ 
are the mass of two colliding TNOs, and $m_{12} = m_1 + m_2$.  When \qc\ is 
much smaller than $Q_b$, the collision produces a more massive large object. 
However, if \qc\ is much larger than $Q_b$, then the collision is 
{\it catastrophic} and leaves behind a `largest remnant' with a mass less 
than half of $m_{12}$ \citep[e.g.,][and references therein]{weiden1974,
green1978,weth1980}. 

Although there are several distinct but complementary approaches to calculating 
$Q_b$ \citep[e.g.,][]{green1984,davis1985,weth1989,housen1990,weth1993,hols1994,
davis1997, weiden1997a,benavidez2009,obrien2003,campo2012}, it is convenient to 
define the collision energy required to disperse half of $m_{12}$ beyond the
gravitational reach of the colliding pair:
\begin{equation}
\label{eq: qdstar}
\qdstar\ = Q_s r^{e_s} + Q_g \rho r^{e_g} ~ ,
\end{equation}
where $\rho$ is the mass density of the TNOs and $(Q_s, Q_g, e_s, e_b)$ 
are model parameters \citep[e.g.,][]{benz1999,lein2012}. In this expression, 
the first (second) term is the strength (gravity) component of the binding
energy. These two elements are comparable for $r \approx$ 0.1~km.  When 
\qc\ is smaller (larger) than \qdstar, collisions between equal mass objects
($\theta \approx 1$) are accretive (destructive) and large objects grow 
(diminish) with time.
When $\theta \ll 1$ and $Q_c = \theta v^2 / 2$, high velocity collisions typically 
remove a mass from the higher mass `target' that exceeds the mass in the
much smaller `projectile'. In these `cratering' collisions, massive particles
gradually lose mass over time.

Setting \qc\ = \qdstar\ for $m_1$ = $m_2$ ($q$ = 1) yields a relation between 
$v$, $r$, $\rho$, $Q_g$, and $e_g$ for large objects \citep{kb2004c,pan2005}:
\begin{equation}
\label{eq: rbreak}
r_b = \left ( \frac{v^2}{8 \rho Q_g} \right )^{1/e_g} ~ .
\end{equation}
Collisions between pairs of equal mass objects with $r \le r_b$ generate
fragments; those with $r > r_b$ produce a larger merged object. Adopting
typical values $Q_g \approx$ 0.1--3~erg~~g$^{-2}$~cm$^{3-e_g}$ and 
$e_b \approx$ 1.2--1.4 \citep{benz1999,lein2012}, the collision velocity 
required to match the observed $r_b \approx$ 50~km is $v \approx$ 
1~\kms\ for $\rho \approx$ 
1.0--1.5~\gcmc. With orbital velocities, $v_K \approx$ 4--5~\kms\ and
eccentricities, $e \approx$ 0.05--0.2, TNOs at $a \approx$ 30--50~au have 
typical relative velocities comparable to the required $v$. Thus, the 
observed $r_b$ is reasonably consistent with theoretical expectations.

Several recent numerical calculations support this result 
\citep[e.g.,][]{krivov2005,charnoz2007,benavidez2009,fraser2009b,
campo2012,kb2012,schlicht2013}.  When 
the collision velocities are appropriate for present-day TNOs, the shape 
of the TNO size distribution depends on (i) the initial $n(r)$,
(ii) the \qdstar\ parameters, and (iii) time. Calculations extending over
0.5--4.5~Gyr result in $r_b \approx$ 50~km for nominal values of $Q_g$ and
$e_g$. Plausible ranges in $Q_g$ and $e_b$ yield factor of two variations
in $r_b$.  Large TNOs with $r \gtrsim r_b$ maintain their initial size 
distribution.  Smaller TNOs with $r \lesssim r_b$ have wavy size 
distributions with peaks and valleys that depend on $v$, $Q_g$, and $e_b$ 
\citep[see also][]{campo1994a,obrien2003,krivov2005,wyatt2011}.  At the 
smallest sizes, $r \lesssim$ 0.1--1~km, $n(r)$ follows a power law with 
$q \approx$ 3--4.

Despite the general agreement on how the break radius and the size 
distribution change with input parameters, the calculations derive 
very different size distributions for $r \approx$ 0.1--100~km. Several
display clear divots \citep[e.g.,][]{benavidez2009,fraser2009b}; others
do not \citep[e.g.,][]{campo2012,schlicht2013}. Although waviness at
$r \approx$ 1--100~km is characteristic, none obviously match the size
distribution derived from the \nh\ cratering record. 

\vskip 6ex
\section{COLLISIONAL CASCADES: ANALYTICAL RESULTS}
\label{sec: casc-an}

\subsection{Background}
\label{sec: casc-an-back}

To explore the possibilities for understanding the \nh\ observations, we 
begin with an analysis of equilibrium size distributions for the TNO 
population.  In any model, $n(r)$ depends on the rates collisions remove 
objects with a range of sizes $r \lesssim r_b$ and add debris with a range 
of sizes $r \ll r_b$.  When these rates are equal, $n(r)$ reaches a steady 
state where the shape remains fixed and the absolute level slowly declines 
with time. Analytic results show how the equilibrium shape of $n(r)$ depends 
on \qc\ and \qdstar\ \citep[see the discussion in][]{wyatt2011}. Numerical 
calculations designed to achieve a steady state generate wavy size distributions 
that resemble the analytic shapes reasonably well \citep[e.g.,][]{kb2016a,kb2017a}. 
Because all of the recent numerical simulations of the TNO populations suggest 
steady-state size distributions over some range of sizes, it is reasonable 
to investigate whether analytic steady-state size distributions can match 
the \nh\ observations.

For nearly two decades, analytic theories of collisional cascades have 
been developed to explain the long-term evolution of debris disks 
\citep[e.g.,][]{wyatt2002,dom2003,wyatt2007a,wyatt2007b,wyatt2008,kw2011a,
wyatt2011,kb2016a,knb2016,kb2017a}.  
Defining \rmax\ as the radius of the largest object in a swarm of solids,
a cascade of catastrophic collisions produces a flow of material from 
\rmax\ to \rmin\ the radius of the smallest object in the swarm. Sometimes
\rmin\ = 0 \citep[e.g.,][]{dohn1969,will1994}. More often, \rmin\ is the
minimum size stable against radiation pressure from the central star
\citep[e.g.,][]{burns1979,arty1988,wyatt2008}. For the Sun, 
\rmin\ $\approx$ 1~\mum. 

As the cascade proceeds, \rmax\ and the mass in solids \md\ decline with 
time $t$ \citep{wyatt2002,dom2003,wyatt2008,kb2017a}:
\begin{equation}
\label{eq: rmax}
\rmax = \frac{\r0}{(1 + t/\tau_0)^\gamma} ~ 
\end{equation}
\begin{equation}
\label{eq: M0}
M_d = \frac{\M0}{(1 + t/\tau_0)^\gamma} ~ ,
\end{equation}
where \r0\ is the initial radius of the largest object, \M0\ is the initial 
mass in solids with $r \le r_0$, and $\gamma \ge 1$ is a constant 
\citep{kb2017a}. In these expressions, $t_0$ sets the collision time 
\begin{equation}
\label{eq: tcoll}
t_0 = \frac{r_0 \rho P}{12 \pi \Sigma_0}
\end{equation}
for solids with semimajor axis $a$, orbital period $P$, and initial 
surface density $\Sigma_0 = \M0 / 2 \pi a \Delta a$ within an annulus
of width $\Delta a$. Defining $\alpha = -(\M0/t_0) \dot{M_d}^{-1}$, 
$\tau_0 = (\gamma + 1) \alpha t_0$. 

In eqs.~\ref{eq: rmax}--\ref{eq: M0}, the decline of \rmax\ and $M_d$
depends on $\alpha$ and $\gamma$, which encode the evolution of the 
cascade as a function of \vsqd, the ratio of the center-of-mass 
collision energy to \qdstar. When \vsqd\ $\approx$ 8, $\tau_0$ is large;
the cascade slowly reduces \rmax\ and \M0. Increasing \vsqd\ decreases
$\tau_0$. More energetic collisions produce more rapid changes in 
\rmax\ and \M0.

The simplest analytic theory assumes that catastrophic collisions set 
the equilibrium $n(r)$.  Cratering collisions with $Q_c < \qdstar$ are 
ignored \citep{wyatt2011}. The model also assumes that all particles have 
the same collision velocity $v$.  Within a logarithmic grid of particles 
with indices $k$ = 1 to $k$ = N where $r_1 < r_N$, the mass loss rate of 
particles is assumed to be independent of the bin. The mass contained in 
the bin is then \citep{wyatt2011}:
\begin{equation}
M_k = C_0 R_k^{-1} ~ ,
\label{eq: mks}
\end{equation}
where $C_0$ is an arbitrary constant \citep[see also eq. 15 of][]{wyatt2011}.
The sum $R_k$ is the rate of collisions which disperse at least half of the 
combined mass of a pair of colliding particles,
\begin{equation}
R_k = C_1 \sum_{i=1}^{i=k} \epsilon_{ik} M_i (r_i + r_k)^2 / r_i^3 ~ ,
\label{eq: rk}
\end{equation}
where $C_1$ is another constant. When catastrophic collisions dominate
\begin{equation}
\epsilon_{ik} = \begin{cases}
                0 & Q_c < \qdstar\ \\
                1 & Q_c \ge \qdstar\ \\
                \end{cases}
\label{eq: eps}
\end{equation}

Although it is straightforward to develop an iterative technique to solve this 
set of equations \citep{wyatt2011}, \citet{kb2016a} proposed a recursive solution. 
When $k$ = 1, $R_1 = 4 C_1 M_1 / r_1$.  Thus, $M_1^2 = C_0 / 4 C_1 r_1$.  For 
$k \ge$ 2, $R_k$ is a sum over terms with known $M_i$ and one term with $M_k$:
\begin{equation}
R_k = 4 C_1 M_k / r_k + C_1 \sum_{i=1}^{i=k-1} \epsilon_{ik} M_i (r_i + r_k)^2 / r_i^3 ~ .
\label{eq: rk2}
\end{equation}
Setting $R_{ki}$ equal to the second term in eq.~\ref{eq: rk2}, the 
quadratic equation
\begin{equation}
4 C_1 M_k^2 + R_{ki} M_k - C_0 = 0 ~ ,
\label{eq: mk2}
\end{equation}
has only one possible solution with $M_k > 0$.

\begin{figure}[t!]
\begin{center}
\includegraphics[width=4.5in]{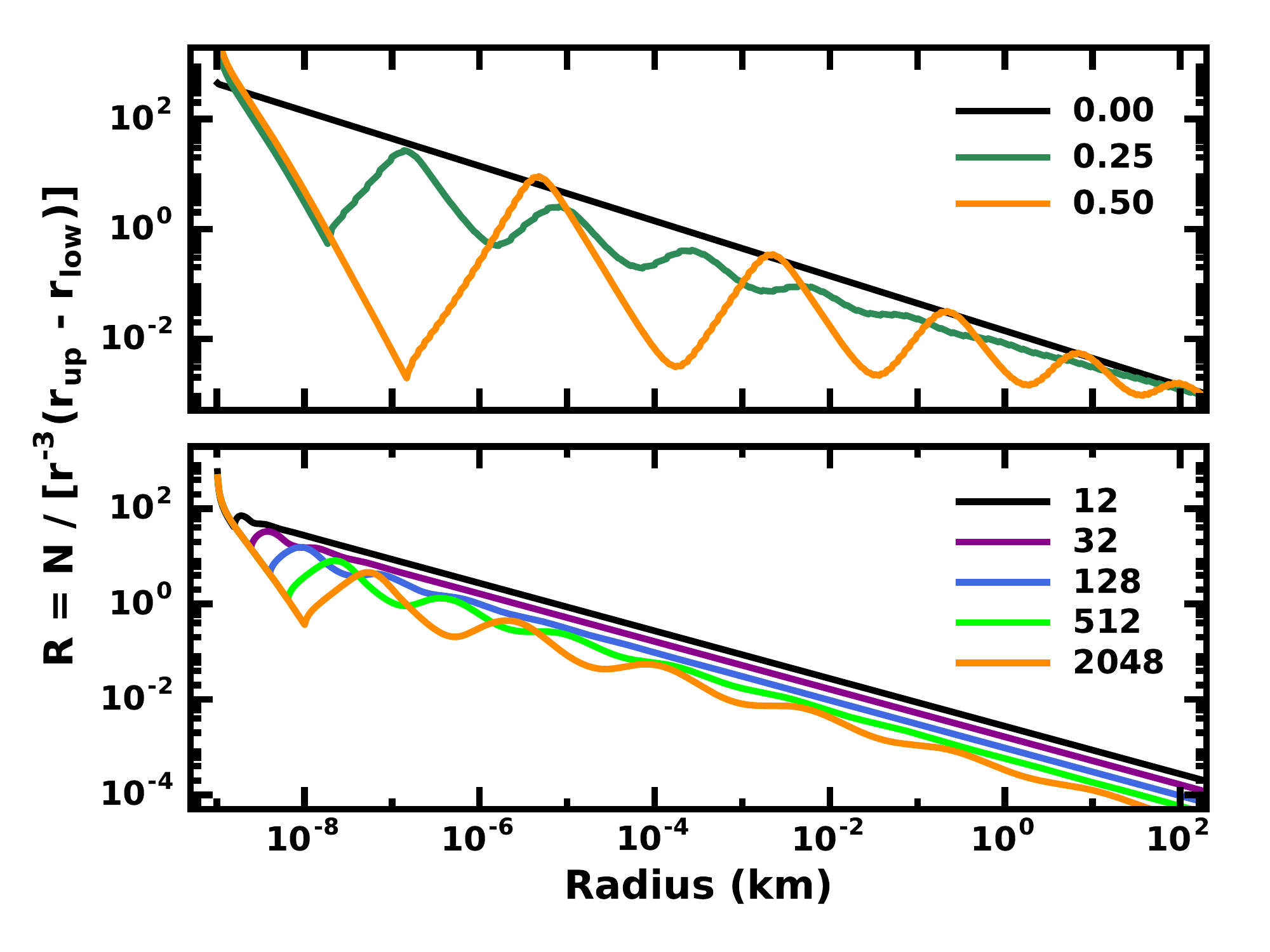}
\vskip -2ex
\caption{\label{fig: wave1}
Equilibrium $R(r)$ derived from the analytic model described in the text.
{\it Lower panel:} Results for constant \qdstar\ and the values 
of \vsqd\ listed in the legend. Models with larger \vsqd\ have 
wavier size distributions.
{\it Upper panel:} Results for constant \qdstar\ = 8 (black curve) 
and $\qdstar\ \propto r^{e_q}$ with $e_q$ = 0.25 (green curve) and 
$e_q$ = 0.50 (orange curve).  Systems with larger $e_q$ have shallower
slopes and wavier size distributions than those with smaller $e_q$.
}
\end{center}
\end{figure}

\subsection{Simple Examples}
\label{sec: casc-an-ex1}

To calculate an equilibrium $n(r)$, we have several options. Setting \vsqd\ = 
constant for all sizes allows us to recover previous analytic results
\citep[e.g.,][]{dohn1969,will1994,obrien2003,pan2005}. With \rmin\ = 0,
$n(r)$ has the standard power-law slope $q$ = 3.5. When \rmin\ = 1~\mum, 
$n(r)$ is wavy with an overall slope $q$ = 3.5 \citep[Fig.~\ref{fig: wave1},
lower panel; see also][]{campo1994a,obrien2003,pan2005,wyatt2011}.  
The amplitude of the wave grows with increasing collision energy, from a 
modest wave when \vsqd\ = 12 to a pronounced wave when \vsqd\ = 2048.

When \qdstar\ is a function of particle radius, the characteristic slope of the 
size distribution and the waviness change dramatically \citep[e.g.,][]{obrien2003}.
For a simple power-law function $\qdstar \propto r^{e_q}$, the power-law slope is 
$q = (21 + e_q) / (6 + e_q)$ \citep{obrien2003,pan2012}.  In the strength regime 
of eq.~\ref{eq: qdstar}, standard values for $e_s \approx 0$ to $-0.5$ yield 
$q$ = 3.50--3.72.  In the gravity regime, $e_g \approx$ 1.2--1.4 implies $q \approx$ 
3.13--3.03; $n(r)$ is then shallower than the standard $q$ = 3.5 power-law.

The upper panel of Fig.~\ref{fig: wave1} illustrates the changing properties of 
$R(r)$ when $e_q$ is positive. For each example, we normalize the ratio of $Q_c$ 
to \qdstar\ at \vsqd\ = 8 for collisions between equal mass objects with $r = \rmax$. 
For all $r < \rmax$, \vsqd\ $>$ 8; collisions between smaller equal-mass objects 
always produce catastrophic collisions. The black curve in the Figure has $e_q$ = 0 
and the standard slope $q$ = 3.5. When $e$ = 0.25 (green curve), $n(r)$ is shallower 
with an average $q$ = 3.4. A system with a larger $e_q$ = 0.5 has an even shallower 
size distribution with an average $q$ = 3.3.  For $e_q > 0.5$, $n(r)$ is wavier and 
has a smaller average slope.  Once $e_q \gtrsim$ 1, $n(r)$ is dominated by large 
amplitude waves; the average slope only crudely characterizes the shape of the
size distribution.

The waviness of $n(r)$ clearly depends on $e_q$.  When \vsqd\ = 8 and $e_q$ = 0, waviness 
due to the small-size cutoff is negligible.  At larger $e_q$, \vsqd\ monotonically 
increases from 8 at $r = \rmax$ to $\vsqd\ = 8 (\rmax/\rmin)^{e_q}$ at $r = \rmin$. 
Adopting \rmax\ = 500~km and \rmin\ = 1~\mum, $\vsqd \approx$ 6700 ($5.7 \times 10^6$) 
at $r = \rmin$ for $e_q$ = 0.25 (0.50).  Compared to the examples where $\vsqd$ is 
a constant (e.g., \vsqd\ = 12--2048; Fig.~\ref{fig: wave1}, lower panel), the waves 
in the upper panel of Fig.~\ref{fig: wave1} are much larger. With large ranges in 
\vsqd\ likely in real systems, these examples show how the parameters in the relation 
for \qdstar\ impact the size distribution for constant collision velocity.

The slope and waviness of $n(r)$ also depend on how $v$ varies with particle radius 
\citep{pan2012}. When the collision velocity has a power-law relation, $v \propto r^{e_v}$, 
imposing mass conservation yields a slope $q = (21 + e_q - 2 e_v) / (6 + e_q - 2 e_v)$. 
With \vsqd\ as the basic parameter setting the slope, the twin factors of two in this 
expression come from the $v^2$ term; the negative sign results from $v_c^2$ in the 
numerator instead of the denominator of \vsqd. In the upper panel of Fig.~\ref{fig: wave1}, 
a system with \vsqd\ = 8 at $r = \rmax$ and $e_v$ = $e_q$ = 0 recovers $n(r)$ with the 
standard slope $q$ = 3.5.  Setting $e_v$ = $-e_q / 2$ = $-0.125$ (green curve) or = 
$-0.250$ (orange curve) yields the same wavy $n(r)$ and a shallower general slope from 
\rmin\ to \rmax. Adopting a positive $e_v$ results similarly wavy $n(r)$ with steeper 
slopes from \rmin\ to \rmax.

In these examples, the waves result from the lack of small particles with $r < \rmin$.
Relative to a size distribution that extends to $r$ = 0, particles at or just larger 
than the cutoff experience fewer destructive collisions and are relatively overabundant 
\citep{campo1994a,obrien2003}. In turn, the higher frequency of these overabundant 
particles results in the destruction of more particles much larger than the cutoff, 
generating a minimum in $n(r)$. As these maxima and minima alternate among progressively 
larger particles, the amplitude of the waviness gradually decreases.  For any \qdstar, 
larger \vc\ results in a larger overabundance at the cutoff, which in turn creates larger 
waves among larger particles.

\subsection{Physical Examples}
\label{sec: casc-an-ex2}

In most applications, relations for \qdstar\ and $v(r)$ are rarely single power-laws.
Physically plausible expressions for \qdstar\ consist of at least two power-laws
(e.g., eq.~\ref{eq: qdstar}) and may have an additional component that depends on the
collision velocity \citep[e.g.,][]{benz1999,lein2012}. Throughout calculations of 
planet formation, the relative velocities of particles often depend on particle radius 
in a complicated way \citep[e.g.,][]{kb2015a,kb2017a}. During the late stages of planet
formation, the relative velocities are often much simpler, with small particles having
somewhat larger velocities than large particles \citep[e.g.,][]{gold2004,kb2008,kb2010,
kb2017a}.  Here we consider how the equilibrium $n(r)$ reacts to the double power-law 
expression for \qdstar\ with $v$ independent of $r$.

To explore how the equilibrium $n(r)$ depends on the \qdstar\ relation, we consider 
variations on the standard parameters. For these examples, we adopt \rmax\ = 100~km and 
$v$ = 1--2~\kms; $Q_s \approx 10^3 - 10^8$~\ergg~cm$^{-e_s}$, $e_s \approx$ $-0.5$--0.0, 
$Q_g \approx$ 0.1--3~erg~g$^{-2}$~cm$^{3-e_g}$, and $e_g \approx$ 1.1--1.5 \citep[e.g.,][and 
references therein]{asphaug1996,benz1999,obrien2003,lohne2008,wyatt2008,kb2008,kb2010,
weiden2010b,kb2012,lein2012}. We construct a large grid of equilibrium size distributions 
and examine how the shape depends on the collision velocity and the \qdstar\ parameters. 

\begin{figure}[t!]
\begin{center}
\includegraphics[width=4.5in]{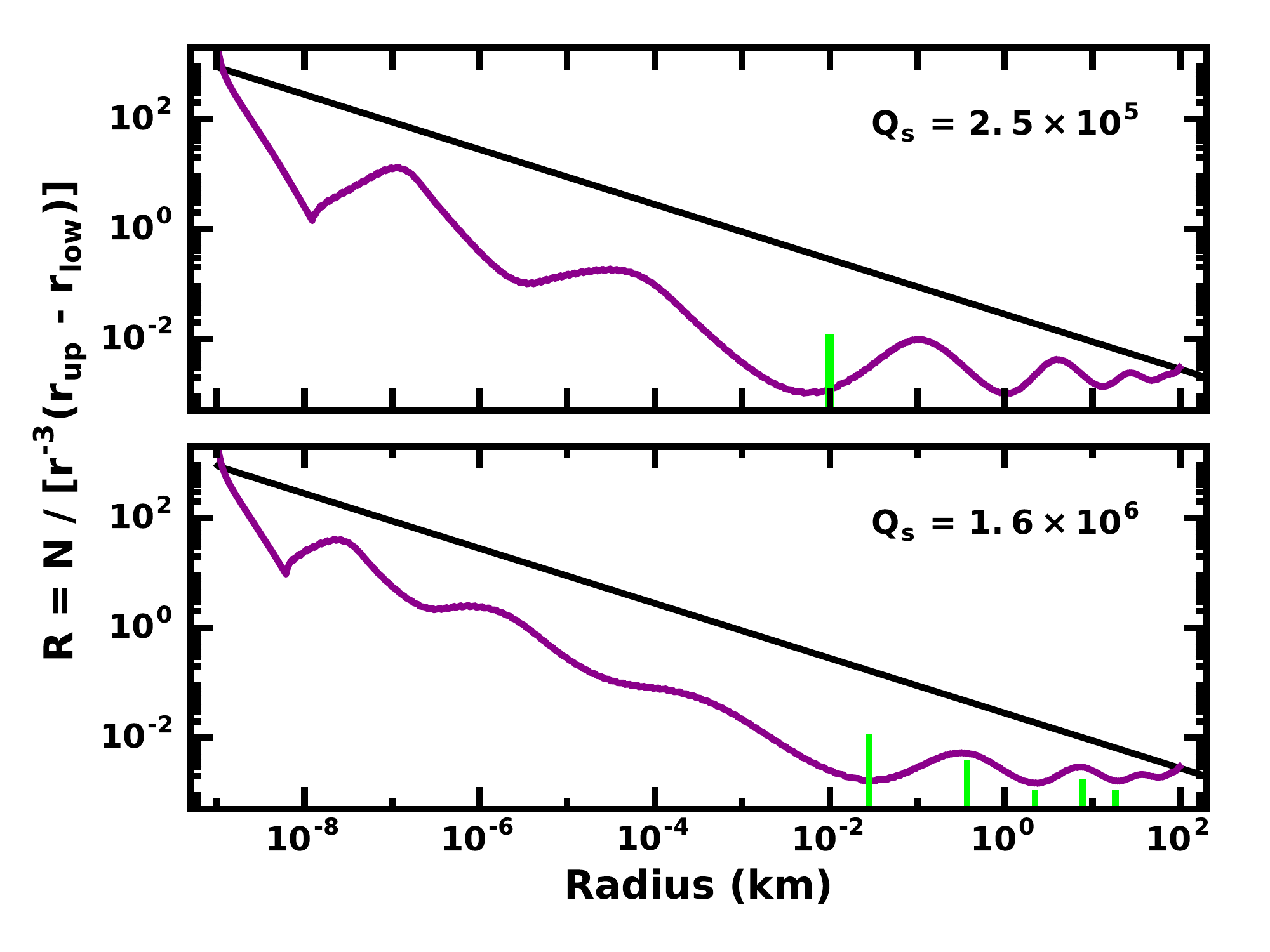}
\vskip -2ex
\caption{\label{fig: wave2}
As in Fig.~\ref{fig: wave1} for analytic models with $v$ = 1~\kms\ and 
a physically plausible $\qdstar(r)$ (eq.~\ref{eq: qdstar}, $e_s = -0.4$, 
$Q_g$ = 0.3~\ergg, and $e_g$ = 1.35).  Compared to models with constant 
\qdstar\ = 8 (black line in each panel), systems with a standard relation 
for \qdstar\ have wavier size distributions (purple curves). Waviness 
increases as $Q_s$ decreases. The long vertical green line in each panel shows
the position of the transition radius. Shorter green lines in the lower panel
indicate the predicted positions of peaks and valleys \citep[e.q., 
Eqs.~\ref{eq: r-peak}--\ref{eq: r-valley};][]{obrien2003}.
}
\end{center}
\end{figure}

Within our grid, the shape of $R(r)$ is most sensitive to $Q_s$ (Fig.~\ref{fig: wave2}).  
In the lower panel, the wave from the small size cutoff at 1~\mum\ is pronounced: there 
is a clear minimum in $R(r)$ at 6~\mum\ followed by a maximum at 22~\mum. The next set of 
features -- with a minimum at 300~\mum\ and a maximum close to 1~mm -- have a smaller 
amplitude and a longer wavelength.  Although the third set -- minimum close to 3~cm and 
a maximum at 30~cm -- continues the progression of smaller amplitude and longer wavelength,
the final set of features at 0.01--100~km has a larger amplitude and a shorter 
wavelength (in log space). 

At the large sizes in Fig.~\ref{fig: wave2}, the minimum in \qdstar\ at the
transition radius $r_t$, $Q_t(r = r_t)$, generates the second set of waves 
in $n(r)$ \citep[e.g.,][]{obrien2003}.  Generally, the bulk strength is 
constant or decreases with radius ($e_s \le$ 0).  The transition radius is then 
\begin{equation}
r_t = \left ( \frac{-e_s ~ Q_s}{e_g ~ \rho ~ Q_g} \right )^{1/(e_g - e_s)} ~ . 
\label{eq: rtrans}
\end{equation}
For particles with $r \approx r_t$, the minimum in \qdstar\ produces a minimum (or
`valley') in $R(r)$ where particles are easiest to break \citep{obrien2003}. 
Particles with these sizes remove less material in collisions with larger particles, 
generating a local maximum (or `peak') in $R(r)$. In their analytic derivation, 
\citet{obrien2003} express the positions of the peaks $r_p$ and valleys $r_v$ in 
terms of $Q_t$, $r_t$, and \vc:
\begin{equation}
r_p = \left ( \frac{2 Q_t}{v^2} \right )^{-1 / (e_g + 3)} ~ r_t^{e_g / (e_g + 3)} ~ r_v^{3 / (e_g + 3)} ~ ,
\label{eq: r-peak}
\end{equation}
and 
\begin{equation}
r_v = \left ( \frac{2 Q_t}{v^2} \right )^{-1 / (e_g + 3)} ~ r_t^{e_g / (e_g + 3)} ~ r_p^{3 / (e_g + 3)} ~ .
\label{eq: r-valley}
\end{equation}
With a valley at $r_v = r_t$,
the first relation establishes the first peak with $r > r_t$ at 
$r_{p,1} = (2 Q_t / v^2)^{-1 / (e_g + 3)} ~ r_t$. Eq.~\ref{eq: r-valley} 
then sets the first valley at 
$r_{v,1} = (2 Q_t / v^2)^{-2 / (e_g + 3)} ~ r_t^{5 / (e_g + 3)}$. 
Switching between Eqs.~\ref{eq: r-peak} and \ref{eq: r-valley} yields a sequence of
positions for peaks and valleys with $r > r_t$.

In the lower panel of Fig.~\ref{fig: wave2}, the long vertical green line at $r$ = 28~m 
marks the transition radius for the adopted fragmentation parameters. The minimum at 
33~m in the analytic $R(r)$ matches this prediction. Following this minimum, the 
analytic size distribution has a set of maxima (0.4~km, 7~km, and 35~km) that 
closely match the peaks at 0.4~km, 7.7~km, and 33~km predicted from Eq.~\ref{eq: r-peak}.
Valleys at 2.3~km, 19~km, and 53~km similarly lie close to the predictions of 2.2~km,
18~km, and 49~km. Comparisons of sequences for other analytic $R(r)$ with 
$Q_s \gtrsim 10^6$~erg~g$^{-1}$~cm$^{0.4}$ yield an excellent correspondence between 
the peaks and valleys in analytic size distributions with the predictions of 
Eqs.~\ref{eq: r-peak}--\ref{eq: r-valley}.

When $Q_s \lesssim 10^6$~erg~g$^{-1}$~cm$^{0.4}$, the radius of the minimum in the 
analytic $R(r)$ is generally smaller than the transition radius $r_t$
(Fig.~\ref{fig: wave2}, upper panel). The positions of peaks and valleys at $r > r_t$
also fall at smaller radii than predicted from Eqs.~\ref{eq: r-peak}--\ref{eq: r-valley}.
When $Q_s$ is small, the smaller $Q_t$ results in a longer wavelength between peaks and
valleys on either side of $r_t$ (Eqs.~\ref{eq: r-peak}--\ref{eq: r-valley}). There is 
then an interference between the waves generated by the small-size cutoff and those 
caused by the minimum in \qdstar. This interference displaces the waves from the 
predictions of the \citet{obrien2003} model.  

\begin{figure}[t!]
\begin{center}
\includegraphics[width=4.5in]{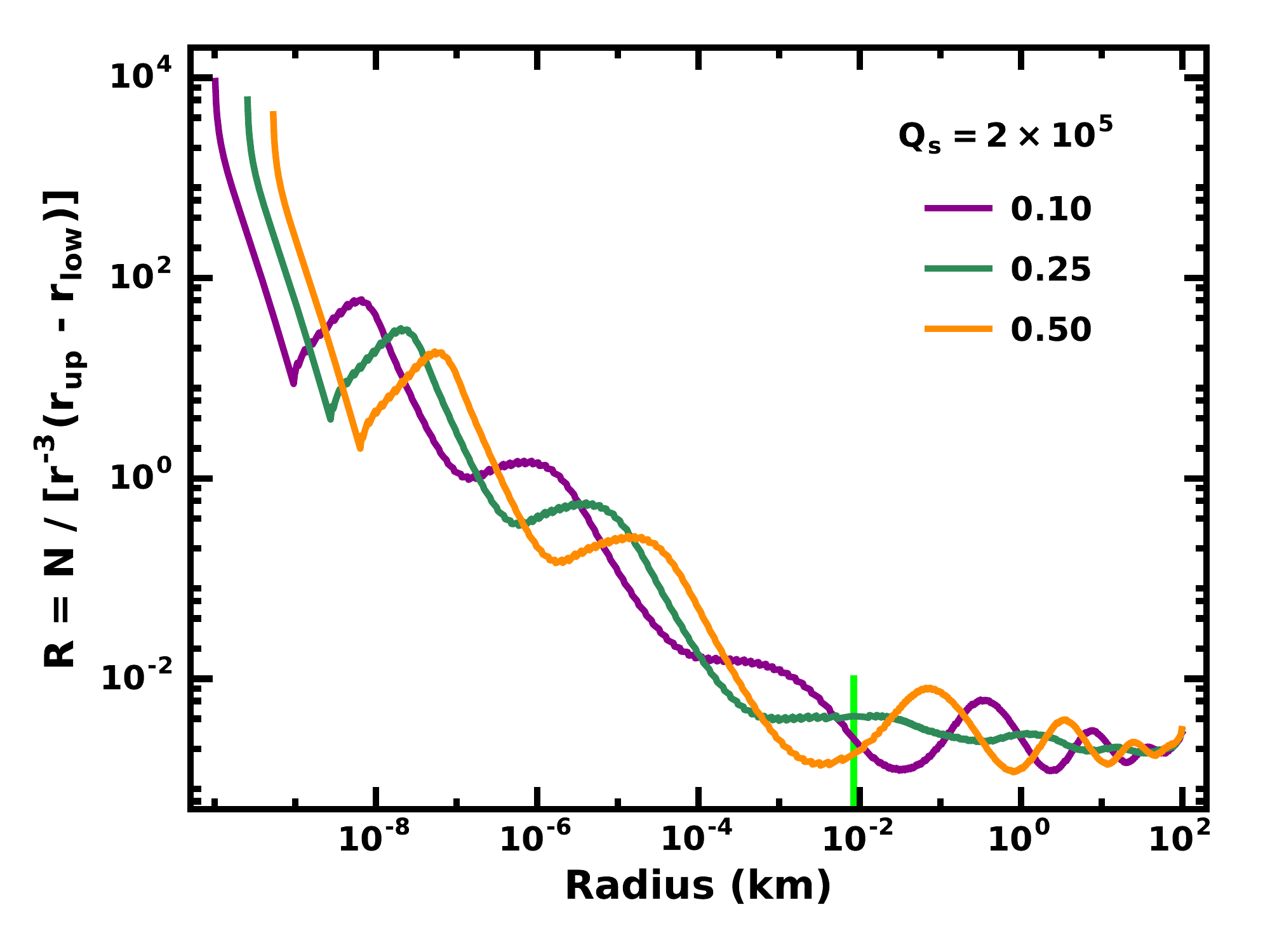}
\vskip -2ex
\caption{\label{fig: wave3}
Comparison of $R(r)$ for analytic models with 
$Q_s = 2 \times 10^5$~erg~g$^{-1}$~cm$^{-0.4}$, $e_s = -0.4$, 
$Q_g$ = 0.3~\ergg, and $e_g$ = 1.35 for various \rmin\ (in microns)
as listed in the legend. The vertical green line indicates the position
of the transition radius, $r_t$ = 8.4~m. Despite the similarity in the size 
distributions at small sizes, interference between the small-size cutoff 
and the transition radius produce small waves when \rmin\ = 0.25~\mum.
For $r \gtrsim r_t$, different choices for \rmin\ yield larger waves.
}
\end{center}
\end{figure}

For some combinations of \rmin\ and the \qdstar\ parameters, the waves from \rmin\ and 
\rt\ interfere destructively, yielding size distributions with slopes similar to those 
in Fig.~\ref{fig: wave2} but with much smaller waves (Fig.~\ref{fig: wave3}). In this
example, $Q_s = 2 \times 10^5$~erg~g$^{-1}$~cm$^{0.4}$, $e_s = -0.4$, 
$Q_g$ = 0.3~erg~g$^{-2}$~cm$^{1.65}$, and $e_g$ = 1.35. Setting \rmin\ = 0.1~\mum\ yields
a steep slope and a large-amplitude wave from the small-size cutoff. After reaching a 
minimum at $r$ = 30~m, $R(r)$ is fairly flat; waves from the transition in \qdstar\ are 
obvious. Aside from a shift to larger sizes, setting \rmin\ = 0.5~\mum\ results in a nearly 
identical size distribution: the morphology of the waves is identical for small and larger 
sizes. Although choosing \rmin\ = 0.25~\mum\ makes little difference in $R(r)$ at $r \lesssim$ 30~cm, 
the amplitude of waves at larger sizes is much smaller than for other \rmin. In this example, 
the waves from the small-size cutoff nearly cancel those from the transition in \qdstar\ at 
large $r$; $R(r)$ is then nearly flat from $r$ = 1~m to $r$ = 100~km.

Understanding the origin of the vanishing waves with \rmin\ = 0.25~\mum\ is 
straightforward.  For the adopted \qdstar\ parameters in Fig.~\ref{fig: wave3}, 
the transition radius is $r_t$ = 8.4~m (eq.~\ref{eq: rtrans}) independent of \rmin. 
When \rmin\ = 0.5~\mum\ (Fig.~\ref{fig: wave3}, orange curve), waves from the small-size 
cutoff produce a minimum (maximum) in $R(r)$ at 3--4~m (60--70~m).
For \rmin\ = 0.1~\mum\ (Fig.~\ref{fig: wave3}, purple curve), these features occur at 
10--100~cm. While the small-size cutoff clearly places minima on either size of $r_t$, 
maxima are well-displaced from $r_t$.  Comparing the maxima at $\sim$ 1~m (purple curve) 
and at 60--70~m (orange curve), it is clear that a model with some \rmin\ between 
0.1~\mum\ and 0.5~\mum\ will yield a maximum close to the transition radius at 8.4~m.
Having a maximum from the small-size cutoff at the transition radius approximately 
`cancels' the minimum produced from the transition radius, yielding a size distribution
with little waviness for $r \gtrsim r_t$.

\begin{figure}[t!]
\begin{center}
\includegraphics[width=4.5in]{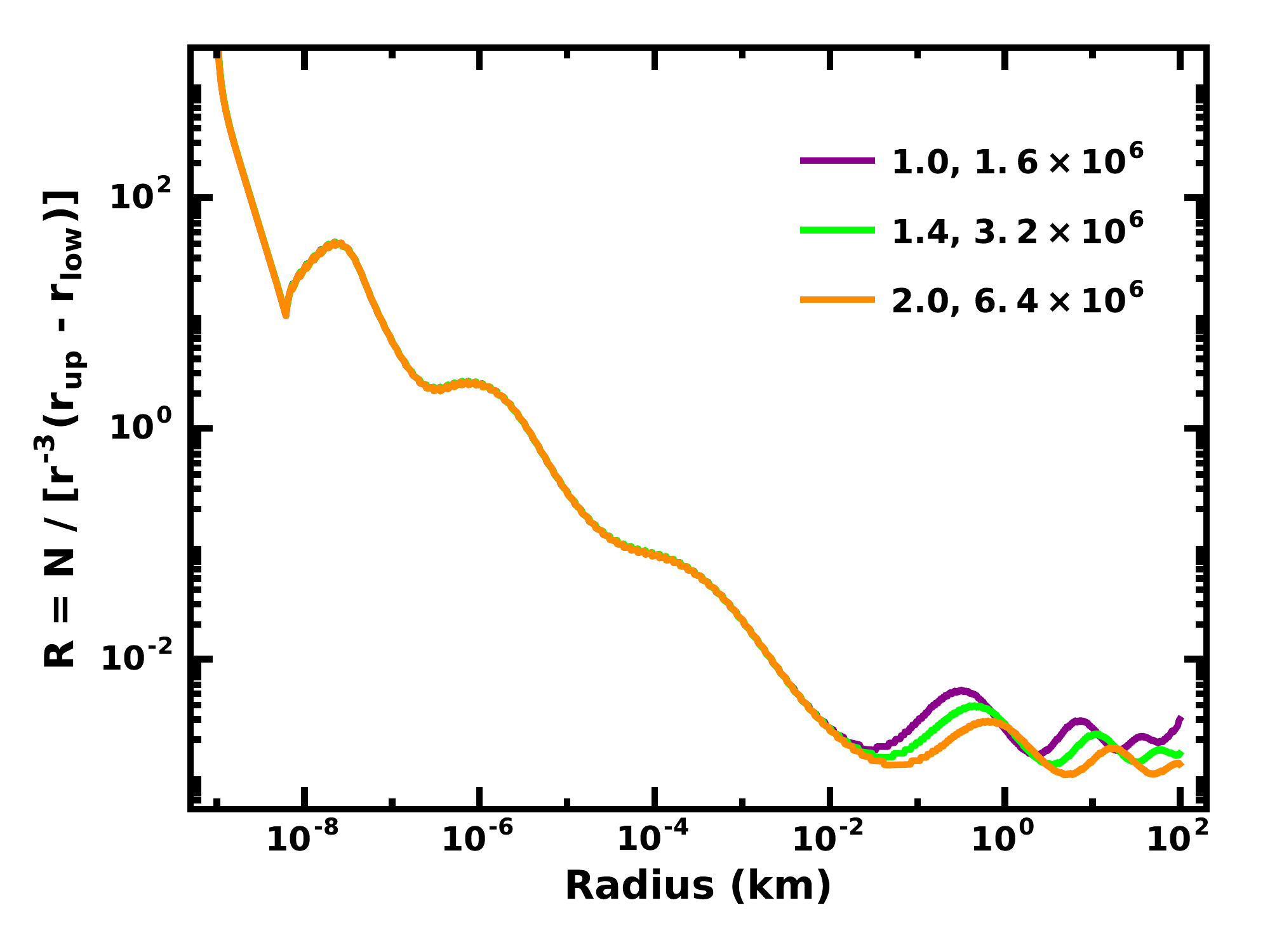}
\vskip -2ex
\caption{\label{fig: wave4}
As in Fig.~\ref{fig: wave3} for analytic models with $e_s = -0.4$, 
$Q_g$ = 0.3~\ergg, and $e_g$ = 1.35 for various \vc\ and $Q_s$ as listed 
in the legend. For small particles, it is possible to produce identical 
$R(r)$  with different $v$ and $Q_s$.  The displacements of the peaks and 
valleys in $R(r)$ grow at larger sizes.
}
\end{center}
\end{figure}

Fig.~\ref{fig: wave4} illustrates how the morphology of the size distribution depends
on \vsqd. For fixed \qdstar, raising \vc\ from 1~\kms\ to 1.4~\kms\ to 2~\kms\ increases 
the amplitude and wavelength of the waves.  At small sizes, it is possible to scale the
bulk component of the strength to compensate for the larger collision velocity and generate
identical $R(r)$ from 0.1~\mum\ to $r_t$. In these examples, the transition
radius moves from $r_t \approx$ 30~m ($Q_s = 1.6 \times 10^6$~erg~g$^{-1}$~cm$^{0.4}$) to
$r_t \approx$ 40~m ($Q_s = 3.2 \times 10^6$~erg~g$^{-1}$~cm$^{0.4}$) to
$r_t \approx$ 60~m ($Q_s = 6.4 \times 10^6$~erg~g$^{-1}$~cm$^{0.4}$), shifting the peaks
and valleys to larger $r$. With a smaller $Q_t$, models with the smaller \vc\ have the
smaller wave amplitude and wavelength at $r \gtrsim r_t$. For the parameters in 
Fig.~\ref{fig: wave4}, however, the difference in amplitudes and wavelengths is less than
10\%.

For the \qdstar\ parameters we consider, the shape of the equilibrium size distribution 
is fairly insensitive to $e_s$ and the gravity component of \qdstar.  Because the shape
depends on the small size cutoff and the transition radius, it is usually possible to 
select a $Q_s, e_s$ pair that yields a similar $R(r)$ to one with our `standard' choice, 
$e_s = -0.4$, and a slightly different $Q_s$.  
When $r \gtrsim$ 1~km, the \citet{benz1999} choices for the gravity component,
$Q_g$ = 2.1~erg~g$^{-2}$~cm$^{1.81}$ and $e_q$ = 1.19, yield similar binding energies as
the \citet{lein2012} choices,
$Q_g$ = 0.3~erg~g$^{-2}$~cm$^{1.65}$ and $e_q$ = 1.35. Near typical transition radii,
$r_t \approx$ 0.01--0.1~km, the \citet{lein2012} parameters yield somewhat smaller 
\qdstar\ than the \citet{benz1999} parameters and thus somewhat smaller $r_t$ and $Q_t$.
Although the shapes of size distributions using these two sets of parameters are then
different for fixed bulk strength, the differences are small compared to the variations
illustrated in Figs.~\ref{fig: wave2}--\ref{fig: wave4}. 

Equilibrium $R(r)$ appropriate for TNOs are also insensitive to \rmax. 
For \rmin\ = 1~\mum\ and \rmax\ = 50--200~km, the shape of the size distribution 
from 1~\mum\ to 40--50~km is independent of \rmax. For larger sizes, the shape and
degree of waviness depend on the \qdstar\ parameters and the collision velocity. 
However, the changes are relatively small compared to the magnitude of the waviness
at $r \lesssim$ 30--50~km and the variations as a function of \vc\ and the parameters
for \qdstar.

\subsection{Application to TNOs}
\label{sec: casc-an-app}

To apply the analytic model to observations of TNOs, we make a slight adjustment.
Planetesimals with radii $r \le r_b$ have the equilibrium size distribution; those 
with $r > r_b$ follow a power-law with index $q$ = 5--6. With this modification, we 
include large TNOs whose size distribution is fixed over the lifetime of the Solar 
System. 

\begin{figure}[t!]
\begin{center}
\includegraphics[width=4.5in]{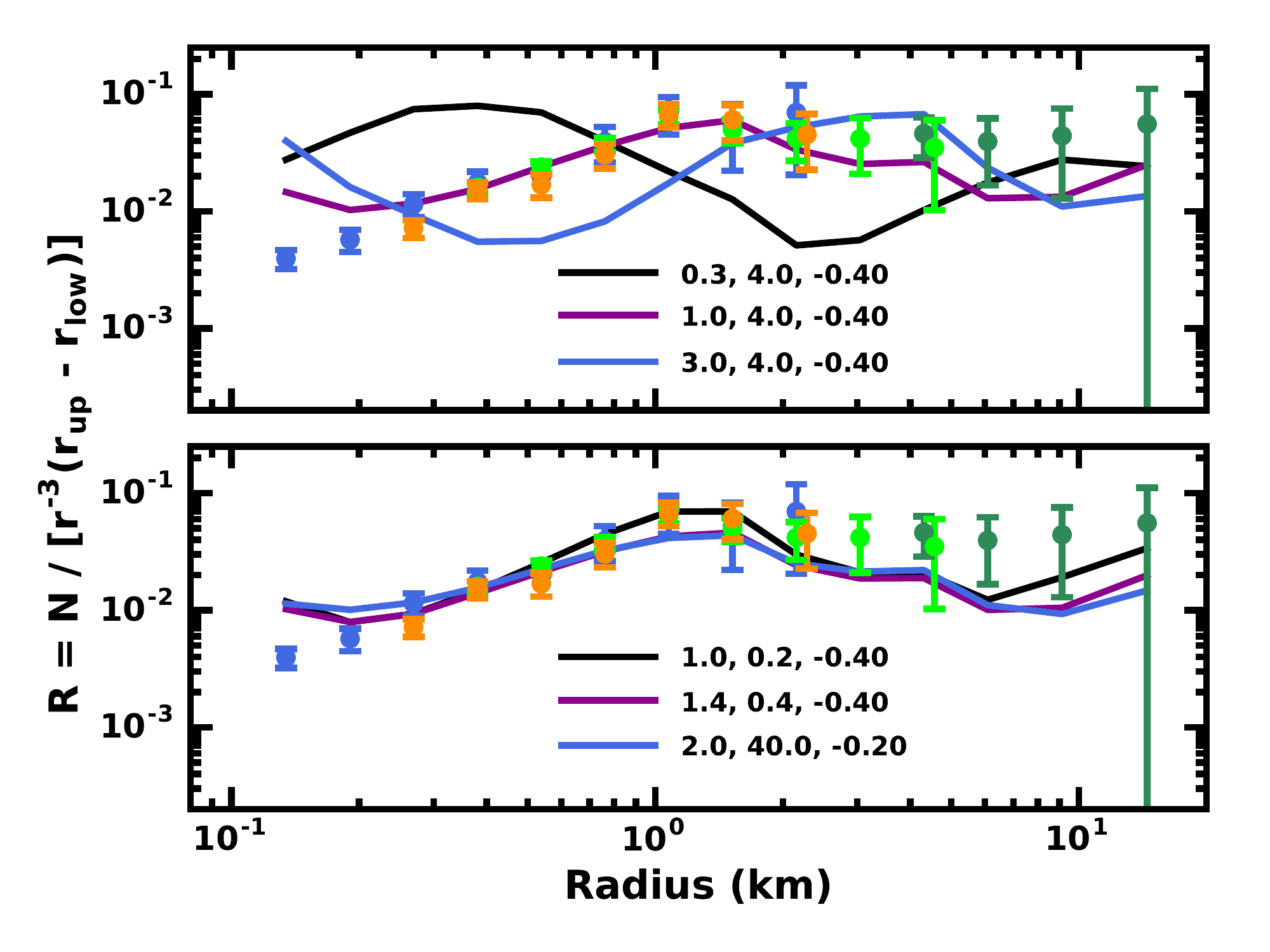}
\vskip -2ex
\caption{\label{fig: obs2}
Comparison of the observed size distribution inferred from \nh\ cratering data 
\citep[data color-coded as in Fig.~1, ][]{singer2019} with equilibrium size 
distributions derived from an analytic model where \qdstar\ is a function of 
radius. Some models have been adjusted vertically for clarity. For all models, 
$Q_g$ = 0.3~erg~cm$^{1.65}$~g$^{-2}$ and $e_g$ = 1.35.
{\it Lower panel:} legend indicates numerical values for \vc\ in \kms, 
$Q_s$ in units of $10^5$~erg~g$^{-1}$~cm$^{e_s}$, and $e_s$.
{\it Upper panel:} for calculations with \vc\ = 1.4~\kms, legend indicates numerical 
values for \rmin\ in \mum, $Q_s$ in units of $10^4$~erg~g$^{-1}$~cm$^{0.4}$, and $e_s$.
}
\end{center}
\end{figure}

Finding the best match to the \nh\ observations requires a search algorithm. 
We consider each model within the grid described in the previous section, where
\vc\ = 1--2~\kms, \rmin\ = 0.1--10~\mum, $Q_s = 10^3 - 10^8$~\ergg~cm$^{-e_s}$, 
$e_s$ = $-0.5$--0.0, $Q_g$ = 0.3~erg~g$^{-2}$~cm$^{1.65}$, and $e_g$ = 1.35.
With little change in the shape of $R(r)$ as a function of \rmax, we fix $r_b$ = 100~km 
and derive equilibrium size distributions for $r < r_b$.  For each set of parameters, 
we adjust a scale factor to derive the best match to the \nh\ data. Among the complete 
set of models, we search for those that minimize $\chi^2$ using the quoted errors from 
\citet{singer2019}. 
The best models have a typical $\chi^2$ per degree of freedom
of 3--4. While these fits do not attain a $\chi^2$ per degree of freedom of $\sim$ 1,
the best fits are significantly better than the worst fits with $\chi^2$ per degree of
freedom $\gtrsim$ 100--1000.

Fig.~\ref{fig: obs2} compares six equilibrium size distributions with the \nh\ data.
In the lower panel, calculations with collision velocities, \vc\ = 1 (black curve) 
and 1.4~\kms\ (purple curve), and small values of $Q_s$ yield size distributions that 
provide satisfactory matches to the data at 0.2--10~km. When the collision velocity 
is 2~\kms\ (blue curve), models with larger $Q_s$ and smaller $e_s$ match the data as
well as those with smaller $Q_s$.  Although higher velocity models match better at small 
sizes, they generate a somewhat larger valley at $r$ = 10--20~km which may not be present 
in the data.

For all collision velocities considered in the lower panel of Fig.~\ref{fig: obs2}, we
searched for reasonable matches with significantly different values of $Q_s$ and $e_s$
than those listed in the panel. When \vc\ = 1--1.4~\kms, model size distributions with 
smaller $Q_s$ have a minimum at 0.3--0.5~km and a steep rise to smaller sizes; at larger 
sizes, the waviness is larger than observed.  Systems with factor of 100 larger $Q_s$ 
have a fairly flat $R(r)$ with negligible waves compared to the \nh\ data. When \vc\ = 
2~\kms, size distributions with small $Q_s$ rarely have a minimum at 0.1~km and are 
too wavy to match the \nh\ data.  Strong ice particles have equilibrium size distributions 
that are not wavy enough to match the \nh\ data.

Varying \rmin\ from the standard 1~\mum\ does not allow better matches to the data 
(Fig.~\ref{fig: obs2}, upper panel). For a broad range of $Q_s$ and $e_s$, reducing 
\rmin\ to 0.1--0.3~\mum\ eliminates the drop in $R(r)$ from 0.5--0.6~km to 0.1~km. 
Similarly, increasing \rmin\ to 3--10~\mum\ also makes it harder to match the data at 
the smallest sizes. 

The amplitude and position of the wave at $r$ = 0.1--1~km are sensitive to $Q_s$ and $e_s$. 
In any model, 10\% changes to $Q_s$ and 5\% changes to $e_s$ have a small impact on the 
wave.  Larger modifications either make the amplitude of the wave smaller or shift it 
to smaller or larger $r$. For collision velocities \vc\ = 1--2~\kms\ and the range of 
\qdstar\ parameters we studied, making the wave amplitude or wavelength larger is 
impossible. Models with smaller and larger values for \vc\ also tend 
to provide poorer matches to the data.

Adopting velocity laws with shallow power-laws, $v \propto r^{e_v}$ with $e_v <$ 0.01--0.02,
changes these results insignificantly. In our experiments, larger variations in 
collision velocity from the smallest to the largest objects in the grid generate 
much poorer matches to the \nh\ data. While it may be possible to identify $v(r)$ 
relations that allow better matches to the \nh\ data \citep[e.g., the possibilities
discussed in][]{pan2012}, our analysis suggests a constant velocity among TNOs provides
a better match to the \nh\ observations.

Given the uncertainties in and the simplicity of the analytic model, it provides a 
reasonable match to the \nh\ data. Compared to standard collisional disruption models
where the slope for $r \lesssim$ 100~km is 3.5, the match to the \nh\ data at 0.1--10~km
with the analytic equilibrium model is impressive (compare with Fig.~\ref{fig: obs1}).
Despite this success, the analytic model does not include cratering collisions which 
remove less than half the mass from the target. In most numerical simulations, cratering 
enables significant mass loss from the system and sometimes competes with catastrophic 
collisions in generating the collisional cascade. With no analytic model for cratering, 
we rely on numerical simulations to consider whether including cratering in a collision 
algorithm can maintain the reasonable match between the analytic model and the \nh\ data.

\vskip 6ex
\section{COLLISIONAL CASCADES: NUMERICAL RESULTS}
\label{sec: casc-num}

\subsection{Methods}
\label{sec: casc-num-meth}

To calculate the evolution of KBOs with different sizes, we run a series of numerical 
simulations with \orch, an ensemble of computer codes designed to track the accretion, 
fragmentation, and orbital evolution of solid particles ranging in size from a few microns 
to thousands of km \citep{kenyon2002,bk2006,kb2008,bk2011a,bk2013,kb2016a,knb2016}.  
Using the coagulation component of \orch, we start with an ensemble of solids with minimum
radius \rmin\ and maximum radius \rmax\ orbiting the Sun within a single annulus having an
inner radius $a_{in}$ and outer radius $a_{out}$. The solids have mass density $\rho$, 
initial total mass \M0, and initial surface density $\Sigma_0$. 

To evolve the size and velocity distributions of solids in time, \orch\ derives collision
rates and outcomes with standard particle-in-a-box algorithms. Systems start with an 
initial size distribution $n(r) \propto r^{-q}$ in discrete bins with a mass spacing factor
$\delta = m_{i+1}/m_i$ between adjacent bins. When a pair of solids collide, the mass of 
the merged object is \begin{equation}
m = m_1 + m_2 - \mesc ~ .
\label{eq: msum}
\end{equation}
The mass of debris ejected in a collision is
\begin{equation}
\mesc =0.5 ~ (m_1 + m_2) \left ( \frac{Q_c}{Q_D^*} \right)^{b_d} ~ .
\label{eq: mesc}
\end{equation}
The exponent $b_d$ is a constant of order unity \citep[e.g.,][]{davis1985,weth1993,kl1999a,
benz1999,obrien2003,koba2010a,lein2012}. 

To place the debris in the grid of mass bins, we set the mass of the
largest collision fragment as
\begin{equation}
\mmaxd = m_{l,0} ~ \left ( \frac{Q_c}{Q_D^*} \right)^{-b_l} ~ \mesc ~ ,
\label{eq: mlarge}
\end{equation}
where $m_{l,0} \approx$ 0.01--0.5 and $b_l \approx$ 0--1.25
\citep{weth1993,kb2008,koba2010a,weid2010}. When $b_l$ is large,
catastrophic (cratering) collisions with $Q_c \gtrsim \qdstar$ 
($Q_c \lesssim \qdstar$) crush solids into smaller fragments.  
Lower mass objects have a differential size distribution 
$N(r) \propto r^{-q_d}$. After placing a single object with mass 
\mmaxd\ in an appropriate bin, we place material in successively 
smaller mass bins until (i) the mass is exhausted or (ii) mass is 
placed in the smallest mass bin. Any material left over is removed 
from the grid.

In most calculations, we assume that the orbital $e$ and $i$ are constant 
with time.  Otherwise, we derive orbital evolution due to collisional damping 
from inelastic collisions and gravitational interactions.  For inelastic and 
elastic collisions, we follow the statistical, Fokker-Planck approaches of 
\citet{oht1992} and \citet{oht2002}, which treat pairwise interactions (e.g., 
dynamical friction and viscous stirring) between all objects.  We also compute 
long-range stirring from distant oligarchs \citep{weiden1989}.

Our solutions to the evolution equations conserve mass and energy to machine
accuracy. Typical calculations require several 12~hr runs on a system with 
56 cpus; over the $10^6$--$10^8$ timesteps in a typical 2--4~Gyr run, calculations 
conserve mass and energy to better than one part in $10^{10}$.

Although nearly all other numerical treatments of KBO evolution adopt a variant of 
the particle-in-a-box algorithm for collision rates \citep[e.g.,][and references 
therein]{benavidez2009,fraser2009b,schlicht2011,campo2012,schlicht2013}, the details 
of deriving collision outcomes often differ from one investigation to the next. 
To make clear connections with previous calculations, we first consider a set of 
calculations with standard starting conditions.  This analysis also allows us to 
understand the relationships between the features in the size distributions and 
the initial conditions and various input parameters.  We then examine how to build 
a numerical cascade model that generates the size distribution derived from Charon 
impactors with \nh.

\subsection{Evolution of the Size Distribution for Standard Parameters}
\label{sec: casc-num-std}

In models 1--4 (Table~\ref{tab: model-pars}), we explore parameter spaces considered in 
previous publications \citep{obrien2003,kb2008,fraser2009b,benavidez2009,kb2010,kb2012,
campo2012,schlicht2013}. 
In a single annulus at 30--60~au that spans the Kuiper belt, the large initial mass 
(45~\mearth) guarantees evolution towards an equilibrium size distribution in 0.5--1~Gyr. 
The choice for the mass resolution, $\delta$ = 1.12, should yield smooth size distributions 
with relatively little noise \citep[e.g.,][]{kb2016a}.  For a system with an initial \rmax\ = 
500~km and collision velocity \vc\ = 1~\kms, we choose four sets of fragmentation parameters 
that result in destructive collisions for objects with mass density $\rho$ = 1.5~\gcmc\ and 
$1~\mum\ \lesssim r \lesssim$ 100--200~km.  The initial size distribution has a steep slope, 
$q$ = 5.5, for $r \ge r_l$ with $r_l$ = (a) 1~km, (b) 3~km, (c) 10~km, (d) 30~km, or (e) 100~km. 
In all calculations, $r_l = r_s$; there is no intermediate size population between the steep
power-law at large sizes ($r \gtrsim r_l$) and the shallower power-laws at small sizes 
($r \lesssim r_s$).  At smaller sizes, we consider initial slopes with integer values between 
$-3$ and $3$ inclusive. Most of the initial mass is concentrated in size bins with $r \approx r_l$; 
destructive collisions initiate a robust cascade for $r \le$ 100~km. Although growth is possible 
for $r \gtrsim$ 200~km, the large collision velocity and small mass guarantees modest evolution 
in the population of the largest objects over 1--5~Gyr.

\begin{deluxetable}{lcccccccccc}
\tablecolumns{11}
%\tablewidth{10cm}
\tabletypesize{\small}
\tablenum{1}
\tablecaption{Input Parameters for Coagulation Calculations\tablenotemark{a}}
\tablehead{
  \colhead{Model} &
  \colhead{$v$ (\kms)} &
  \colhead{~~$b_d$~~} &
  \colhead{~~$m_{l,0}$~~} &
  \colhead{~~$b_l$~~} &
  \colhead{$Q_s$} &
  \colhead{$e_s$} &
  \colhead{$Q_g$} &
  \colhead{$e_g$} &
  \colhead{$r_t$ (km)} &
  \colhead{$Q_t$ (\ergg)}
}
\label{tab: model-pars}
\startdata
1 & 1.0 & 1.0 & 0.2 & 0.0 & ~~$7 \times 10^7$~~ & ~~$-0.45$~~ & ~~2.10~~ & ~~1.19~~ & ~0.1668~~ & ~~$1.23 \times 10^6$~~ \\
2 & 1.0 & 1.0 & 0.2 & 0.0 & ~~$4 \times 10^6$~~ & ~~$-0.45$~~ & ~~2.10~~ & ~~1.19~~ & ~0.0291~ & ~~$1.54 \times 10^5$~~ \\
3 & 1.0 & 1.0 & 0.2 & 0.0 & ~~$2 \times 10^5$~~ & ~~$-0.45$~~ & ~~2.10~~ & ~~1.19~~ & ~0.0047~ & ~~$1.76 \times 10^4$~~ \\
4 & 1.0 & 1.0 & 0.2 & 0.0 & ~~$2 \times 10^5$~~ & ~~$-0.40$~~ & ~~0.30~~ & ~~1.35~~ & ~0.0084~ & ~~$1.78 \times 10^4$~~ \\
5 & 1.0 & 1.0 & 0.2 & 0.0 & ~~$4 \times 10^6$~~ & ~~$-0.40$~~ & ~~0.30~~ & ~~1.35~~ & ~0.0467~ & ~~$1.79 \times 10^5$~~ \\
6 & 1.0 & 1.0 & 0.2 & 0.0 & ~~$1 \times 10^4$~~ & ~~$-0.40$~~ & ~~0.30~~ & ~~1.35~~ & ~0.0015~ & ~~$1.76 \times 10^3$~~ \\
7 & 1.0 & 1.0 & 0.2 & 0.0 & ~~$1 \times 10^3$~~ & ~~$-0.40$~~ & ~~0.30~~ & ~~1.35~~ & ~0.0004~ & ~~$2.99 \times 10^2$~~ \\
8 & 1.0 & 1.0 & 0.2 & 0.0 & ~~$4 \times 10^6$~~ & ~~$-0.20$~~ & ~~0.30~~ & ~~1.35~~ & ~0.0887~ & ~~$7.51 \times 10^5$~~ \\
9 & 1.0 & 1.0 & 0.2 & 0.0 & ~~$4 \times 10^6$~~ & ~~$-0.00$~~ & ~~0.30~~ & ~~1.35~~ & \nodata  & ~~$4.00 \times 10^6$~~ \\
10 & 1.4 & 1.0 & 0.2 & 0.0 & ~~$4 \times 10^6$~~ & ~~$-0.40$~~ & ~~0.30~~ & ~~1.35~~ & ~0.0467~ & ~~$1.79 \times 10^5$~~ \\
11 & 2.0 & 1.0 & 0.2 & 0.0 & ~~$4 \times 10^6$~~ & ~~$-0.40$~~ & ~~0.30~~ & ~~1.35~~ & ~0.0467~ & ~~$1.79 \times 10^5$~~ \\
\enddata
\tablenotetext{a}{
The columns list $v$ the collision velocity;
$b_d$ the exponent in the relation between ejected mass and impact energy 
(Eq.~\ref{eq: mesc}); $m_{l,0}$ and $b_l$, parameters in the relation between the size of
the large remnant and the impact energy (Eq.~\ref{eq: mlarge}); 
$Q_s$, $e_s$, $Q_g$, and $e_g$, the parameters in the relation for \qdstar\ (Eq.~\ref{eq: qdstar});
$r_t$, the transition radius where \qdstar\ is a minimum (Eq.~\ref{eq: rtrans}; and
$Q_t$, the minimum \qdstar\ at $r_t$. 
All calculations are performed in an annulus with inner radius a$_{in}$ = 30~au, outer radius 
$a_{out}$ = 60~au, and total mass $M_0$ = 45~\mearth. Particles have minimum sizes \rmin\ =  1~\mum, 
initial maximum sizes \rmax\ = 500~km, and mass density $\rho$ = 1.5 \gcmc.
\ collide at velocity 
$v$ = 1 \kms.  
The initial size distribution is a power-law with slope $q_l = 5.5$ for radii 
$r \gtrsim r_l$ and slope $q_s$ for $r \lesssim r_s$.  The mass grid has $\delta$ = 1.12. In these
calculations, $r_s = r_l$, with $r_l$ = 1, 3, 10, 30, or 100~km; $q_s = -3$ to 3 inclusive. 
}
\end{deluxetable}

\begin{figure}[t!]
\begin{center}
\includegraphics[width=4.5in]{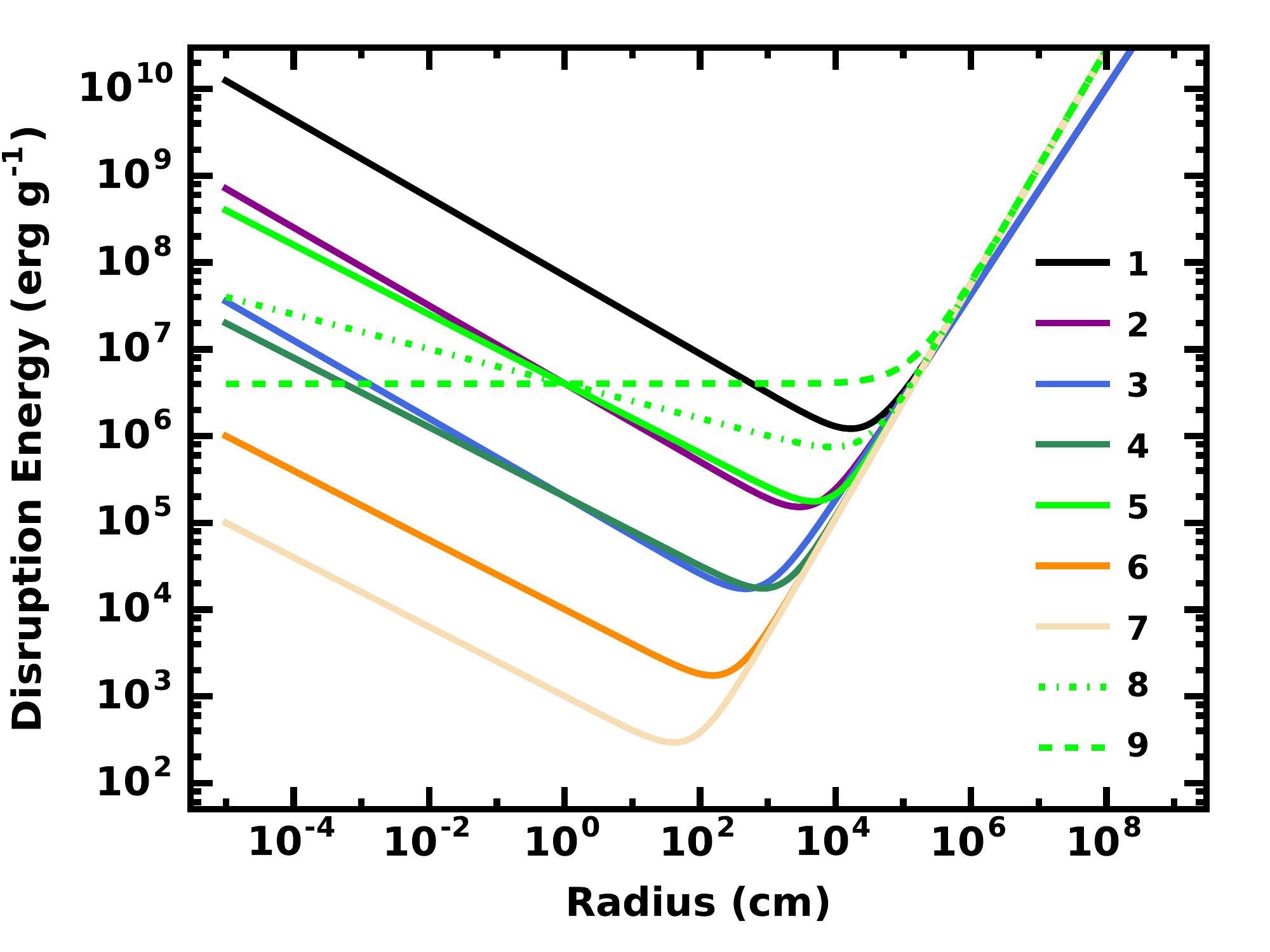}
\vskip -2ex
\caption{\label{fig: qdstar}
Variation of \qdstar\ as a function of radius for the fragmentation models 
summarized in Table~\ref{tab: model-pars}. For $r \gtrsim$ 1~km, the 
\citet{benz1999} and \citet{lein2012} parameters yield similar results for \qdstar.  
At smaller radii, the minimum in \qdstar\ at $r_t$ depends on $Q_s$ and $e_s$. In
the examples shown, $r_t \approx$ 1--100~m.
}
\end{center}
\end{figure}

The sets of fragmentation parameters in the first four rows of Table~\ref{tab: model-pars} span the 
range of possibilities derived from numerical simulations of high velocity collisions between solids
with an icy composition.  In the `strong ice' formulation of \citet{benz1999} for model (1), the 
binding energy in the strength and gravity regimes is comparable with basalt (Fig.~\ref{fig: qdstar},
black line). The `weak ice' parameters derived by \citet{lein2012} for model (4) yield similar 
results in the gravity regime ($r \gtrsim$ 1~km), but the binding energy of 1~cm objects is 350 
times smaller (Fig.~\ref{fig: qdstar}, dark green line). 
The `normal ice' regime of model (2) adopted in \citet{schlicht2013} follows the gravity regime 
of \citet{benz1999} and places the binding energy of 1~cm objects in between the strong and weak 
ice models (Fig.~\ref{fig: qdstar}, purple line). Model (3) is a composite of the \citet{benz1999} 
and \citet{lein2012} approaches (Fig.~\ref{fig: qdstar}, blue line), setting the gravity component 
as in model (1) and the strength component as in model (4).

Before considering the results of the numerical calculations, we place the starting conditions
in the context of analytic models for collisional cascades \citep[e.g.][and references 
therein]{wyatt2002,dom2003,wyatt2008,wyatt2011,kb2017a}. For the initial mass 
$M_0$ = 45~\mearth\ at 45~au, the time scale for collisions between equal-mass objects is 
\begin{equation}
\label{eq: tc45}
t_0 \approx {\rm 8~Myr} \left ( \frac{r_l}{{\rm 1~km}} \right ) ~ \left ( \frac{45~\mearth}{M_0} \right ) ~ .
\end{equation}
Systems with $r_l$ = 1~km evolve on short time scales, allowing the system to reach an 
approximate equilibrium over the 4.5~Gyr age of the Solar System. Other initial conditions,
such as $q_s$, probably have little impact on the equilibrium. With 100~times longer collision 
times, systems with most of the mass in 100~km objects cannot evolve into an equilibrium. 
After 4.5~Gyr, their size distributions probably depend on $q_s$. 

Aside from $t_0$, the evolution of the mass in a collisional cascade depends on the ratio of
the collision velocity to \qdstar. When \vsqd\ is large, collisions between unequal mass objects
produce much more debris than when \vsqd\ is small. The mass in the system then declines much
more rapidly. In our approach, the time scale for the mass to decline is 
$\tau_0 \approx 1.13 \alpha t_0$ \citep{kb2017a} where $\alpha$ is a function of \vsqd. 
When \vsqd\ $\approx$ 5--10, $\alpha \approx$ 5. Equal-mass collisions barely shatter the 
objects; the time scale to reduce the system mass by a factor of two is roughly five times
larger than $t_0$. When \vsqd\ $\approx$ 1000--3000, $\alpha \approx$ 0.1--0.04. Equal-mass
collisions completely shatter the objects and leave behind low mass remnants; the time scale 
for mass reduction is 10--20 times faster than $t_0$.

\begin{deluxetable}{lcccccccccc}
\tablecolumns{11}
\tablewidth{0pt}
\tabletypesize{\footnotesize}
\tablenum{2}
\tablecaption{Collision Time Scales\tablenotemark{a}}
\tablehead{
& \multicolumn{2}{c}{$r_l$ = 1~km} &
  \multicolumn{2}{c}{$r_l$ = 3~km} &
  \multicolumn{2}{c}{$r_l$ = 10~km} &
  \multicolumn{2}{c}{$r_l$ = 30~km} &
  \multicolumn{2}{c}{$r_l$ = 100~km}
\\
  \colhead{Model} &
  \colhead{\vsqd} &
  \colhead{$\tau_0$ (Myr)} &
  \colhead{\vsqd} &
  \colhead{$\tau_0$ (Myr)} &
  \colhead{\vsqd} &
  \colhead{$\tau_0$ (Myr)} &
  \colhead{\vsqd} &
  \colhead{$\tau_0$ (Myr)} &
  \colhead{\vsqd} &
  \colhead{$\tau_0$ (Myr)}
}
\label{tab: timescales}
\startdata
1 & ~2468.7 & 0.4 & ~720.6 & 3.2 & 174.3 & 33.5 & ~47.2 & 296.4 & 11.3 & 3352.9 \\
2 & ~2690.9 & 0.4 & ~731.3 & 3.1 & 174.7 & 33.5 & ~47.3 & 296.3 & 11.3 & 3352.8 \\
3 & ~2705.0 & 0.4 & ~732.0 & 3.1 & 174.7 & 33.5 & ~47.3 & 296.3 & 11.3 & 3352.8 \\
4 & ~2891.3 & 0.3 & ~656.4 & 3.4 & 129.2 & 42.9 & ~29.3 & 443.0 & ~5.8 & 6049.1 \\
5 & ~2862.6 & 0.3 & ~655.5 & 3.4 & 129.2 & 42.9 & ~29.3 & 443.0 & ~5.8 & 6049.2 \\
6 & ~2892.8 & 0.3 & ~656.5 & 3.4 & 129.2 & 42.9 & ~29.3 & 443.0 & ~5.8 & 6049.1 \\
7 & ~2892.8 & 0.3 & ~656.5 & 3.4 & 129.2 & 42.9 & ~29.3 & 443.0 & ~5.8 & 6049.1 \\
8 & ~2605.0 & 0.4 & ~643.5 & 3.5 & 128.8 & 43.0 & ~29.3 & 443.2 & ~5.8 & 6049.6 \\
9 & ~1341.1 & 0.6 & ~519.9 & 4.1 & 122.9 & 44.7 & ~29.0 & 447.4 & ~5.8 & 6061.6 \\
10 & ~5610.8 & 0.2 & 1284.7 & 2.0 & 253.2 & 24.7 & ~57.5 & 251.6 & 11.3 & 3344.0 \\
11 & 11450.5 & 0.1 & 2621.8 & 1.1 & 516.8 & 13.8 & 117.3 & 139.3 & 23.1 & 1809.2\\
\enddata
\tablenotetext{a}{
The columns list \vsqd = $v^2 / \qdstar$ for collisions between equal mass 
objects with the listed radius, $v$ = 1~\kms, and parameters for \qdstar\ in 
Table~\ref{tab: model-pars}; and the evolution time for a collisional cascade 
with an equilibrium size distribution,
$\tau_0 = 1.13 \alpha t_0$, where $t_0$ is defined in Eq.~\ref{eq: tcoll} and
$\alpha = 13.0 (v^2 / qdstar)^{-1.237} ~ + ~ 20.9 (v_c^2 / \qdstar)^{-0.793}$
\citep{kb2017a}.
}
\end{deluxetable}

Table~\ref{tab: timescales} lists \vsqd\ and $\tau_0$ for the parameters in models 1--4.
Ensembles of 1~km objects colliding at 1~\kms\ have large \vsqd\ $\approx$ 3000 and short
evolution time scales, $\tau_0 \approx$ 0.3--0.4~Myr. With the system mass,
$M \propto M_0 / (1 + t / \tau_0)^{1.13}$, these systems lose most of their initial mass on 
time scales of 100~Myr. Significant mass loss allows the system to approach an equilibrium
state. In contrast, the small \vsqd\ and large $\tau_0$ for a ring of 100~km objects implies
little evolution on interesting time scales. Over 4.5~Gyr, these systems retain 40\% to 
75\% of their initial mass and have little time to reach equilibrium. For the largest
objects in the swarm ($r \gtrsim$ 100~km), their size distributions will change little.

In the next sub-sections, we review the evolution of calculations with the parameters of
models 1--4. Because the evolution is repetitive, we discuss model (1) in detail, 
summarizing how the size distribution changes in time as a function of the starting 
conditions and comparing how $R(r)$ at 4.5~Gyr depend on $r_l$ and $q_s$.  For models 
(2)--(4), we illustrate differences between the results of these calculations and 
those of model (1), concentrating on (i) whether $R(r)$ reaches an equilibrium and 
(ii) how these equilibria depend on initial conditions.

\subsubsection{Calculations with Strong Ice}
\label{sec: casc-num-gen-strong}

\begin{figure}[t!]
\begin{center}
\includegraphics[width=4.5in]{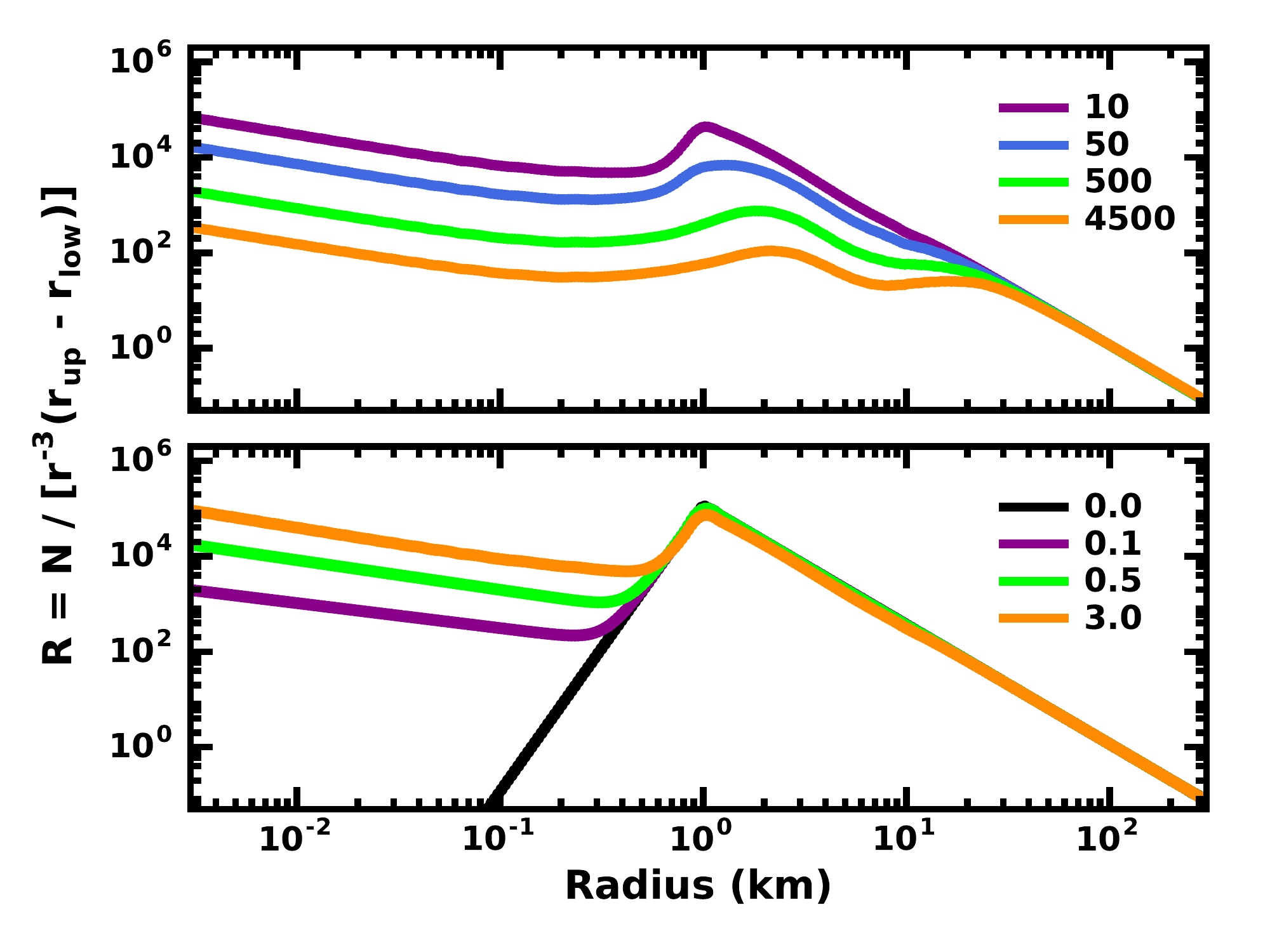}
\vskip -2ex
\caption{\label{fig: cc1}
Evolution of $R(r)$ for a collisional cascade with initial mass
\M0 = 45~\mearth, initial size distribution parameters
$r_s = r_l$ = 1~km, $q_l = 5.5$, and $q_s = -3.0$, and fragmentation parameters
$Q_b = 7 \times 10^7$~erg~g$^{-1}$~cm$^{0.45}$, $e_b = -0.45$, 
$Q_g$ = 2.1~erg~g$^{-2}$~cm$^{1.81}$, and $e_g$ = 1.19. Evolution times in
Myr are listed in each panel.
{\it Lower panel:} Collisions generate a fragmentation tail for $r <$ 1~km. 
Although the amount of material in the tail grows with time, the size 
distribution for $r \gtrsim$ 1~km is unchanged for the first 3~Myr of 
evolution.
{\it Upper panel:} As the evolution proceeds, collisions deplete material
from the peak of the size distribution at 1~km; the peak gradually moves 
to larger radii. 
}
\end{center}
\end{figure}

Fig.~\ref{fig: cc1} illustrates the evolution of $R(r)$ for a system with $r_l$ = 1~km, 
$q_s = -3$, and strong ice \citet{benz1999}. 
Starting from a very peaked size distribution at $t$ = 0, the cascade gradually 
removes km-sized objects and generates a pronounced debris tail at $r \lesssim$ 0.1~km.
After 1--3~Myr (Fig.~\ref{fig: cc1}, lower panel), the debris tail contains as much 
mass as all objects with $r \gtrsim$ 3~km. With a few low amplitude waves at 
$r \lesssim$ 1~cm (not shown) and an overall slope $q \approx$ 3.70, the shape of $R(r)$ 
for $1~\mu m \lesssim r \lesssim$ 0.1~km follows the general expectation for a collisional 
cascade. 

As the calculation proceeds (Fig.~\ref{fig: cc1}, top panel), catastrophic collisions 
gradually eliminate most of the material at 1~km and shift the peak to larger and 
larger sizes. At the same time, cratering collisions continually erode the population
of 10--100~km objects. Over 4.5~Gyr, the system loses 98.7\% of its initial mass. The
remaining material has a steep size distribution at $r \gtrsim$ 30~km with a slope 
similar to the initial $q_l$ = 5.5, which flattens out at $r \approx$ 0.1--10~km and
then rises at much smaller radii with a slope $q \approx$ 3.7.

In this calculation, the amount of mass in the two `peaks' at 1--5~km and at 5--30~km
depends on the evolution time. With more mass initially in km-sized
objects, the time scale for the cascade to destroy half of these objects is 
$t_0 \approx$ 8~Myr (Eq.~\ref{eq: tc45}).  After many destructive collisions over the
first 1--10~Myr, cratering by small particles in the debris tail accelerates the loss of 
km-sized objects.  Over the next 40 Myr, this population drops by a factor of seven, 
turning a very peaked $R(r)$ with a maximum at 1~km into a rounded size distribution 
with a peak close to 1~km. As the cascade proceeds, the typical collision time grows due 
to the falling surface density and the concentration of mass into larger and larger objects.  
By the end of the calculation at 4.5~Gyr, the surface density is 100 times smaller; 
10--20~km objects contain most of the remaining mass.  With $t_0$ now $\sim$ 1~Gyr, the 
peak in $R(r)$ will continue to drop by a factor of two in number every 1--2~Gyr.

Among objects with $r \gtrsim$ 5~km, collision outcomes depend on the relative masses of
the projectiles and the targets. For projectiles at the peak of the size distribution 
($r \approx$ 1--2~km), collisions with 5~km objects are catastrophic. However, cratering
events with much less numerous 10~km objects produce debris and leave behind a slightly 
smaller remnant.  Thus, 5~km objects are removed from the swarm more rapidly than 10--20~km 
objects, creating a pronounced dip in $R(r)$ after 4.5~Gyr of evolution.  Continuing the 
calculation beyond 4.5~Gyr would continue the evolution visible in the top panel of 
Fig.~\ref{fig: cc1}: the peak at 2--3~km would gradually shift to larger radii, possibly 
eliminating the dip at 5~km and generating a fairly flat $R(r)$ at 1--30~km.  At the same 
time, the shoulder in $R(r)$ at 20--30~km would gradually shift to larger radii. Throughout 
this evolution, the fragmentation tail at $r \lesssim$ 0.1--0.3~km would have some modest 
waves superimposed on a constant slope $q \approx$ 3.7.

\begin{figure}[t!]
\begin{center}
\includegraphics[width=4.5in]{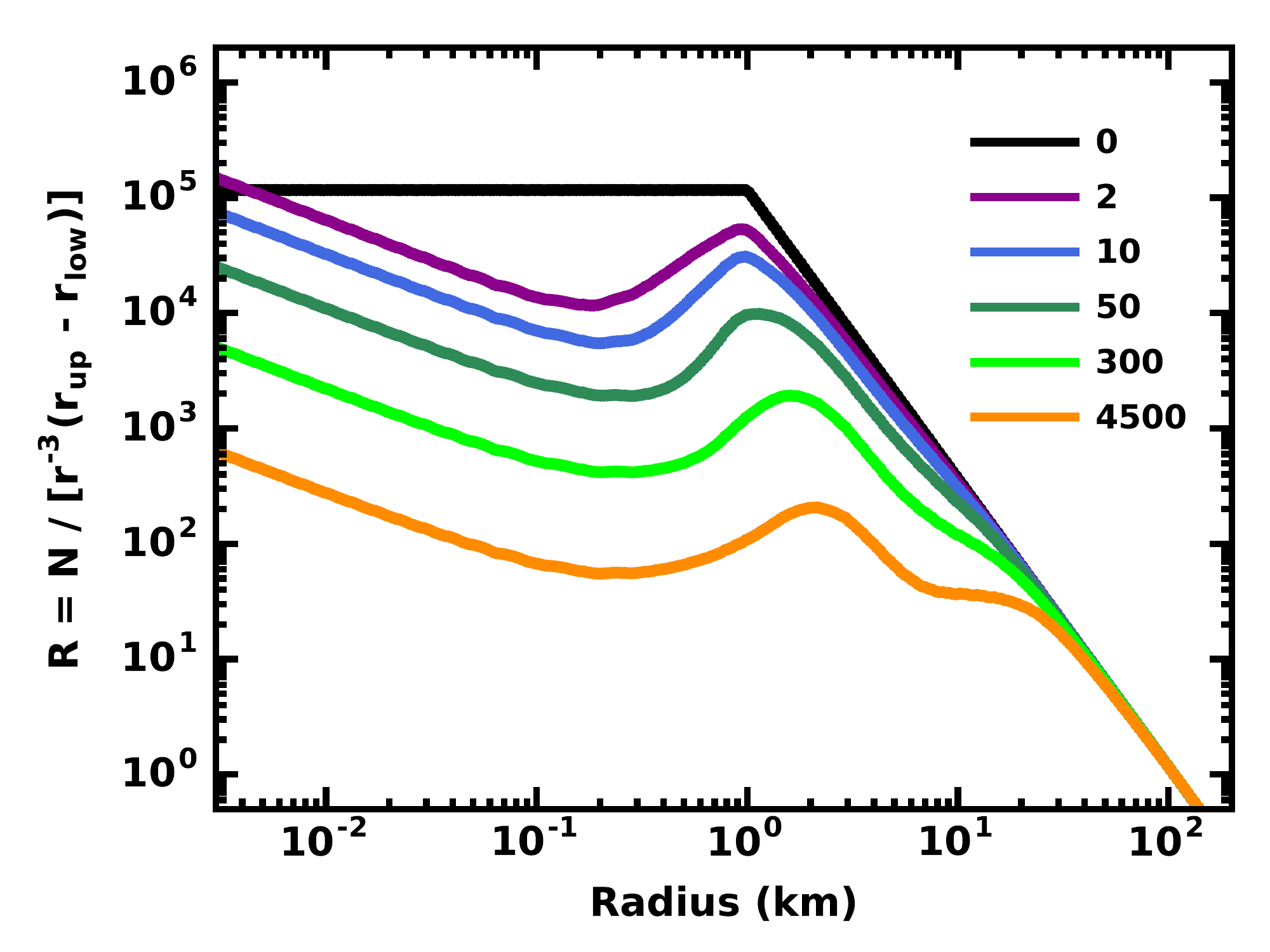}
\vskip -2ex
\caption{\label{fig: cc2}
As in Fig.~\ref{fig: cc1} for $q_s = 3.0$.
Collisional depletion of 1~km objects generates a pronounced wave in the size 
distribution, with a valley near the transition radius, $r_t$ = 0.17~km, a peak
at 1--2~km, and a second valley at 5--10~km. Although the valley at the $r_t$ 
remains fixed, peaks and valleys at larger radii gradually move to larger radii 
with time. 
}
\end{center}
\end{figure}

Starting with a different size distribution for the smaller objects leads to similar results
(Fig.~\ref{fig: cc2}).  When $q_s = 3$, objects with $r \lesssim$ 1~km contain a modest amount
of the total mass. Unlike a calculation with $q_s = -3$, this extra material rapidly destroys
1~km objects through numerous cratering collisions. With $r_t \approx$ 0.15~km, catastrophic and
cratering collisions also remove objects rapidly from bins with $r \approx$ 0.1--0.2~km. In 
$\sim$10~Myr, the number of 1~km (0.1--0.2~km) objects drops by a factor of seven (thirty), 
dramatically changing $R(r)$ at 0.01--1~km. 

In the first 10--50~Myr of this calculation, cratering also removes substantial mass from
particles as large as 10~km. Although collisions among 10--20~km objects are rare, these objects 
are continually peppered by much smaller objects. As a result of this cratering, the slope of
the size distribution slowly declines from the original steep slope $q$ = 5.5 to $q \approx$ 4.6. 
Material from these collisions flows into the debris tail, which robustly maintains the standard 
slope, $q \approx$ 3.7, from 1--10~\mum\ to 0.01~km. 

After 300~Myr, the size distribution in this calculation starts to resemble the calculation 
in Fig.~\ref{fig: cc1} (Fig.~\ref{fig: cc2}, light green curve). While the evolution produces 
a minimum in $R(r)$ at 0.1--0.3~km, the peak slowly shifts from 1~km to 2--3~km. 
As this peak shifts to larger sizes, catastrophic collisions generate a dip in the size 
distribution at 5~km. For larger particles, $R(r)$ consists of a fairly flat (10--30~km) 
section that merges with the steep $q$ = 5.5 section at the largest sizes. After 4.5~Gyr, 
this system loses 99.1\% of its initial mass, slightly more than systems with smaller $q_s$.

\begin{figure}[t!]
\begin{center}
\includegraphics[width=4.5in]{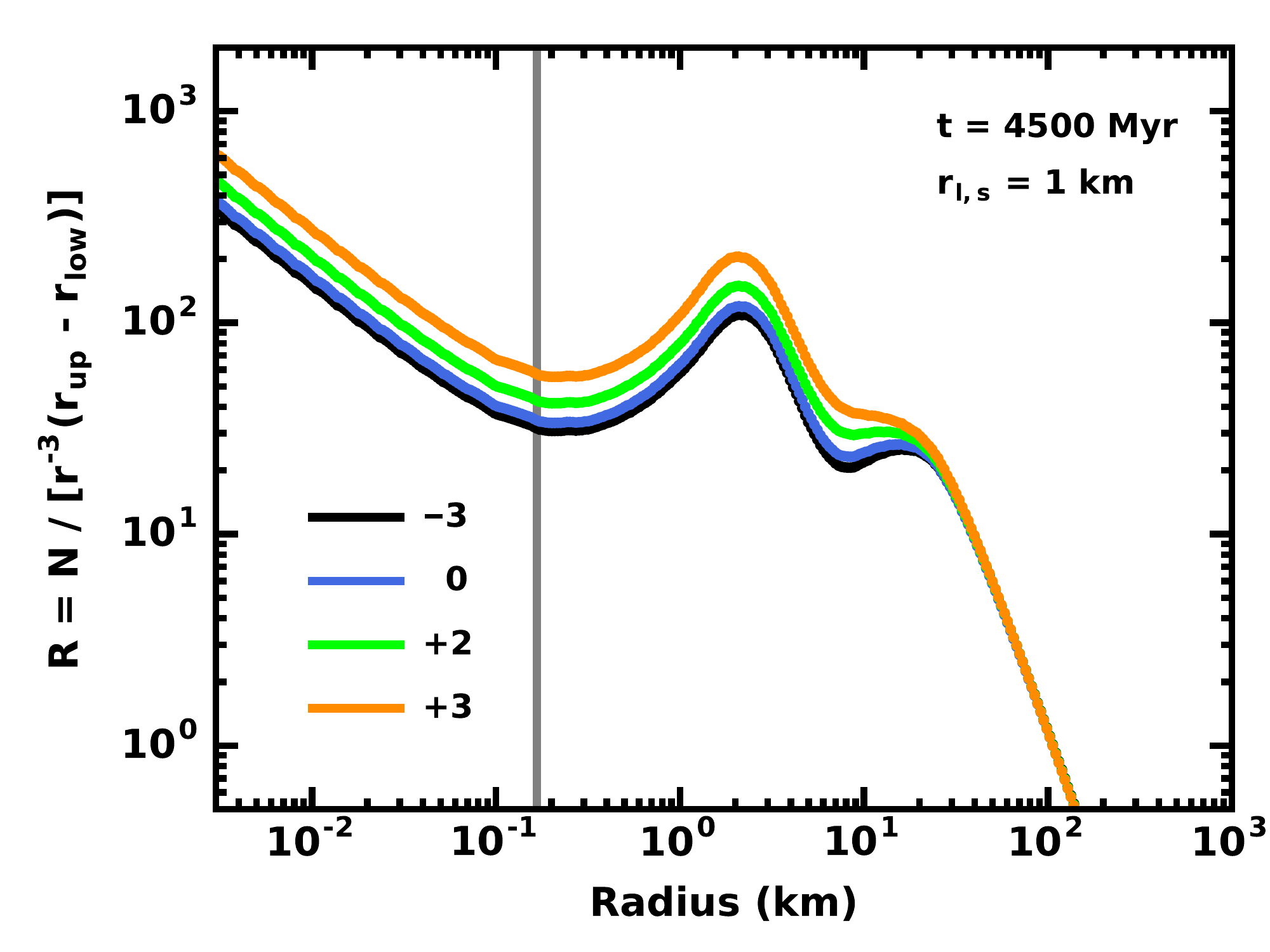}
\vskip -2ex
\caption{\label{fig: cc3}
Size distribution at 4.5 Gyr for collisional cascades with initial mass
\M0 = 45~\mearth, initial size distribution parameters
$r_s = r_l$ = 1~km and $q_l = 5.5$, and fragmentation 
parameters $Q_b = 7 \times 10^7$~erg~g$^{-1}$~cm$^{0.45}$, $e_b = -0.45$, 
$Q_g$ = 2.1~erg~g$^{-2}$~cm$^{1.81}$, and $e_g$ = 1.19 for various slopes
$q_s$ of the initial size distribution as listed in the legend.  The vertical 
grey line indicates the transition radius for this set of \qdstar\ parameters.
The shapes of $R(r)$ are nearly independent of $q_s$.
}
\end{center}
\end{figure}

Despite starting with differing amounts of material with $r \lesssim$ 1~km, all calculations 
with $r_l$ = 1~km and $q_s$ = $-3$ to 3 approach nearly identical size distributions after 
4.5~Gyr of collisional evolution (Fig.~\ref{fig: cc3}). Starting at the smallest sizes we 
consider (1~\mum), all $n(r)$ follow a power law with $q \approx$ 3.7 for $r \lesssim$ 0.05--0.1~km.
Among the smallest particles with $r$ = 1--10~\mum, the ratio \vsqd\ $\approx$ 10. Thus the degree 
of waviness is negligible at the smallest sizes (see Fig.~\ref{fig: wave1}). The first major wave
occurs near the transition radius, $r_t \approx$ 0.17~km, where particles are weakest. After this 
minimum, the shape of the first peak at 2--3~km is insensitive to $q_s$: normalizing the four
curves in Fig.~\ref{fig: cc3} at the same $R$ value at 2--3~km yields nearly indistinguishable 
$R(r)$  for smaller sizes. 

For larger particles, the shape of $R(r)$ at 4.5~Gyr depends on $q_s$. Systems with 
$q_s \approx -3$ have a deeper valley at 5~km than those with $q_s \approx 3$. Because the
valley is deeper, the local peak at 10--20~km is more pronounced in systems with $q_s \approx -3$
than those with $q_s \approx 3$. Over 4.5~Gyr, cratering produces this difference. Systems with
$q_s \approx 3$ initially have more mass in objects with $r \lesssim$ 1~km than those with 
$q_s \approx -3$. With this extra mass, substantial cratering begins at the start of the 
calculation. Over time, this cratering removes slightly more mass at 5--20~km compared to
calculations with $q_s \approx -3$. At 5--20~km, $R(r)$ has a smaller valley at 5~km and a 
less obvious peak at 10--20~km.

The final states of these calculations agree with several expectations for equilibrium size
distributions. For all $q_s$, $R(r)$ has a valley near the transition radius, $r_t$ = 0.17~km. 
From Eqs.~\ref{eq: r-peak}--\ref{eq: r-valley}, equilibrium models have an expected peak at 
1.2~km and a valley at 5.1~km. Results from the numerical calculations yield $r_p \approx$ 2~km 
and $r_v \approx$ 8~km. In these numerical results, the shifts are not from waves emanating 
from the small-size cutoff. The waves at small sizes are too small to impact $R(r)$ at larger 
sizes. Instead, several aspects of the calculation shift the waves to somewhat larger sizes: 
(i) after 4.5~Gyr, the system has not quite reached equilibrium, (ii) our use of the full 
expression for \qdstar\ instead of two separate power laws displaces peaks and valleys 
\citep{obrien2003}, and (iii) our starting condition with a steep slope $q$ = 5.5 at the 
largest sizes creates an inflection point in $n(r)$ which impacts the amount of debris lost 
in cratering collisions.  Fig.~\ref{fig: cc3} demonstrates that the depth of the valley at 
8~km also depends on the initial slope of the size distribution at $r \lesssim$ 1~km. More 
material initially at smaller sizes leads to a shallower valley. Together, these features 
of the calculation shift peaks and valleys from analytic expectations.
 
Among previously published calculations, only \citet{fraser2009b} considers a starting 
point with most of the mass in small objects with $r \approx$ 1~km and the \citet{benz1999}
fragmentation parameters.  The evolution in Figs.~\ref{fig: cc1}--\ref{fig: cc3} generally 
agrees with these results. Using identical fragmentation parameters, \citet{fraser2009b} 
derives size distributions with a valley near $r_t$, a peak at 2--3~km, and a valley (`divot') 
at 10~km. This shape is fairly insensitive to small changes in the fragmentation parameters. 

\begin{figure}[t!]
\begin{center}
\includegraphics[width=4.5in]{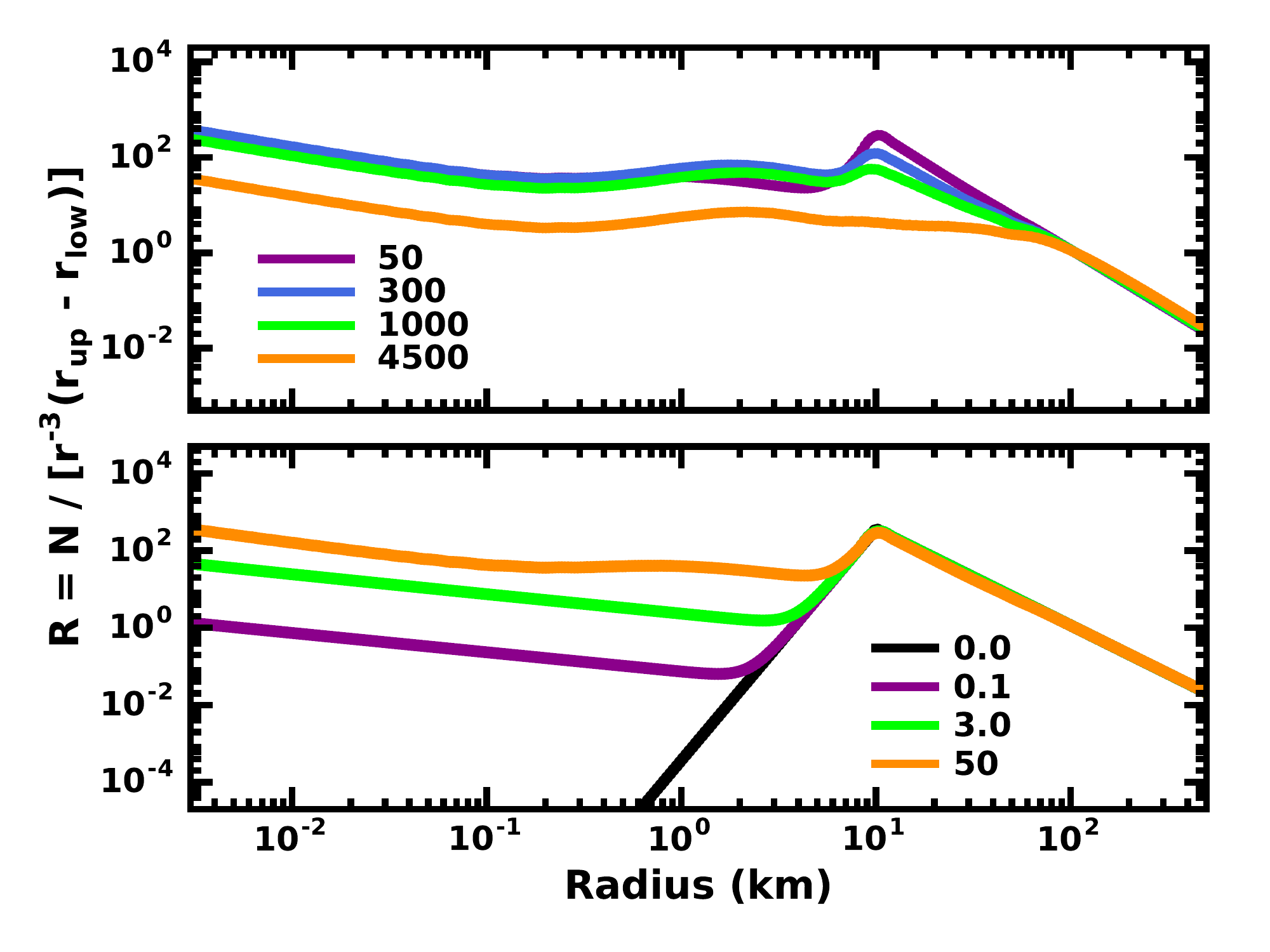}
\vskip -2ex
\caption{\label{fig: cc4}
As in Fig.~\ref{fig: cc1} for $r_l = r_s$ = 10~km; evolution times in Myr appear in the legend for 
each panel. Although the collision time is ten times longer than for systems with $r_l$ = 1~km, 
the shape of the size distribution evolves in a similar fashion. After 4.5~Gyr, $R(r)$ develops 
a clear minimum near the transition radius, $r_t$ = 0.17~km, a modest peak (valley) at 2~km 
(5--6~km), a small peak at 10~km, a gradual decline to 100~km, and a steep drop to 500~km. 
}
\end{center}
\end{figure}

Aside from the longer evolution time, placing most of the initial mass in larger objects has a 
modest impact on $R(r)$ at 4.5~Gyr. When $r_l$ = 10~km, the collision time at the start of each 
calculation is roughly ten times longer than when $r_l$ = 1~km (Eq.~\ref{eq: tc45}).  During the 
first 50~Myr in the evolution of a calculation with $q_s = -3$ (Fig.~\ref{fig: cc4}, lower panel), 
catastrophic collisions of 10~km objects build a prominent debris tail. The deep valley in $R(r)$
advances from 2~km at 0.1~Myr to 6~km at 50~Myr. Early on in this sequence, the debris tail has a 
small wave at 1--10~\mum\ superimposed on a smooth distribution with slope $q \approx$ 3.7. By 50~Myr, 
$R(r)$ flattens at 0.1--3~km and develops a second small wave. At larger sizes ($r \gtrsim$ 10~km), 
$R(r)$ is essentially constant in time.

From 50~Myr to 4.5~Gyr, catastrophic collisions and cratering remove 91\% of the initial mass and 
modify $R(r)$ considerably at 0.1--100~km (Fig.~\ref{fig: cc4}, top panel).  For $r \lesssim$ 0.1~km, 
the system maintains a fairly smooth debris tail with $q \approx$ 3.7 and a small wave at 1--10~\mum.  
As catastrophic collisions remove the sharp peak at 10~km, they establish a distinct valley in the 
debris tail near the transition radius, $r_t$ = 0.17~km.  A wave in $R(r)$ at larger sizes has a 
modest peak at 2~km and a valley at 6~km.  Beyond a tiny residual peak at 10~km, $n(r)$ has a slope 
$q \approx$ 3.7--3.8 at 10--100~km; at 100--500~km, $q \approx$ 5.4.

The features in $R(r)$ have several physical sources. At 10~km, catastrophic collisions are 
numerous enough to destroy most of the objects after 1--2~Gyr. As with the calculations of 
Figs.~\ref{fig: cc1}--\ref{fig: cc3}, catastrophic collisions and cratering combine to create 
the wavy size distribution with valleys at 0.17~km and 6~km and a peak at 2~km. The second 
valley lies closer to the analytic prediction of 5.1~km, but the peak is still off the 
prediction of 1.2~km. Among the largest objects with $r \approx$ 100--500~km and initial 
$q$ = 5.5, collisions between equal mass objects are extremely rare.  Cratering dominates. 
Because cratering can remove material more easily from 100~km objects than 500~km objects, 
the population of 100~km objects decreases slightly, enough to decrease the slope from
$q$ = 5.5 to $q \approx$ 5.4. Debris from these collisions maintains a debris tail with the 
standard slope, $q \approx$ 3.7, from 10--100~km.  Remarkably, the largest objects accrete 
enough from the debris tail to maintain their population over 4.5~Gyr. Some manage to grow 
marginally larger.

\begin{figure}[t!]
\begin{center}
\includegraphics[width=4.5in]{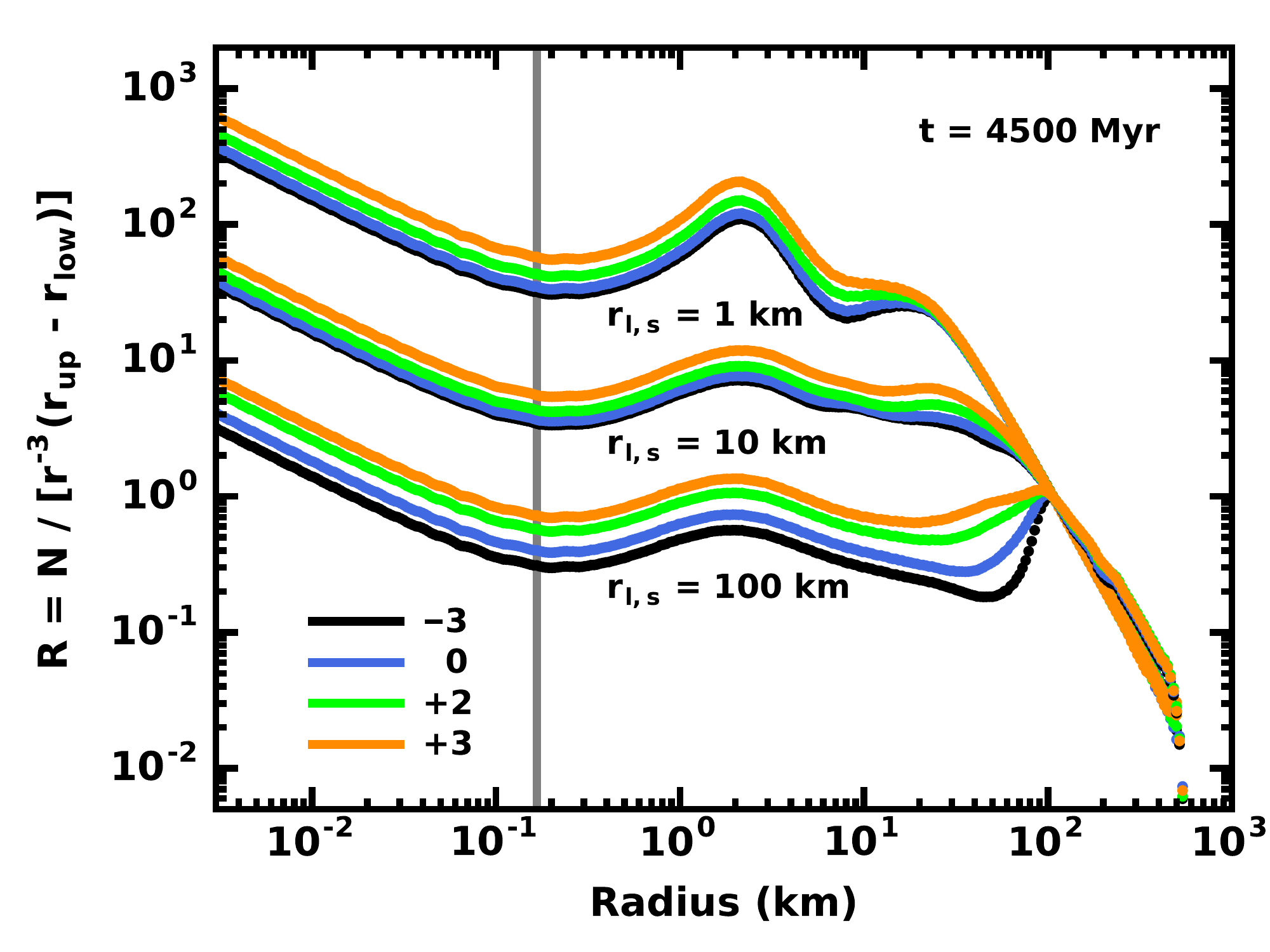}
\vskip -2ex
\caption{\label{fig: cc5}
Size distributions at 4.5~Gyr for strong ice calculations with various $r_{l,s}$ = 1, 10, and
100~km as indicated under each set of curves and various $q_s$ as indicated in the legend in 
the lower left corner.  
The vertical grey line indicates the transition radius for this set of \qdstar\ parameters.
Common features of $R(r)$ include (i) an extended debris tail with slope $q$ = 3.7 at 
$r \lesssim$ 0.1~km, (ii) a valley near the transition radius $r_t$ = 0.17~km, and (iii) a peak
at $r \approx$ 2~km. The height of the 2~km peaks depends on $r_l$ but not $q_s$. Features in 
$R(r)$ for $r \gtrsim$ 5~km -- including a valley at 5--10~km and peaks at 20--100~km -- also 
depend on $r_l$. The slope $q_s$ is only important for calculations with $r_l$ = 100~km, where 
systems with small $q_s$ have a much deeper valley at 40--50~km than those with larger $q_s$.
}
\end{center}
\end{figure}

\begin{figure}[t!]
\begin{center}
\includegraphics[width=4.5in]{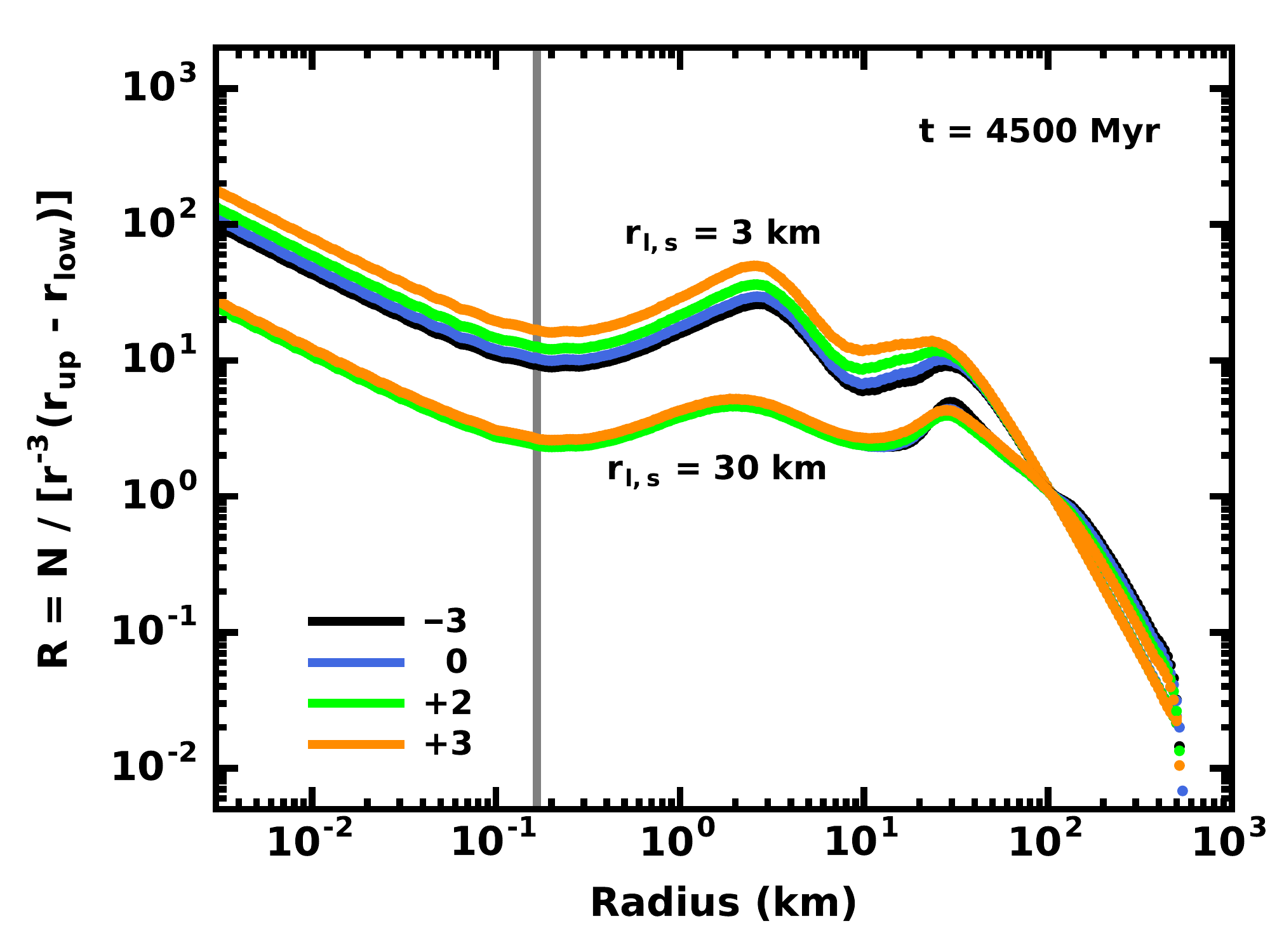}
\vskip -2ex
\caption{\label{fig: cc6}
As in Fig.~\ref{fig: cc5} for $r_l$ = 3 and 30~km.
Common features of $R(r)$ include (i) an extended debris tail with slope $q$ = 3.7
at $r \lesssim$ 0.1~km, (ii) a valley near the transition radius $r_t$ = 0.17~km, and (iii) a peak
at $r \approx$ 2~km. The height of the 2~km peaks depends on $r_l$ but not $q_s$. Features in the
size distribution for $r \gtrsim$ 5~km -- including a valley at 5--10~km and peaks at 20--100~km --
also depend on $r_l$. The slope $q_s$ is only important for calculations with $r_l$ = 100~km,
where systems with small $q_s$ have a much deeper valley at 40--50~km than those with larger $q_s$.
}
\end{center}
\end{figure}

To conclude this sub-section, we compare the final size distributions (at 4.5~Gyr) for systems with 
$r_l$ = 1, 3, 10, 30, and 100~km and various $q_s$ (Fig.~\ref{fig: cc5} and Fig.~\ref{fig: cc6}). 
When $r_l$ = 100~km and $q_s \approx -3$ (Fig.~\ref{fig: cc5}, lowest set of curves), populating the 
debris tail requires relatively rare collisions of massive objects. In calculations with progressively 
larger $q_s$, additional mass in the debris tail allows more cratering, which fills in the deep valley 
(or divot) at 50~km.  Although $R(r)$ at 10-100~km reflects the starting $q_s$, the size distribution 
at smaller $r$ has many of the same features as the size distributions discussed above: 
(i) a debris tail with a small wave at 1--10~\mum\ and a slope $q$ = 3.7 from 1~\mum\ to 0.1~km, 
(ii) a clear valley near the transition radius, $r_t$ = 0.17~km, and 
(iii) a clear peak at 2--3~km. For $r \approx$ 2--20~km, the shape of $R(r)$ clearly depends on 
$r_l$: systems with smaller $r_l$ have more of a valley at 5--10~km. In systems with smaller $r_l$, 
the transition from the steep slope, $q \approx$ 5.5, at large radius to a much shallower slope 
occurs at a smaller radius.

Size distributions generated in calculations with $r_l$ = 3~km have the same features as those with
$r_l$ = 1~km (Fig.~\ref{fig: cc6}, upper group of plots). In both sets of calculations, the debris 
tail has a small wave at 1--10~\mum\ and a slope $q$ = 3.7 from 1--10~\mum\ to the transition radius
at 0.15~km. After the valley near the transition radius, size distributions for $r_l$ = 3~km have a 
shallower rise to a peak at 2--3~km than those with $r_l$ = 1~km. The evolution of this peak is also
different. In systems with $r_l$ = 1~km, the peak shifts from 1~km to 2--3~km over 4.5~Gyr. When 
$r_l$ = 3~km, this peak remains at 3~km throughout the evolution. At larger radii, systems with
$r_l$ = 3~km have a valley at 10~km instead of 8~km and a shallow rise to a second and somewhat
more prominent peak at 25--30~km instead of a more rounded shoulder at 20~km. At the largest sizes,
$r \gtrsim$ 30~km, both sets of calculations maintain a very steep $R(r)$, $q \approx$ 5.5 when 
$r_l$ = 1~km and $q \approx$ 5.3 when $r_l$ = 3~km. Systems with $r_l$ = 3~km lose $\sim$ 97\% of 
their initial mass in 4.5~Gyr, compared to 99\% when $r_l$ = 1~km.  

In a slight contrast to the other calculations, the evolution of systems with $r_l$ = 30~km is almost
completely independent of $q_s$ (Fig.~\ref{fig: cc6}, lower set of curves). Aside from a slight offset,
size distributions for any $q_s$ are identical from 1~\mum\ to 3--4~km. As for calculations with other
$r_l$, there is a small wave at 1--10~\mum\ and a smooth power-law slope $q$ = 3.7 from 1~\mum\ to the
transition radius, a deep valley near $r_t$, and a broad, shallow peak at 2~km. For larger sizes, the
shallow valley at 10~km has a small variation with $q_s$: systems with smaller $q_s$ have a somewhat
deeper valley than those with larger $q_s$. Unlike systems with smaller $r_l$, the collision rates in
these calculations are insufficient to modify the sharp peak in $R(r)$ at 30~km, whose height is fairly 
independent of $q_s$. This behavior is similar to the evolution of systems with $r_l$ = 100~km, but all 
of the $r_l$ = 30~km systems have nearly identical $R(r)$ on either side of the peak. 

Systems with $r_l$ = 30~km show clear signs of evolution in $R(r)$ at the largest sizes. 
For $r \approx$ 30--100~km, the power-law slope evolves from the initial $q$ = 5.5 to $q \approx$ 4
at 4.5~Gyr. Although the slope for $r \approx$ 300--500~km remains constant at $q$ = 5.5, the slope for 
$r \approx$ 100--300~km becomes shallower, $q \approx$ 4.9. Despite the unchanging slope at 300--500~km,
several 500~km objects accrete debris and grow to sizes of 550--600~km. Overall, these systems lose only
$\sim$ 75\% of their initial mass; nearly all of the remaining mass is in objects with $r \gtrsim$ 1~km.

In all of these examples, systems with the same initial mass and $q_s \approx 3$ lose more mass 
than those with $q_s \approx -3$. When $q_s \approx -3$, the swarm has more mass at the peak of 
$R(r)$ and somewhat more mass among the largest objects. Although catastrophic encounters
among equal mass objects at the peak are then more common, the lack of small particles results in 
many fewer cratering collisions. Compared to catastrophic collisions, cratering generates a faster 
flow of material from the largest objects to the smallest. Thus systems with more material initially
in small objects ($q_s \approx$ 3) lose more total mass than systems with more mass initially in 
large objects ($q_s \approx -3$).

Comparisons with published results indicate common features among the calculations. Using similar
fragmentation parameters, \citet{obrien2003} start with a shallower power-law size distribution,
$q \approx$ 3.5, for all sizes. Nevertheless, their calculation generates a valley near the transition
radius; subsequent peaks and valleys lie close to the analytic predictions as in the calculations
described above.  \citet{benavidez2009} consider the evolution of solids with similar initial masses 
and fragmentation parameters, but larger collision velocities. For their fragmentation parameters, 
$r_t \approx$ 0.08--0.15~km. Curiously, their calculations do not show a distinct valley at $r_t$; 
in between deep valleys at 1--2~km and 40--60~km, their Fig. 4 has an obvious peak at 10~km 
\citep[see also][]{campo2012}. In these calculations, the larger $v$ shifts other features in 
$n(r)$ to larger sizes compared to our results; the lack of a valley at the transition radius may 
be due to (i) differences in the fragmentation algorithm, which distributes debris among lower mass 
bins, (ii) the treatment of the evolution of small particles and a small-size cutoff at
1--10~cm instead of 1~\mum, (iii) an inflection point in the initial $n(r)$ at 100~km, 
which might initiate a set of waves not considered in the analytic model and not established in our 
calculations, (iv) the inclusion of populations with different collision velocities, which might 
wash out peaks and valleys in the size distribution, or (v) the design of their figures which may
understate peaks and valleys at small sizes. 

\citet{campo2012} follow the evolution of systems with a power law $n(r)$ at $r \ge$ 100~km and 
no smaller solids. As in our calculations with $r_l$ = 100~km and $q_s = -3$, collisions among 
100~km and larger objects are too infrequent to fill in the valley (divot) at 50--100~km. Unlike our 
results, their size distributions are featureless power-laws over 0.1--30~km with little evidence for 
significant peaks and valleys as in Fig.~\ref{fig: cc5}.  

\subsubsection{Calculations with Normal Strength Ice}
\label{sec: casc-num-gen-ice}

In calculations with normal ice, small objects have less tensile strength than strong ice
particles. The transition radius, $r_t$ = 0.029~km, and the minimum strength, 
$Q_t \approx 1.5 \times 10^5$~\ergg, are 6--8 times smaller than values with the strong ice
parameters (Table~\ref{tab: model-pars}). The difference in \qdstar\ grows to a maximum of 17.5 
for $r \lesssim$ 1~m.  Among particles with $r \gtrsim r_t$, the difference in the binding energy 
gradually diminishes; for $r \gtrsim$ 1~km, the binding energies in the two models are nearly identical.

\begin{figure}[t!]
\begin{center}
\includegraphics[width=4.5in]{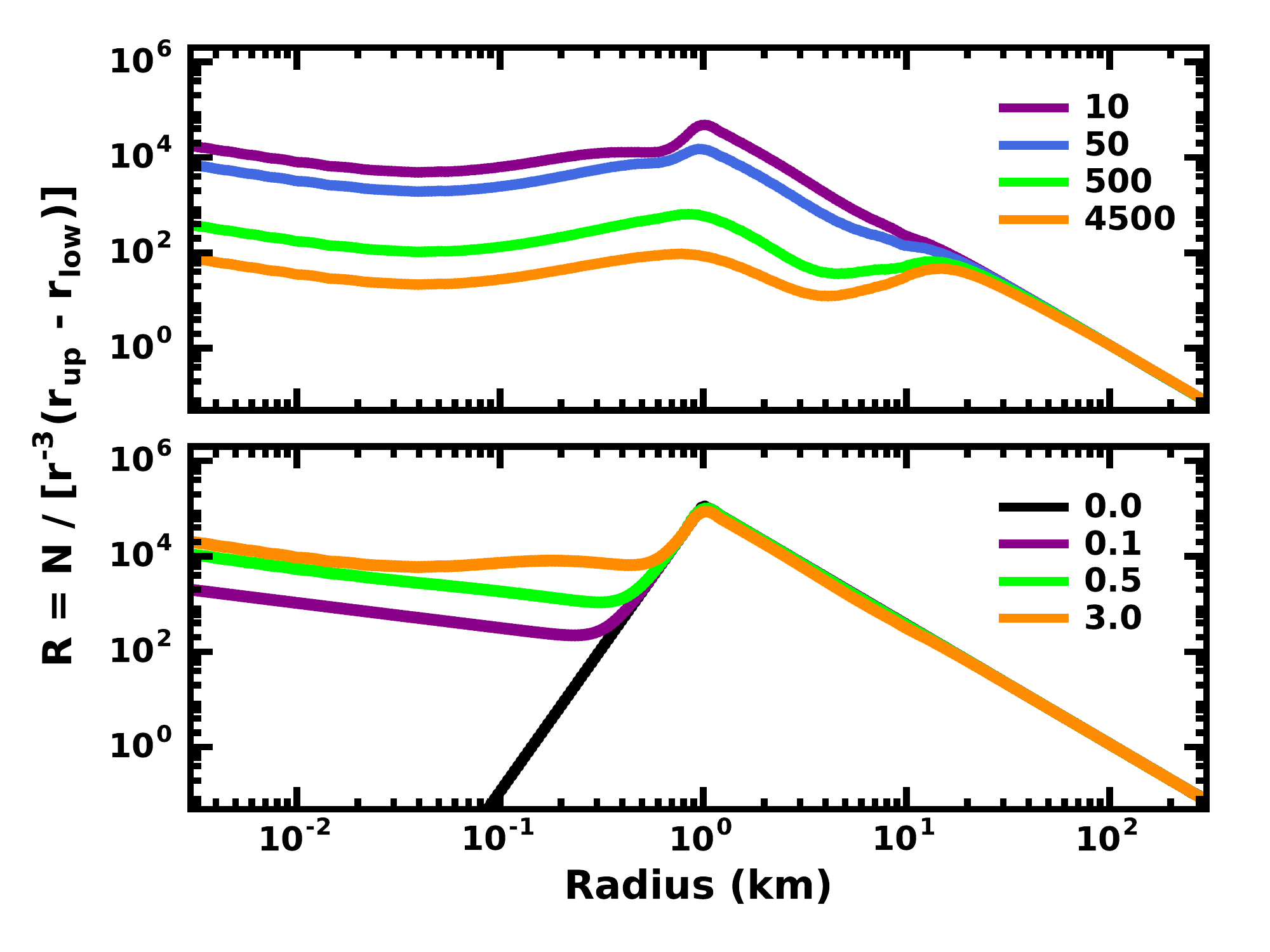}
\vskip -2ex
\caption{\label{fig: cc7}
Evolution of $R(r)$ for a ring with initial mass \M0 = 45~\mearth, $r_s = r_l$ = 1~km, 
$q_l = 5.5$, and $q_s = -3.0$, and the normal ice parameters.  
Evolution times in Myr are listed in each panel.
{\it Lower panel:} During the first 3~Myr, collisions generate a substantial fragmentation 
tail which fills in much of the deep valley at $r \lesssim$ 1~km in the initial state.
{\it Upper panel:} After 10~Myr, collisions deplete material from the peak of the size 
distribution at 1~km, generating a wavy $R(r)$ with two distinct valleys, 0.01--0.1~km and 
3--5~km, and two peaks, 1~km and 10--20~km.
}
\end{center}
\end{figure}

From the analytic model, we expect the evolution of normal ice calculations to follow closely 
those of model (1). Collisions among 1--100~km planetesimals generate substantial debris which 
drives a robust cascade. As the cascade proceeds, $R(r)$ should be somewhat wavier, with a clear 
valley close to the transition radius. Despite the smaller bulk strengths, the rates collisions 
remove large objects from the grid should be roughly similar as in the strong ice calculations.  
Thus, these systems should lose roughly the same amount of material over 4.5 Gyr. 

Fig.~\ref{fig: cc7} confirms these expectations. Starting from a swarm with most of the mass in
1~km objects, catastrophic collisions produce substantial debris in a few Myr. For $r \lesssim$ 
0.01~km, the size distribution follows a smooth power-law with $q \approx$ 3.7; at 1--10~\mum,
a modest wave steepens the slope to $q \approx$ 4.7. By 3--10~Myr, the system has a clear valley 
at 0.03--0.04~km, a slight peak at 0.2--0.4~km, and a second valley at 0.7~km. The second valley
resembles the divots identified in previous simulations \citep[e.g.,][]{fraser2009b,campo2012}. 

As the cascade proceeds, material lost in the catastrophic collisions of large objects and 
numerous cratering collisions of pebbles with 1~km objects wash out the features at 0.1--1~km.
At 500~Myr, $R(r)$ exhibits a characteristic shape, with a distinct valley at 0.03--0.04~km, 
a rounded peak at 1~km, and a second valley at 5~km. Despite the clear evolution for 
$r \lesssim$ 10~km, $R(r)$ at $r \gtrsim$ 20~km is unchanged: collisions remove a few objects 
with $r \approx$ 10--40~km, but do not have enough time to destroy much larger objects.

Compared to the results of model (1) in Fig.~\ref{fig: cc1}, the final size distribution at 
4.5~Gyr for the normal ice calculation in Fig.~\ref{fig: cc7} shows many common features.
Starting at the largest sizes, the evolution of both calculations has negligible impact on 
objects with $r \gtrsim$ 20--30~km. For the starting mass (45~\mearth) and slope ($q$ = 5.5),
the time scale to remove a substantial amount of mass from the largest objects is much longer 
than 4.5~Gyr. At smaller sizes, dramatically shorter collision times allow significant evolution. 
The combination of catastrophic and cratering collisions builds a valley in $R(r)$ close to $r_t$, 
a peak at larger radius, and a second valley between the peak and the unchanged portion of $R(r)$ 
at $r \gtrsim$ 20--30~km. In model (1), the larger $r_t$ results in a main peak at 3~km; the 
six times smaller transition radius with the normal ice parameters maintains a peak close 
to the original peak of 1~km.  In turn, the locations of these peaks set the radius of the 
second valley, roughly 8~km for model (1) and only 4~km for model (2). 

After 4.5~Gyr, the collision cascade with the normal ice parameters removes slightly
less mass (98.6\%) from the grid than with the strong ice parameters (98.7\%). With a 
smaller $r_t$, the calculation shown in Fig~\ref{fig: cc7} always has somewhat less mass in 
solids with $r \lesssim r_t$ than the calculation with the strong ice parameters. 
A smaller mass generates fewer cratering collisions and a smaller flow of mass from the largest
objects to the smallest objects. With somewhat more mass in the largest objects, this calculation
also generates more debris in catastrophic collisions. Overall, these differences almost precisely
balance, yielding a total mass loss nearly identical to mass loss with the strong ice parameters.

\begin{figure}[t!]
\begin{center}
\includegraphics[width=4.5in]{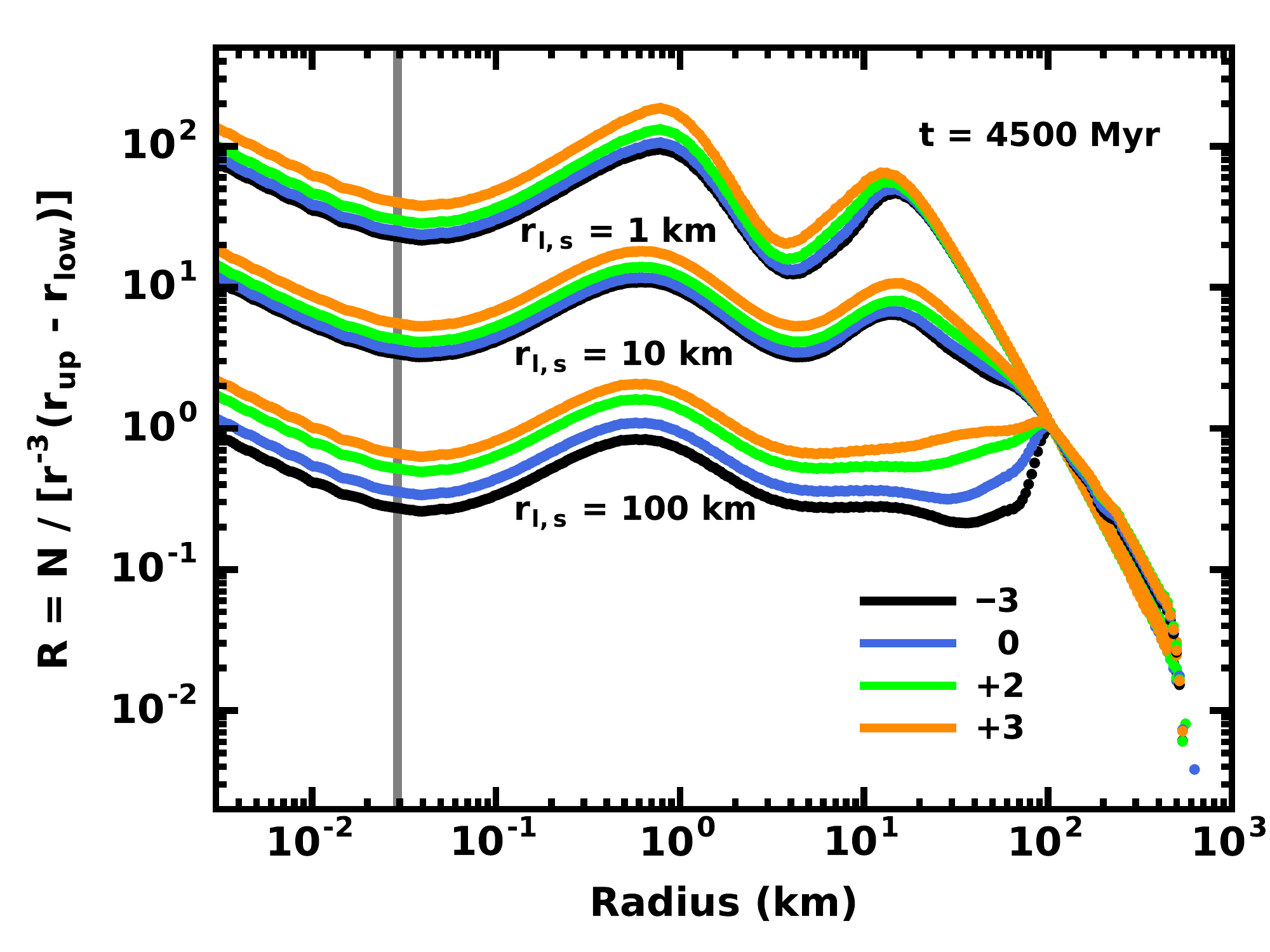}
\vskip -2ex
\caption{\label{fig: cc8}
Final size distributions at 4.5~Gyr for systems with $r_s = r_l$ = 1~km (upper set of curves),
10~km (middle set of curves), and 100~km (lower set of curves), values of $q_s$ listed in the
legend, and the normal ice fragmentation parameters. The vertical line indicates the transition 
radius, $r_t$ = 0.029~km. Compared to results with the strong ice parameters, these systems have 
wavier $R(r)$ with more pronounced peaks and valleys.
}
\end{center}
\end{figure}

Calculations with different initial $r_l$ and $q_s$ yield similar outcomes. When $r_l$ = 1~km and 
$q_s > -3$ (Fig.~\ref{fig: cc8}, upper set of curves), systems begin the calculation with more
solids in bins with $r \lesssim$ 1~km. From the start of the calculation, this material creates
a flurry of cratering collisions, which remove substantial amounts of mass from the largest size
bins. Over time, the increased flow of mass from the largest objects to the smallest objects (and
then out of the grid entirely) results in a somewhat smaller mass at the end of the calculation. 
The total mass removed ranges from 98.6\% of the initial mass for $q_s = -3$ to 98.7\% ($q_s$ = 0)
to 99.1\% ($q_s = 3$). 

Despite the different final masses, all calculations with $r_l$ = 1~km have nearly identical final
$R(r)$. For mass bins with $r \gtrsim$ 175~km, the size distribution at 4.5~Gyr is identical to 
the starting point. Smaller size bins suffer losses, ranging from 1--2 objects at $r \approx$ 
150--170~km to a reduction by more than three orders of magnitude for $r \approx$ 1~km. 
At 20--200~km, the slope of $R(r)$ falls from the initial $q_l$ = 5.5 to $q$ = 5.  Among smaller 
particles, $R(r)$ then drops to a valley at 4~km, rises back to a peak at 0.8--1~km, falls to another 
valley just short of the transition radius at 0.04~km, and then rises with a slope $q$ = 3.7 to 
10--20~\mum. The small-size cutoff generates a much steeper slope, $q = 4.7$, at 1--10~\mum. 

Results for systems with $r_l$ = 10~km follow a similar pattern (Fig.~\ref{fig: cc8}, middle set of 
curves). With more mass tied up in the largest objects, more mass bins become involved in the evolution.
Here, all bins with $r \lesssim$ 300~km lose at least one particle. Thus, the slope of the size
distribution remains constant at $q$ = 5.5 for $r \approx$ 300--500~km, declines to $q$ = 5.2 for 
$r \approx$ 100--300~km, and then falls to $q$ = 4.4 for $r \approx$ 20--100~km. For $r \lesssim$ 20~km, 
$R(r)$ is identical to calculations with $r_l$ = 1~km, with deep valleys at 0.4~km and 4~km
surrounding a peak at 0.7--0.8~km. These size distributions are less wavy than those with $r_l$ = 1~km.

With more mass initially in the largest objects, these systems retain much more mass. A ring with 
$M_0$ = 45~\mearth\ and
$q_s = -3$ loses 90\% of its mass in 4.5~Gyr. Larger $q_s$ yields a ring with less mass after 4.5~Gyr:
9.6\% of the initial mass for $q_s$ = 0 and 7.3\% for $q_s$ = 3. As in other calculations described
earlier, systems with larger $q_s$ have more mass initially in small objects. Aside from being closer
to the small-size cutoff at 1~\mum, small objects start to remove mass from the largest objects sooner,
resulting in more mass loss overall.

In systems with $r_l$ = 100~km, the evolution has a different character (Fig.~\ref{fig: cc8}, lower
set of curves). When $q_s = -3$ (black curve), nearly all of the mass starts in 100~km objects with
long collision times ($t_0 \approx$ 800~Myr; Eq.~\ref{eq: tc45}) and even longer evolution times
($\tau_0 \approx$ 3~Gyr, Table~\ref{tab: timescales}). Over $\sim$ 1~Gyr of evolution, most of the 
100~km objects experience at least one catastrophic collision which generates significant debris. 
Cratering collisions from the debris then begin to remove mass from the swarm of 100~km objects. By
the end of the calculation, $R(r)$ of the debris at $r \lesssim$ 2--3~km looks almost identical to 
$R(r)$ with $r_l$ = 1~km or 10~km: a valley at 3--4~km, a rounded peak at 1~km, another valley near 
the transition radius, a smooth power law at $r \approx$ 10~\mum\ to 0.01~km with $q$ = 3.7, and a 
steeper power law with $q$ = 4.7 at 1--10~\mum. 

Among larger objects, the size distribution is very different. Initially, these systems have no 
objects with $r \gtrsim$ 500~km; after 4.5~Gyr, they have more than 300. When $r_l$ = 1~km or 10~km, 
the population of objects with $r \approx$ 200--500~km is nearly unchanged after 4.5~Gyr. In systems 
with $r_l$ = 100~km, nearly all particles with $r \approx$ 300--500~km accrete some material from the 
rest of the swarm in 4.5~Gyr. Smaller particles lose mass: after 4.5~Gyr, the number of 100~km (200~km) 
objects is 50\% (40\%) smaller than in the initial state. 

The disparate evolution among large and small objects creates a wavy $R(r)$ among the largest objects. 
For $r \approx$ 100--500~km, $n(r)$ is a wavy power law with typical $q \approx$ 5.  The number of 
objects is roughly constant at $r \approx$ 30--100~km, generating a deep divot in the $R$-plot shown 
in Fig.~\ref{fig: cc8}. The number of objects then grows with decreasing radius, producing a small
peak at 10--20~km and an equally tiny valley at 4--5~km.

\begin{figure}[t!]
\begin{center}
\includegraphics[width=4.5in]{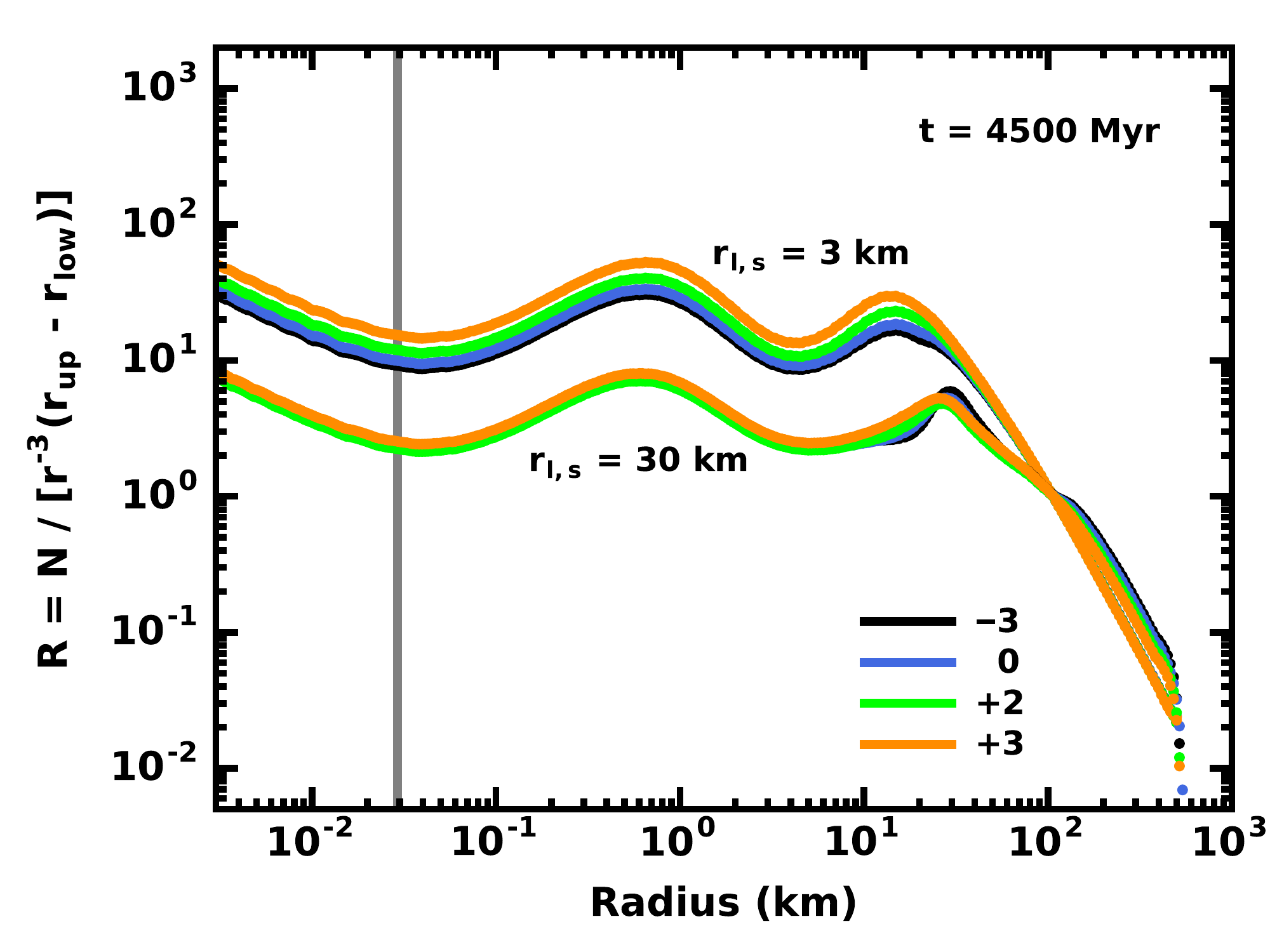}
\vskip -2ex
\caption{\label{fig: cc9}
As in Fig.~\ref{fig: cc8} for $r_s = r_l$ = 3~km (upper set of curves) and 30~km (lower set
of curves). 
}
\end{center}
\end{figure}

Systems with larger $q_s$ have fewer features in the sizes distribution. When $q_s \ge -2$, growth of the 
largest objects is restrained; collisions generate only 200 (50) objects with $r \gtrsim$ 500~km when 
$q_s$ = 0 (3). Size distributions from 100~km to 500~km are then smoother, with smaller waves for larger
$q_s$. Larger $q_s$ also yields smaller divots at 30--100~km; instead of a flat $R(r)$ at 4.5~Gyr with 
$q \approx$ 0 (for $q_s = - 3$), systems with $q_s$ = 0--3 have smoother power-laws with $q \approx$ 2.

Independent of $q_s$, size distributions for $r \lesssim$ 3~km have the same features. As in calculations
with $r_l$ = 1~km or 10~km, $R(r)$ rises from a valley at 3~km to a rounded peak at 1~km.  All systems 
display a deep valley at 0.03--0.04~km, just larger than the transition radius, $r_t$ = 0.029~km.  The 
distribution than rises steeply from 0.01~km to 10~\mum, with power-law slope $q$ = 3.7, and concludes 
with a steeper rise from 10~\mum\ to 1~\mum\ with slope $q$ = 4.7.

As with the strong ice models, results for $r_l$ = 3~km and 30~km and the normal ice
are nearly independent of $q_s$ (Fig.~\ref{fig: cc9}). When $r_l$ = 3~km, the
locations of peaks and valleys at 0.1--30~km match those for $r_l$ = 1~km and 10~km; the amplitude of 
the waviness lies between the levels derived for 1~km and 10~km. At $r \approx$ 20--500~km, the $q$ = 5
slope of the power-law is smaller than the initial $q_s$ = 5.5. For $r \approx$ 1~\mum\ to 0.01~km,
there is a smooth power-law with $q$ = 4.7 at 1--10~\mum\ that transitions into another smooth power-law 
with $q$ = 3.67 at 10~\mum\ to 0.01~km.

Results for $r_l$ = 30~km follow a similar trend. Aside from a normalization factor, $R(r)$ 
at 4.5~Gyr for systems with $r_l$ = 30~km and any $q_s$ closely follow those with $r_l$ = 1--10~km from
$r$ = 1~\mum\ to $r \approx$ 10~km. At larger sizes, the shape depends on $q_s$. Systems with $q_s \approx -3$
have a narrow and somewhat taller peak at 30~km than those with $q_s \approx$ 3. For $r \approx$ 100-500~km,
all systems have a featureless power-law with $q \approx$ 5 and a few larger objects that have accreted material 
from the rest of the swarm. At these large sizes, the main difference between calculations with $r_l$ = 30~km 
and those with $r_l$ = 1--10~km is the collision time. Over 4.5~Gyr, systems with $r_l$ = 30~km do not suffer 
enough destructive collisions to round-off or shift the initial peak at 30~km. Longer time scale calculations
would probably yield a more rounded peak at 20--30~km with a similar shape for all $q_s$.

\subsubsection{Calculations with Weak Ice}
\label{sec: casc-num-gen-weak}

In weak ice models, we expect wavier $R(r)$ than with strong or normal ice. Among particles 
with $r \lesssim$ 0.1~km, bulk strengths are small.  The minimum \qdstar\ is 
$Q_t \approx 2 \times 10^4$~\ergg\ at a transition radius $r_t \approx$ 5~m (model 3) or 
8~m (model 4).  At $r \approx$ 1--10~\mum, the binding energy is 20--350 times smaller 
than for strong or weak ice.  For a fixed $v$, collisions among particles with smaller 
binding energies should produce size distributions with larger waves at the smallest sizes. 
These waves may extend close to the transition radius, changing the morphology of waves 
generated at the minimum $Q_t$. Thus, we expect more complicated size distributions than 
those in the strong ice or normal ice calculations.

At larger sizes ($r \gtrsim$ 1~km), the similar binding energies in the two models should 
yield similar outcomes as in other models. In the gravity regime, the weak ice models have 
identical \qdstar\ at $r \approx$ 2~km. Although \qdstar\ is larger (smaller) for the 
model (3) fragmentation parameters at smaller (larger) sizes, the differences are minor. 
Thus, these two sets of parameters yield similar collision and evolution times for large 
objects with $r \gtrsim$ 0.1~km (Table~\ref{tab: timescales}). Aside from a wavier size
distribution, we expect a set of peaks and valleys at 1--100~km and the same steep slope
at 100--500~km.

The evolution of collisional cascades derived from calculations with the weak ice fragmentation 
parameters closely follows the evolution described for stronger ice models. Catastrophic 
collisions among objects at the peak of the initial size distribution generate copious 
amount of debris with sizes ranging from 1~\mum\ to $r_l$. Once these collisions produce 
some debris, cratering collisions add material to the cascade. On time scales of 10--20~Myr 
($r_l$ = 1~km) to 1--2~Gyr ($r_l$ = 100~km), the cascade develops an approximately 
equilibrium $R(r)$ at 1~\mum\ to several km, with peaks and valleys that stay fixed for 
the remainder of the calculation. In systems with $r_l$ = 1--10~km, the largest 
objects do not accrete significant mass from the swarm; the largest object always has 
$r$ = 500~km.  When $r_l$ = 30--100~km, large objects grow, reaching sizes of 520--525~km 
($r_l$ = 30~km) to 600~km ($r_l$ = 100~km). 

Despite the weaker bulk strength, swarms of weak ice particles lose less mass than those 
with ice or strong ice.  Differences in mass loss range from a few tenths of a per cent 
for $r_l$ = 1~km to $\sim$ 10\% for $r_l$ = 100~km. In weak ice systems, collisions among 
particles in the debris tail are more destructive. The mass in particles with $r \approx$ 
1~\mum\ to 1--10~m is smaller in weak ice models than in stronger ice models. With less 
mass in the debris tail, cratering collisions remove less mass from more massive particles
with $r \gtrsim$ 1~km. Given the rather small differences among the models, it is clear 
that mass loss is driven by catastrophic collisions among 1~km and larger particles.

\begin{figure}[t!]
\begin{center}
\includegraphics[width=4.5in]{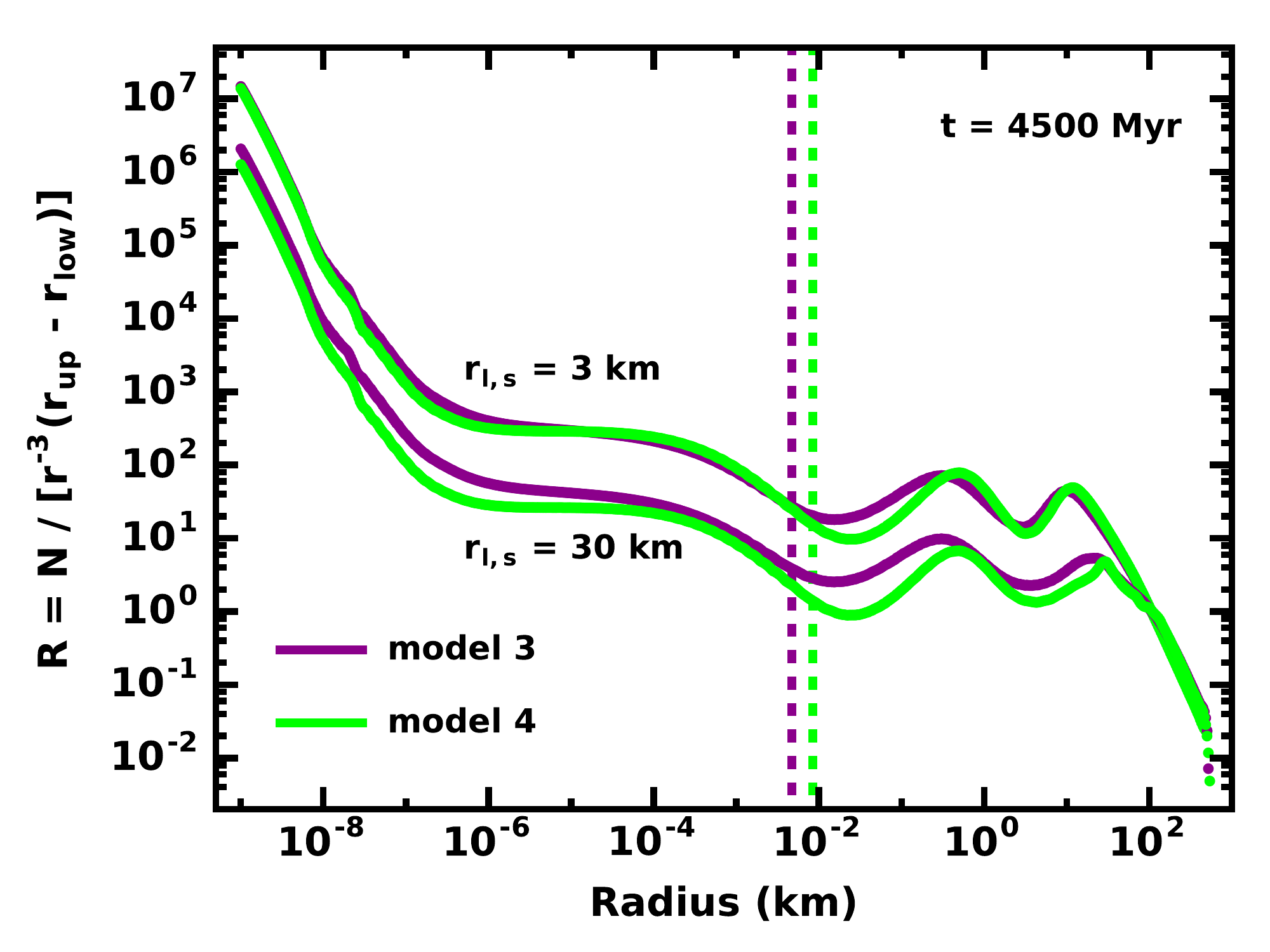}
\vskip -2ex
\caption{\label{fig: cc10}
Final $R(r)$ at 4.5~Gyr for systems with $r_s = r_l$ = 3~km (upper set of curves) or
30~km (lower set of curves), $q_s$ = 3, and the fragmentation parameters indicated in the legend. 
Vertical dashed lines indicate the transition radius for each set of fragmentation parameters. 
}
\end{center}
\end{figure}

At 4.5~Gyr, the size distributions for models with $r_l$ = 3~km and 30~km confirm expectations
(Fig.~\ref{fig: cc10}).  The overall shape of $R(r)$ is nearly independent of $r_l$ and the 
fragmentation parameters.  Particles with $r \approx$ 1--10~\mum\ (10--100~\mum) 
have a steep power-law slope $q \approx$ 5.3 (4.5). These systems have a pronounced wave with 
valleys at 300~\mum, 20--30~m, and 2--5~km; a weak peak at 10~cm; and stronger peaks at 
300--500~m and 1--3~km. The wave from the small-size cutoff clearly impacts the waves due
to the minimum in \qdstar; in all systems, the radius of the deepest valley is 3--10 times 
larger than $r_t$. 

For systems with other $q_s$, the size distributions are similar. The small particles have a
small range of power-law slopes, $q \approx$ 5.3--5.4 at 1--10~\mum\ and 4.4--4.6 at 10--100~\mum.
Calculations with the model (3) parameters produce shallower valleys, with the main valley marginally
closer to the transition radius. The positions and heights of peaks at 5--10~km depend little 
on $q_s$ or the fragmentation parameters; however, peaks at 300--500~m lie at smaller sizes in
model (3) calculations. The size distributions of the largest particles are virtually unchanged. 
The typical power-law slope at 100--500~km, $q \approx$ 5.2--5.4 is close to the initial $q$ = 5.5.

Calculations with $r_l$ = 1~km and 10~km yield results similar to those shown in Fig.~\ref{fig: cc10}.
At small sizes ($r \lesssim$ 10~m, size distributions are nearly independent of $r_l$ and $q_s$,
with similar placement of peaks and valleys and little difference in power-law slopes at 1--100~\mum.
Both sets of calculations have peaks at 400~m and 10~km. Aside from having waves with larger amplitudes,
the model (4) calculations with $r_l$ = 1~km and 10~km have narrower peaks somewhat closer to 1~km 
than the model (3) calculations. Aside from a gradual evolution to a smaller slope, all calculations 
show little change in the power-law $R(r)$ for $r$ = 100--500.

\begin{figure}[t!]
\begin{center}
\includegraphics[width=4.5in]{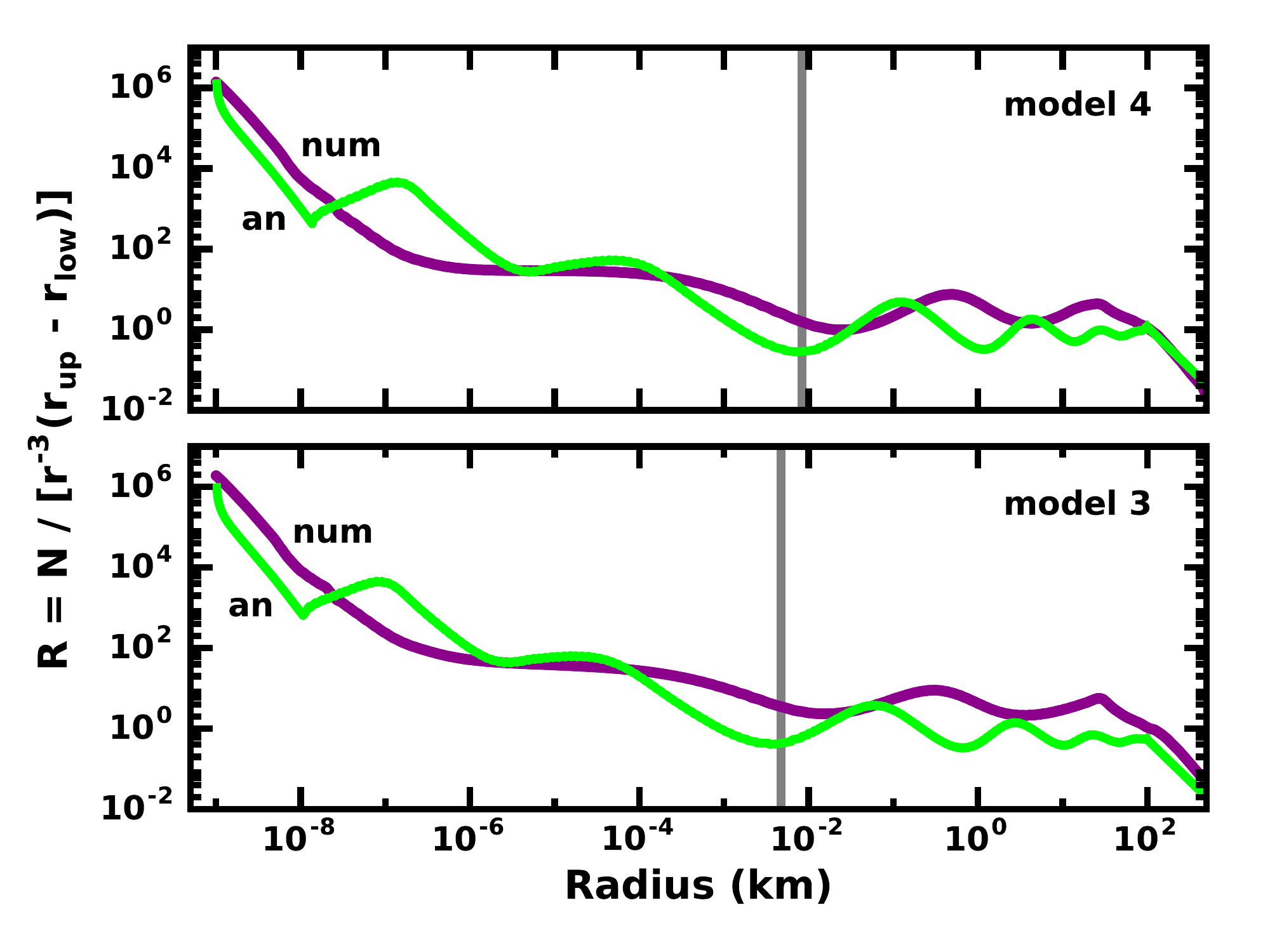}
\vskip -2ex
\caption{\label{fig: eqcc1}
Comparison of theoretical $R(r)$ for the model (3) (lower panel) and the model (4) (upper panel)
fragmentation parameters derived from the analytic model in \S\ref{sec: casc-an-ex2} (`an'; 
green curves) and numerical models with $r_l = r_s$ = 30~km and $q_s$ = 3 (`num'; purple curves). 
The vertical grey line indicates the position of $r_t$. From 1~\mum\ to 100~km, the level of 
waviness and the overall shape of $R(r)$ from the numerical models agrees well with the analytic 
model. In the numerical model, the placement of peaks and valleys is displaced from the 
predictions of the analytic model.
}
\end{center}
\end{figure}

When $r_l$ = 100~km, $R(r)$ for large objects at 4.5~Gyr depends on $q_s$. As in the calculations 
with $r_l$ = 100~km and the stronger ice models (see Figs.~\ref{fig: cc5} 
and \ref{fig: cc8}), a few of the largest objects reach sizes of $\sim$ 600~km. 
The derived $n(r)$ follows a smooth power-law $q \approx$ 5 at 100--500~km and then drops abruptly 
at 50--100~km. Smaller $q_s$ leads to larger drops (divots). From 3--50~km, $R(r)$ 
has a few small amplitude waves superimposed on a power-law with $q \approx$ 2--3. Systems with 
smaller $q_s$ have shallower slopes. Below 2--3~km, $n(r)$ for $r_l$ = 100~km follows the results 
for $r_l$ = 1--30~km: rising to a clear peak at 400--600~m, falling to a valley at 20--30~m, 
curving up and over to a valley at 300~\mum, and finally rising steeply to 1--10~\mum. 

Overall, $R(r)$ derived from calculations with the weak ice parameters match the $R(r)$ 
inferred from the analytic model in \S\ref{sec: casc-an-ex2} (Fig.~\ref{fig: eqcc1}). 
Over roughly 11 orders of magnitude in size -- from 1~\mum\ to 100~km -- the general shapes of 
the two distributions are similar. In both panels, the numerical model does not possess the 
deep valley at 10~\mum\ and the peak at 100~\mum\ in the analytic model.  In the lower panel 
of Fig.~\ref{fig: eqcc1}, the analytic model has a deep valley near the transition radius of 
4.7~m and smaller valleys at 0.5~km and 10~km. In the numerical model, the waviness has a 
smaller amplitude; peaks and valleys are displaced
to larger radius. In the upper panel of Fig.~\ref{fig: eqcc1}, the deepest valley in the analytic
$R(r)$ lies just short of the transition radius; shallower valleys are at larger radii than those in
model (3). The $R(r)$ in the numerical model has a similar degree of waviness with peaks and valleys
at larger radii than the analytic model. 

Several aspects of the calculations prevent a detailed match of the numerical model to the 
analytical model. In the analytical model, the rate material is removed from each mass bin 
sets the size distribution. Requiring a constant flow of material from the largest to the
smallest mass bin allows an analytical solution \citep{wyatt2011}. In the numerical calculations,
every collision removes mass from a pair of bins and redistributes this mass among bins of
lower mass. Redistribution tends to smooth out peaks and valleys in the size distribution. 
Among the largest objects, the requirement that the number of collisions per time step be
an integer creates shot noise from one time step to the next. Redistributed mass from these 
occasional collisions generates pulses in the size distribution that propagate from large to
small objects, further diminishing the strengths of peaks and valleys. Although most objects
suffer catastrophic collisions, cratering events dominate the evolution of particles with
$r \gtrsim$ 300--700~km.  Debris from these collisions also fills in peaks and valleys at the 
smallest sizes. 

\begin{figure}[t!]
\begin{center}
\includegraphics[width=4.5in]{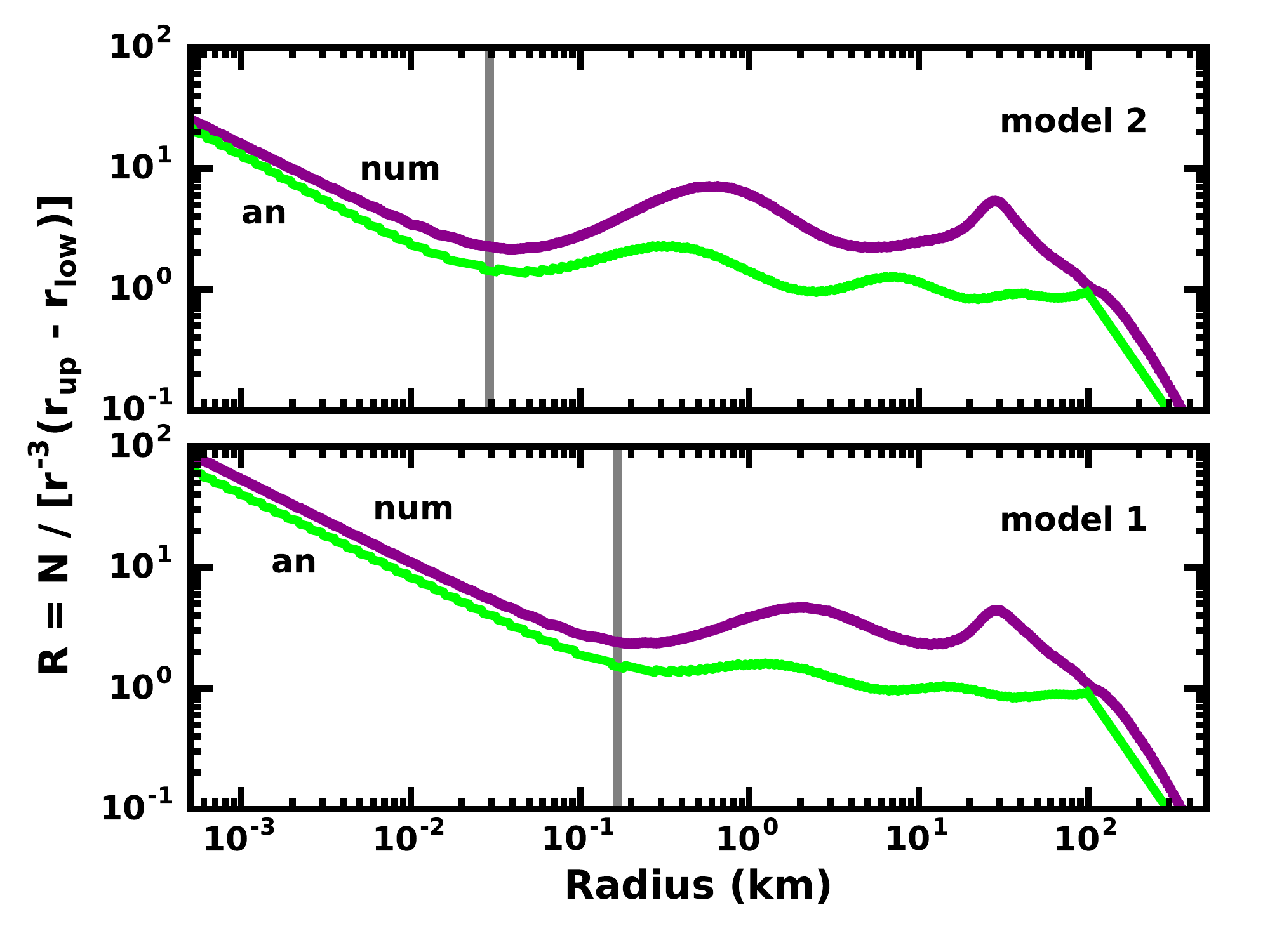}
\vskip -2ex
\caption{\label{fig: eqcc2}
As in Fig.~\ref{fig: eqcc1} for the strong ice (lower panel) and the normal ice (upper panel)
fragmentation parameters.
}
\end{center}
\end{figure}

The lack of equilibrium at the largest sizes also prevents a good match between the 
analytical model and the numerical simulations.
At any size, the time scale to reach an equilibrium is at least 10--20 times larger than the 
collision time.
For $t_0 \approx$ 1~Gyr and $\tau_0 \approx$ 3--4~Gyr, objects with $r \gtrsim$ 100~km do not
have time to reach equilibrium in a 4.5~Gyr calculation. Although this lack of a complete 
equilibrium impacts all of $R(r)$ to some degree, material with $r \gtrsim$ 10~km is much more 
out of equilibrium than solids with $r \lesssim$ 10~km. Calculations with $r_l$ = 30--100~km 
do not have time for collisions to smooth out the initial peak at 30--100~km and thus produce
different waves than predicted by the analytic model. 

The analytic and numerical results for the strong ice and normal ice parameters also agree 
reasonably well (Fig.~\ref{fig: eqcc2}). At the smallest sizes, $r \lesssim$ 0.01--0.1~km,
the two approaches are nearly indistinguishable; with little waviness, the slopes match precisely.
Both approaches yield $R(r)$ with valleys at the transition radius. In the lower panel of 
Fig.~\ref{fig: eqcc2}, the waviness in the analytic $R(r)$ is minimal; the power-law slope is 
$q \approx$ 3. The numerical model maintains the initial peak at 30~km, has more waves, and a 
power-law slope $q \approx$ 3. In the upper panel, the larger waviness in the analytic model
is a result of the smaller $Q_s$. However, the overall slope from 30~m to 100~km is still 
$q \approx$ 3.  The numerical model has a larger waviness, but a similar overall slope 
$q \approx$ 3 from 30~m to 100~km.

In these examples, it is simple to see how the numerical model might evolve towards the analytic
model over a longer evolution time. As collisions remove more mass from the swarm at 0.1--100~km, 
the sharp peak at 30~km will become more rounded and eventually disappear.  The debris tail at 
smaller sizes will maintain the equilibrium $n(r)$ with slope $q \approx$ 3.7. This evolution
will have little impact on solids at 200--500~km, which will retain a size distribution with a steep
slope $q \approx$ 4--5. At 0.1--10~km, collisions will likely preserve the shallow slope $q \approx$ 3
and perhaps develop the same waviness as the analytic model. For the parameters in these calculations,
we estimate the time scale to reach this equilibrium is $\sim$ 30--40~Gyr, much longer than the age
of the Solar System.

\begin{figure}[t!]
\begin{center}
\includegraphics[width=4.5in]{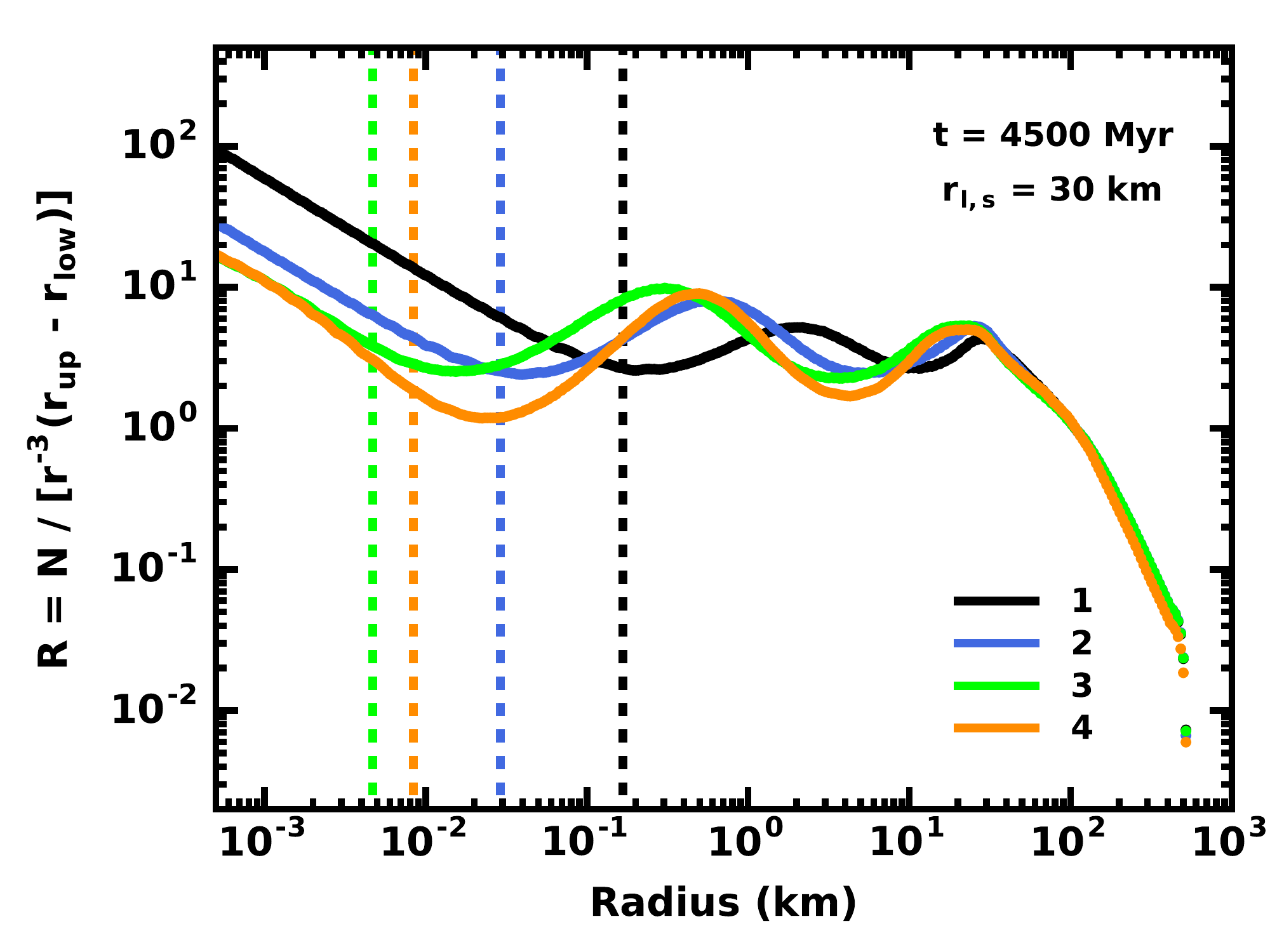}
\vskip -2ex
\caption{\label{fig: cc11}
Comparison of $R(r)$ for $r_l = r_s$ = 30~km at 4.5~Gyr for the model fragmentation
parameters listed in the legend. Vertical dashed lines indicate the position of the transition
radius for the appropriate fragmentation model. Aside from the details of the wavy patterns, all
models share three features: (i) a power-law with $q \approx$ 3.7 at 1~\mum\ to 10--100~m, 
(ii) a wavy distribution superimposed on a power-law with $q \approx$ 3 at 10--100~m to 30--100~km,
and (iii) a steep power-law with $q \approx$ 4.5--5.5 at 100--500~km.
}
\end{center}
\end{figure}

Despite the large difference in fragmentation parameters, size distributions for $r$ = 1~m to 500~km 
at 4.5~Gyr for models (1)--(4) have many similar features (Fig.~\ref{fig: cc11}). For models (2)--(4),
the first valley is at a radius $r \approx$ 10--50~m (Fig.~\ref{fig: cc11}, blue, green, and orange 
curves); this valley is close to the transition radius for model (2) and much larger than $r_t$ for 
models (3) and (4). In systems with the strong particles of model (1), the first valley lies at 300--400~m
(Fig.~\ref{fig: cc11}, block curve), somewhat larger than the transition radius. In all four examples,
the first peak is at a radius roughly ten times larger than the first valley. Systems with weaker 
particles have a larger wave amplitude than those with stronger particles. Following the first peak,
the second valley is at 5--10 times larger radius. In the example shown, there is enough room for a 
second peak and then the sharp power-law decline from 30--40~km to 500~km.

Neglecting the details of the waviness in Fig.~\ref{fig: cc11}, these size distributions share three main
features. At large sizes, $r \approx$ 30--100~km to 500--600~km, $n(r)$ is a power law with a slope $q$
ranging from 4--5 at 30--200~km and 5.0--5.5 at 200--500~km. At intermediate sizes, $r \approx$ 0.01--0.1~km
to 10--30~km, $n(r)$ has a wavy pattern superimposed on a shallow power-law with $q \approx$ 3. Finally,
at the smallest sizes ($r \lesssim$ 0.01~km), $n(r)$ is a steeper power-law with a typical $q \approx$ 3.7.
For ice and strong ice, this steeper power-law has very little waviness. Although waviness dominates the 
power-law in weak ice calculations, the change in $n(r)$ from 1~\mum\ to 1--10~m is the same as a power-law
with $q \approx$ 3.7--3.9.

%The broad features of $R(r)$ in Fig.~\ref{fig: cc11} generally match the observations in 
%Fig.~\ref{fig: obs1}. At large sizes, $r \gtrsim$ 50--100~km, the model $n(r)$ agrees with the data
%by construction. Calculations start with a steep power-law at 100--500~km; collisions times are too long 
%to modify $n(r)$ on 4--5~Gyr time scales. At 1--100~km, however, collisions have time to modify the shape.
%Over 4--5~Gyr, all $R(r)$ develop a characteristic shape that is fairly flat, $q \approx$ 3, from 0.1~km 
%to 100~km. The superficial match between the observed and model size distributions is encouraging. 

These calculations show that the degree of waviness at 0.1--100~km depends on particle strengths. Strong
ice models have waves with an amplitude in $R(r)$ of 2--4 (Figs.~\ref{fig: cc3}, \ref{fig: cc5}, and 
\ref{fig: cc11}). While consistent with the \nh\ data at 1--10~km, these models never develop a factor
of ten wave at 0.1--1~km in any calculation. The normal ice models feature larger waves with amplitudes 
as large as a factor of five (Figs.~\ref{fig: cc7}, \ref{fig: cc8}, and \ref{fig: cc11}), which falls a 
factor of two short of the required amplitude.  While the two weak ice models do not generate a valley 
(peak) at the observed 0.1~km (0.5--1.0~km), the amplitude and wavelength of features in $R(r)$ are 
similar to those in the data. 

\subsection{Variations on the Standard Parameters}
\label{sec: casc-num-vars}

With eight parameters to characterize each collisional cascade model (see Table~\ref{tab: model-pars}),
it is not feasible to make an exhaustive exploration of the available parameter space outside the `standard'
parameters considered above. Here, we consider several variants on models (1)--(4), with the goal of 
understanding how the shape of the size distribution at 0.1--100~km depends on the input parameters.

For fixed collision velocity and fragmentation parameters, calculations with different values of $b_d$, 
$m_{l,0}$, and $b_l$ yield results that are indistinguishable from those with the standard parameters. 
Although $n(r)$ is somewhat sensitive to $b_d$ for collisional cascades at 1~au \citep{kb2016a}, tests
with $b_d$ = 9/8 for solids at 45~au generate nearly identical $R(r)$ as those with $b_d$ = 1. Additional 
tests with $(m_{l,0}, b_l)$ = (0.2, 1.0) and (0.5, 0.75) instead of (0.2, 0.0) suggest $n(r)$ for icy 
objects at 45~au and at 4.5~Gyr is independent of these parameters.

To explore the sensitivity of waves to $Q_s$, we consider calculations with
$Q_s = 10^4$~erg~g$^{-1}$~cm$^{0.4}$ (model (6), Fig.~\ref{fig: qdstar}, orange curve) and 
$Q_s = 10^3$~erg~g$^{-1}$~cm$^{0.4}$ (model (7), Fig.~\ref{fig: qdstar}, wheat curve). 
In calculations with larger $Q_s$, $Q_g$ and $e_g$ have little impact on the shape of the size 
distribution. Thus, we adopt the \citet{lein2012} parameters for the gravity component of \qdstar, 
which generate somewhat larger waves in the size distribution at 1--100~km. To make a tighter 
connection to results with the analytic model, we set $e_s = -0.4$. In the next suite of calculations, we
examine results for different values of $e_s$.

\begin{figure}[t!]
\begin{center}
\includegraphics[width=4.5in]{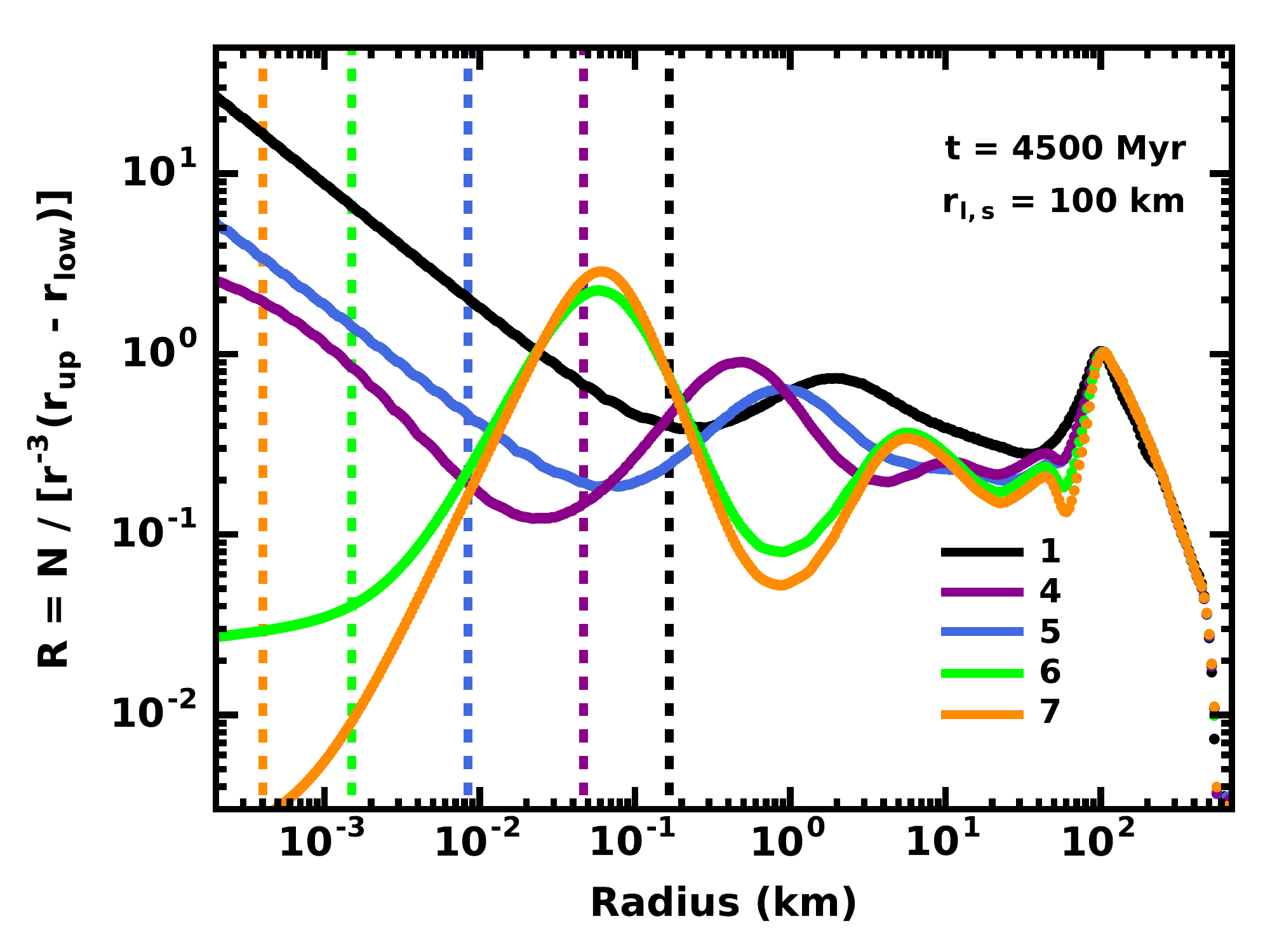}
\vskip -2ex
\caption{\label{fig: cc12}
Comparison of $R(r)$ for $r_l = r_s$ = 30~km at 4.5~Gyr for the model fragmentation
parameters listed in the legend. Vertical dashed lines indicate the position of the transition
radius for the appropriate fragmentation model. Aside from the details of the wavy patterns, all
models share three features: (i) a power-law with $q \approx$ 3.7 at 1~\mum\ to 10--100~m, 
(ii) a wavy distribution superimposed on a power-law with $q \approx$ 3 at 10--100~m to 30--100~km,
and (iii) a steep power-law with $q \approx$ 4.5--5.5 at 100--500~km.
}
\end{center}
\end{figure}

Fig.~\ref{fig: cc12} compares results for calculations with $r_l = r_s$ = 100~km. All calculations 
show a pronounced peak in $R(r)$ at $r$ = 100~km. Within a 45~\mearth\ swarm of solids at 30--60~au,
the time scale for catastrophic collisions to remove a substantial fraction of 100~km objects is 
long, $\gtrsim$ 5--10~Gyr (Table~\ref{tab: timescales}). Thus, the slope of the size distribution
at 100--500~km is remarkably stable over 4.5~Gyr, with $q \approx$ 5.1 for all models.  Several of 
the largest objects with $r \approx$ 400--500~km gradually accrete debris from the rest of the swarm 
and reach sizes of $\sim$ 600--650~km. Although these objects gain significant mass, they are few in 
number: $\sim$ 5--10 with radii as large as 600--650~km.

For radii $r \approx$ 0.1--100~km, the overall slope of $R(r)$ is flat, with $q \approx$ 3 for all 
models. Systems with strong ice have a modest wave, with valleys at 0.2~km and 40--50~km and a peak 
at 2~km. For systems with weaker ice, peaks and valleys lie at smaller radii; the amplitude of the 
wave grows. In models (4) and (5), the amplitude of the wave is small at 5--100~km
and then grows dramatically at smaller radii. In models (6) and (7), there is a significant wave for
$r \lesssim$ 20--30~km. Model (6) has a deep valley at $r$ = 1~km and a peak at 6~km, with a peak to 
valley amplitude of 4.5. For $r \gtrsim$ 0.01~km, model (7) has the same wavy size distribution as
model (6), with valleys and peaks at the same locations but a larger peak-to-valley amplitude. 
Although the waves at 0.1--100~km in model (6) and model (7) have as large an amplitude as the wave
in the \nh\ data, the positions of the peaks and valleys do not match.`

At the smallest sizes, $r \lesssim$ 0.1~km, waviness is a strong function of $Q_s$. In model (1), 
the size distribution follows a power-law with $q \approx$ 3.7--3.8 from $r_t$ to 1~\mum. Reducing
$Q_s$ by a factor of $\sim$ 20--300 for models (5) and (6) results in a similar slope $q \approx$ 
3.6--3.8 from $r_t$ to 10~\mum\ and a much steeper slope $q \approx$ 4.9--5.4 at 1--10~\mum. In
model (6), the overall size distribution is rather flat, $q \approx$ 2.5--3, with a large peak at 
0.05--0.07~km and a very deep valley at 0.01--10~m. For $r \lesssim$ 1~cm, there is a sharp rise 
in $R(r)$. At 1--10~\mum, the slope of $q \approx$ 5.4 is similar to the rise in model (4). However,
the slope at 10~\mum\ to 1~cm, $q \approx$ 4.5, provides a clear measure of the steep slope of the 
size distribution below the deep valley at 0.01--10~m.  In model (7), the $q \approx$ 5.41 at 
1--10~\mum\ follows the slope in model (6); however, the rise from 1~cm to 10~\mum\ is more dramatic,
with $q \approx$ 4.85. 

To understand how the shape of the size distribution depends on the bulk strength component
of \qdstar\ in more detail, we perform calculations with different values for $e_s$
(Fig.~\ref{fig: qdstar}, compare solid, dashed, and dot-dashed lime curves). In models (8) 
and (9), particles have the bulk strength of normal ice and the \citet{lein2012} parameters 
for the gravity component. In model (9), the strength component of \qdstar\ is 
independent of radius, $e_s$ = 0. For model (8), the exponent $e_s = -0.2$ lies midway between
the standard values, $e_s$ = $-0.45$ to $-0.4$, and $e_s$ = 0. 

\begin{figure}[t!]
\begin{center}
\includegraphics[width=4.5in]{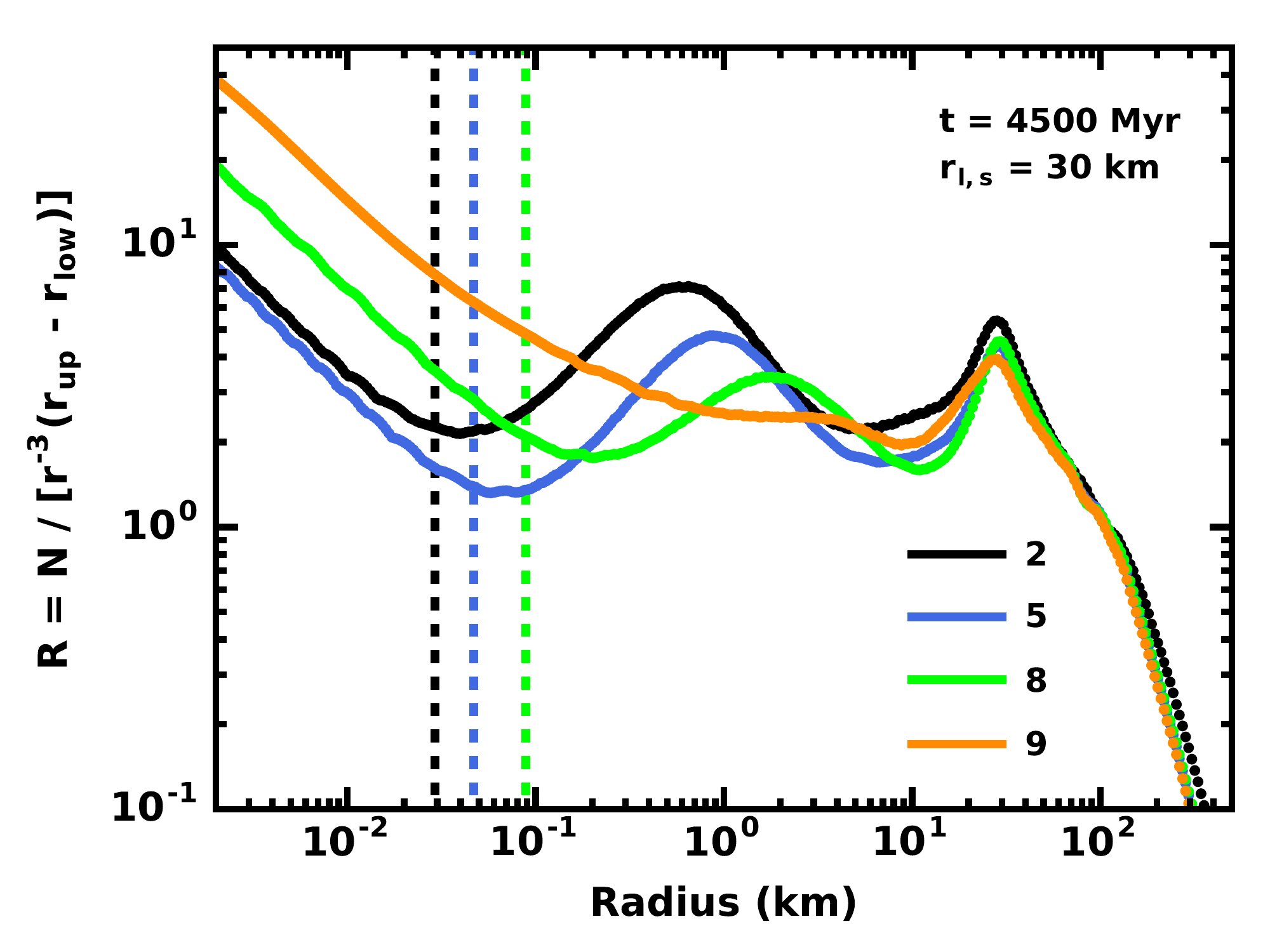}
\vskip -2ex
\caption{\label{fig: cc13}
As in Fig.~\ref{fig: cc12} for $r_l = r_s$ = 30~km at 4.5~Gyr. All calculations have the same
$Q_s$. Aside from the pronounced at 30~km common to all calculations, the degree of waviness 
is more sensitive to $e_s$ (compare blue, green, and orange curves) than to gravity component
of \qdstar\ (compare balck and blue curves). Systems with $e_s$ = 0 have much less wavy $R(r)$ 
than systems with $e_s = -0.2$ or $-0.4$.
}
\end{center}
\end{figure}

Fig.~\ref{fig: cc13} compares a set of $R(r)$ at 4.5~Gyr for calculations with $r_{l,s}$ = 30~km. 
In the standard normal ice models (2) and (5), the size distributions have a sharp peak at 30~km,
where catastrophic collisions have been unable to destroy most of the initial set of objects in
4.5 Gyr. In model (2), calculations with the \citet{benz1999} parameters for the gravity component
of \qdstar\ generate a shallow trough at 3--20~km, a broad peak at 700~m, and a valley near the 
transition radius at 29~m (Fig.~\ref{fig: cc13}, black curve). When we switch to the \citet{lein2012} 
parameters for the gravity component (Fig.~\ref{fig: cc13}, blue curve), the larger transition radius, 
$r_t$ = 47~m, pushes the features in $R(r)$ to larger sizes. However, the amplitude of the waves
is independent of the parameters.

In model (8), the shallower slope of the strength component of \qdstar\ generates smaller waves in
$R(r)$ (Fig.~\ref{fig: cc13}, green curve). When $e_s = -0.2$, $r_t$ = 89~m. Although the first 
valley in $R(r)$ is at a larger radius, 200--300~m, the shape of $R(r)$ from 1~\mum\ to the valley
is similar to models with smaller $e_s$. With the first peak at 2--3~km and a second valley at 10--15~km, 
the waviness in this model has a somewhat smaller wavelength than in the standard models (2) and (5).
The shape between the first peak and the second valley is also different: roughly sinusoidal in
model (8) compared to a flatter trough in models (2) and (5). Despite these differences, the size
distribution in model (8) has the same sharp peak at 30~km, a power-law slope $q \approx$ 4.1 at 
30--100~km, and a power-law slope $q \approx$ 5.3 at 100--500~km. As in models (2) and (5), some
of the largest objects accrete material from the rest of the swarm. Compared to models with 
$r_l = r_s$ = 100~km, growth is limited, reaching sizes of 600~km (instead of 700~km) in 4.5~Gyr.

In model (9) calculations, setting $e_s$ = 0 does not allow a transition radius. Size distributions 
with these parameters are more featureless (Fig.~\ref{fig: cc13}, orange curve). Aside from a power-law
slope $q \approx$ 3.8 (3.7) from 1~\mum\ to 1~cm (10~m), $R(r)$ has an inflection point at $r \approx$
1~km where the size distribution becomes fairly flat, a shallow dip at 10~km, and a sharp rise to a
peak at 20--30~km. Despite these differences, the size distribution at larger $r$ is similar to that
in models (2), (5), and (8): a power-law slope $q \approx$ 4.1 at 30--100~km and a slope $q \approx$
5.3 at 100--500~km. Modest accretion from the rest of the swarm allows the largest objects to reach
sizes $r \approx$ 600~km after 4.5~Gyr.

These examples illustrate the difficulties in generating $R(r)$ with a waviness larger than that 
derived with the standard fragmentation parameters. At the largest sizes, the total mass sets the
collision time; when $r_l = r_s$ = 30--100~km, collisions among the largest objects are rare. The
shape of the size distribution remains close to the initial one over 4.5~Gyr of evolution. At smaller 
sizes, $R(r)$ has a clear waviness superimposed on a fairly flat power-law with $q \approx$ 3 from 
0.1~km to 10~km. Decreasing the slope of the bulk strength component in \qdstar\ leads to smaller waves.

Fragmentation parameters outside the range considered here are probably not physically plausible. 
Several studies of the energy required for catastrophic fragmentation of 1--1000~km asteroids
\citep[e.g.,][]{davis1985,housen1990,love1996} yield values for \qdstar\ intermediate between those 
of \citet{benz1999} and \citet{lein2012}. Although TNOs are icier than typical asteroids, it seems
unlikely that the gravity component of \qdstar\ can be much smaller than the relation derived by 
\citet{lein2012}.  Among investigations for small sizes, the coefficients derived for the bulk 
strength component of \qdstar\ for asteroids are similar to the value adopted for model (1), with 
a slope $e_s \approx -0.2$ to $-0.6$ \citep[e.g.,][]{farinella1982,housen1990,hols1994,housen1999a}.
Our choices allow for a much smaller strength for ice, while retaining `standard' values for the 
variation with radius. 

Aside from these studies, \citet{durda1998} derived much steeper relations for asteroids in both the 
strength and gravity regimes. Their minimum $\qdstar \approx 1 - 2 \times 10^4$~\ergg\ is similar to
the $Q_t \approx 1.8 \times 10^4$~\ergg\ in models (3) and (4), but at a transition radius, 
$r_t \approx$ 70~m, intermediate between the $r_t$ for models (1) and (2). Although inferred to match
properties of the size distribution of asteroid, we considered several test calculations with the
\citet{durda1998} relation for \qdstar. In our approach, the large values for $e_s$ and $e_b$ make
calculations of collisional cascades somewhat delicate. Nevertheless, several tests suggest this 
relation will not yield size distributions for TNOs substantially different from results in models
(1)--(11).

In a final suite of calculations, we consider how the collision velocity shapes the size distribution.
For the normal ice bulk strength and the \citet{lein2012} gravity parameters for \qdstar, model (10) has
$v$ = 1.4~\kms; model (11) has $v$ = 2~\kms. With these parameters, \vsqd\ is twice as large in model
(10) as in model (5); \vsqd\ is four times larger in model (11). Based on results for the analytic 
size distribution (Figs.~\ref{fig: wave1}--\ref{fig: wave4}), systems with larger \vsqd\ should have 
wavier size distributions.

\begin{figure}[t!]
\begin{center}
\includegraphics[width=4.5in]{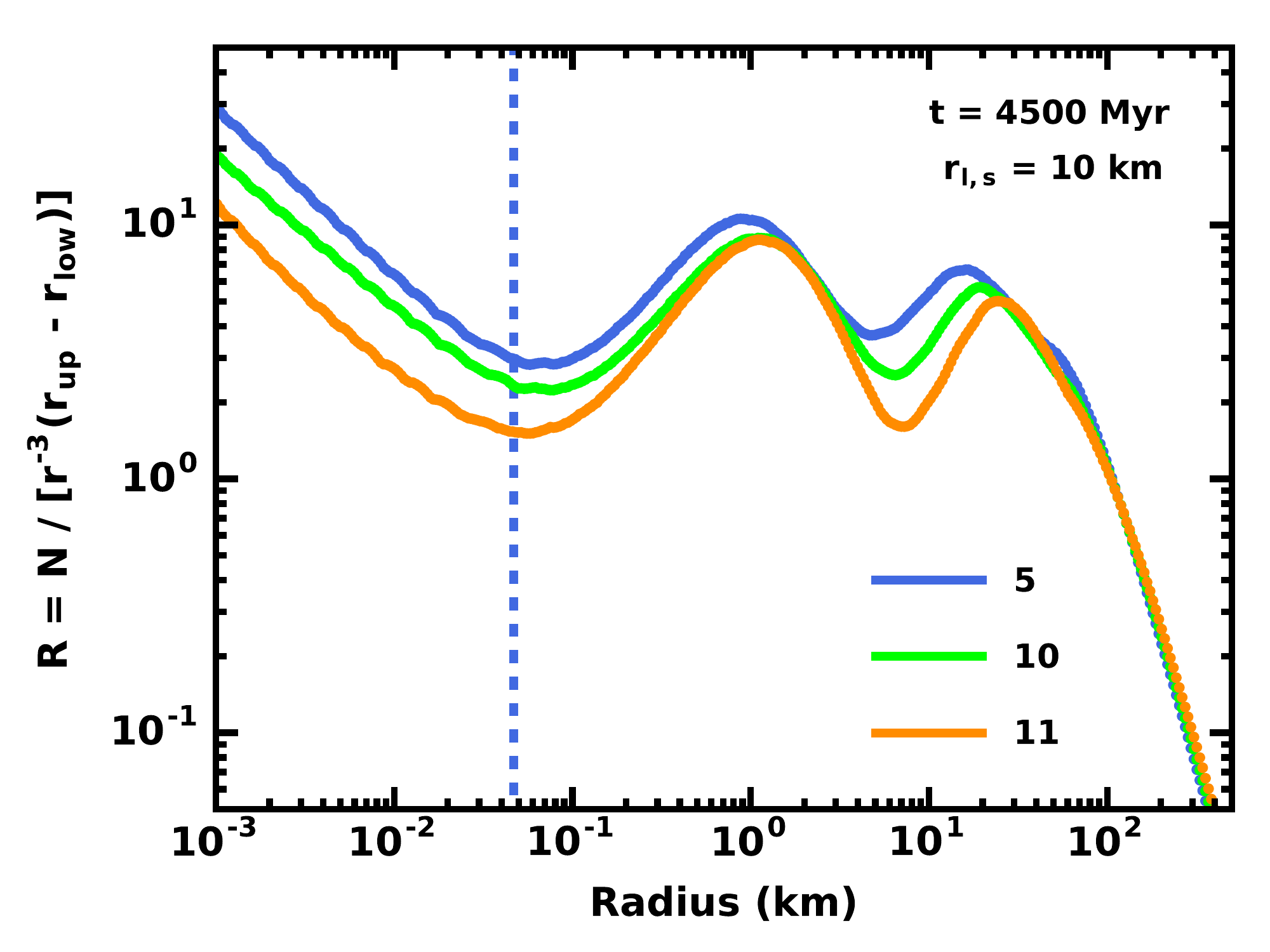}
\vskip -2ex
\caption{\label{fig: cc14}
Comparison of $R(r)$ for $r_l = r_s$ = 10~km at 4.5~Gyr for systems with normal ice and the 
\citet{lein2012} gravity component and collision velocities $v$ = 1~\kms\ (blue curve), 
1.4~\kms\ (green curve), and 2~\kms\ (orange curve). Larger collision velocities eliminate the
shoulder in the size distribution at 40--60~km. At smaller sizes, the amplitudes and wavelengths
of peaks and valleys grow with larger $v$. However, the number of peaks/valleys and the overall
shape of the size distribution is fairly independent of $v$.
}
\end{center}
\end{figure}

Fig.~\ref{fig: cc14} illustrates the impact of the collision velocity on systems with $r_l = r_s$ = 10~km.
All three calculations generate a deep valley near the transition radius, $r_t$ = 47~m. At the smallest 
sizes, the power-law slopes for $R(r)$ in model (10) are $q \approx$ 5.1 at 1--10~\mum, 3.5 at 10~\mum\ to 
1~cm, and 3.6 at 10~\mum\ to 10~m. Power-law slopes in model (11) are nearly identical: $q \approx$ 5.2 
at 1--10~\mum, 3.5 at 10~\mum\ to 1~cm, and 3.6 at 10~\mum\ to 1~m. Compared to model (5), which has 
identical fragmentation parameters and $v$ = 1~\kms, the size distributions of these models have somewhat 
steeper slopes at 1--10~\mum\ ($q \approx$ 5.1--5.2 instead of $q \approx$ 4.9) and somewhat shallower
slopes at large sizes ($q \approx$ 3.5--3.6 instead of $q \approx$ 3.6--3.7). Although the size distributions
are wavy at small sizes, the waves are not large. 

For sizes $r \gtrsim$ 0.1~km, systems with larger collision velocities have wavier size distributions
with longer wavelengths between peaks and valleys. In Fig.~\ref{fig: cc14}, all three models have peaks
at similar levels in $R(r)$ at 1~km. In model (11), valleys at the transition radius and at 8--10~km
(Fig.~\ref{fig: cc14}, orange curve) are much deeper than those in model (5) or model (10). The size
distribution in model (11) has a peak at a larger radius, $\sim$ 30~km, than those in model (5), 
$\sim$ 15~km, or in model (10), $\sim$ 20~km. Despite these differences, the overall shapes of the 
size distributions are fairly similar.

At the largest sizes, $R(r)$ is nearly independent of $v$. Calculations with $v$ = 1~\kms\ have a 
small shoulder in the size distribution at 50~km, which is absent in calculations with larger $v$.
Despite the missing shoulder, the power-law slope at 100--500~km changes little with $v$. At low
velocities, $q \approx$ 5.1--5.6 for $r_l$ = 100--1~km. When $v$ = 2~\kms, $q \approx$ 5.0--5.6.
However, the growth of the largest objects is more sensitive. At $v$ = 1~\kms, the largest objects
reach sizes of 650~km. The largest objects grow only to 575--600~km when $v$ = 2--1.4~\kms.

To facilitate comparisons between these results and other calculations, Table~\ref{tab: sizedist-pars}
summarizes results from the full suite of numerical calculations described above. For each combination
of fragmentation parameters and $r_l$, the Table lists the ratio of the final mass (at 4.5~Gyr) to the
total mass, the radius of the largest object in the grid, the power-law slope for five specific size 
intervals, the radius $r_v$ of the first valley at or larger than the transition radius, the first peak 
$r_p$ with $r > r_v$, and the amplitude of the wave between $r_p$ and $r_v$, $A_{pv} = R(r_p) / R(r_v)$.  
The five power-law slopes span the full range of radii in the size distribution; $q_1$: 1--10~\mum,  
$q_2$: 10~\mum\ to 1~cm,  $q_3$: 0.01--10~m,  $q_4$: 0.01--100~km, and $q_5$: 100--500~km.  
Each slope is calculated as 
$q = ({\rm log}~n(r_2) - {\rm log}~n(r_1)) / ({\rm log}~r_2 - {\rm log}~r_1)$,
where $r_1$ ($r_2$) is the lower (upper) limit of the size range.
For most calculations, the waviness in $R(r)$ is dominant at sizes, $r \approx$ 0.1--100~km. Thus, the 
slopes $q_1$, $q_2$, and $q_3$ provide some measure of the shape of the fragmentation tail at small sizes. 
Similarly, the slope $q_5$ measures the impact of the cascade on the initial slope of $q$ = 5.5 for 
$r \gtrsim$ 100~km. 
At intermediate sizes, $q_4$ allows a comparison between the $q_s$ = $-3$ to 3 and the final slope 
for $r \lesssim$ 100~km.

The parameters $r_v$, $r_p$, and $A_{pv}$ allow an evaluation of the waviness for each model.  Systems 
with $A_{pv} \gtrsim$ 4--5 have a significant waviness correlated with $Q_s$; larger $A_{pv}$ requires 
smaller $Q_s$. Comparison of $r_v$ with $r_t$ in Table~\ref{tab: model-pars} provides a separate 
evaluation of the size distribution: systems with $r_v \approx r_t$ have large $q_s$ and less waviness.
When $Q_s$ is small, $r_v$ is well off $r_t$; the size distribution is then very wavy.

\begin{deluxetable}{lccccccccccc}
\tablecolumns{12}
%\tablewidth{10cm}
\tabletypesize{\tiny}
\tablenum{3}
\tablecaption{Mass and Size Distribution Parameters at 4.5~Gyr\tablenotemark{a}}
\tablehead{
  \colhead{Model} &
  \colhead{$r_l$ (km)} &
  \colhead{~~~~$M / M_0$~~~~} &
  \colhead{$r_m$ (km)} &
  \colhead{~~~~$q_1$~~~~} &
  \colhead{~~~~$q_2$~~~~} &
  \colhead{~~~~$q_3$~~~~} &
  \colhead{~~~~$q_4$~~~~} &
  \colhead{~~~~$q_5$~~~~} &
  \colhead{$r_v$ (km)} &
  \colhead{$r_p$ (km)} &
  \colhead{$A_{pv}$}
}
\label{tab: sizedist-pars}
\startdata
1 &~~~1.0 & 0.013 & 500 & 3.81 &  3.74 &  3.69 &  3.56 &  5.50 &  0.225 &  2.17 &  3.49 \\
1 &~~~3.0 & 0.031 & 500 & 3.81 &  3.74 &  3.69 &  3.42 &  5.44 &  0.225 &  2.62 &  2.94 \\
1 &~~10.0 & 0.089 & 501 & 3.81 &  3.74 &  3.69 &  3.32 &  5.36 &  0.217 &  2.01 &  2.14 \\
1 &~~30.0 & 0.265 & 530 & 3.81 &  3.74 &  3.69 &  3.27 &  5.00 &  0.208 &  1.94 &  1.97 \\
1 & 100.0 & 0.787 & 600 & 3.81 &  3.74 &  3.69 &  3.07 &  5.09 &  0.217 &  1.94 &  1.89 \\
2 &~~~1.0 & 0.013 & 500 & 4.68 &  3.66 &  3.68 &  3.41 &  5.49 &  0.040 &  0.78 &  4.42 \\
2 &~~~3.0 & 0.036 & 500 & 4.68 &  3.66 &  3.68 &  3.34 &  5.46 &  0.038 &  0.65 &  3.75 \\
2 &~~10.0 & 0.094 & 501 & 4.68 &  3.66 &  3.68 &  3.18 &  5.35 &  0.038 &  0.65 &  3.36 \\
2 &~~30.0 & 0.271 & 550 & 4.68 &  3.66 &  3.68 &  3.14 &  5.01 &  0.037 &  0.58 &  3.27 \\
2 & 100.0 & 0.794 & 625 & 4.68 &  3.66 &  3.68 &  2.92 &  5.09 &  0.037 &  0.58 &  3.20 \\
3 &~~~1.0 & 0.016 & 500 & 5.36 &  3.78 &  3.40 &  3.36 &  5.63 &  0.014 &  0.30 &  4.05 \\
3 &~~~3.0 & 0.039 & 500 & 5.36 &  3.78 &  3.40 &  3.26 &  5.40 &  0.014 &  0.30 &  3.94 \\
3 &~~10.0 & 0.094 & 501 & 5.36 &  3.78 &  3.40 &  3.14 &  5.34 &  0.014 &  0.30 &  3.90 \\
3 &~~30.0 & 0.275 & 535 & 5.36 &  3.78 &  3.40 &  3.09 &  4.95 &  0.014 &  0.30 &  3.84 \\
3 & 100.0 & 0.804 & 605 & 5.36 &  3.78 &  3.40 &  2.88 &  5.08 &  0.014 &  0.30 &  3.79 \\
4 &~~~1.0 & 0.016 & 500 & 5.40 &  3.76 &  3.44 &  3.32 &  5.64 &  0.024 &  0.56 &  9.66 \\
4 &~~~3.0 & 0.045 & 501 & 5.40 &  3.76 &  3.44 &  3.23 &  5.44 &  0.024 &  0.48 &  7.85 \\
4 &~~10.0 & 0.117 & 502 & 5.40 &  3.76 &  3.44 &  3.10 &  5.44 &  0.024 &  0.48 &  7.71 \\
4 &~~30.0 & 0.352 & 565 & 5.40 &  3.76 &  3.44 &  3.01 &  5.23 &  0.024 &  0.48 &  7.49 \\
4 & 100.0 & 0.910 & 650 & 5.40 &  3.76 &  3.44 &  2.80 &  5.13 &  0.024 &  0.46 &  7.28 \\
5 &~~~1.0 & 0.014 & 500 & 4.93 &  3.58 &  3.66 &  3.42 &  5.64 &  0.067 &  1.10 &  5.61 \\
5 &~~~3.0 & 0.040 & 501 & 4.93 &  3.58 &  3.66 &  3.30 &  5.44 &  0.067 &  0.98 &  4.01 \\
5 &~~10.0 & 0.114 & 515 & 4.93 &  3.58 &  3.66 &  3.19 &  5.45 &  0.067 &  0.91 &  3.69 \\
5 &~~30.0 & 0.347 & 575 & 4.93 &  3.58 &  3.66 &  3.11 &  5.31 &  0.065 &  0.91 &  3.55 \\
5 & 100.0 & 0.906 & 650 & 4.93 &  3.58 &  3.65 &  2.89 &  5.13 &  0.065 &  0.91 &  3.43 \\
6 &~~~1.0 & 0.021 & 500 & 5.41 &  4.48 &  2.61 &  3.45 &  5.67 &  0.946 &  5.18 & 16.24 \\
6 &~~~3.0 & 0.042 & 501 & 5.41 &  4.48 &  2.61 &  3.31 &  5.47 &  0.910 &  5.58 & 14.96 \\
6 &~~10.0 & 0.114 & 502 & 5.41 &  4.48 &  2.61 &  3.17 &  5.44 &  0.877 &  6.75 &  8.83 \\
6 &~~30.0 & 0.279 & 592 & 5.41 &  4.48 &  2.61 &  3.09 &  5.24 &  0.877 &  6.75 &  5.46 \\
6 & 100.0 & 0.877 & 630 & 5.41 &  4.48 &  2.61 &  2.86 &  5.12 &  0.877 &  5.79 &  4.46 \\
7 &~~~1.0 & 0.021 & 500 & 5.41 &  4.85 &  2.27 &  3.40 &  5.73 &  0.946 &  5.20 & 18.74 \\
7 &~~~3.0 & 0.042 & 501 & 5.41 &  4.85 &  2.27 &  3.27 &  5.53 &  0.910 &  5.63 & 16.20 \\
7 &~~10.0 & 0.114 & 502 & 5.41 &  4.85 &  2.27 &  3.14 &  5.49 &  0.877 &  6.65 &  9.55 \\
7 &~~30.0 & 0.279 & 592 & 5.41 &  4.85 &  2.27 &  3.05 &  5.28 &  0.877 &  6.85 &  7.55 \\
7 & 100.0 & 0.877 & 630 & 5.41 &  4.85 &  2.27 &  2.83 &  5.17 &  0.877 &  5.82 &  6.37 \\
8 &~~~1.0 & 0.014 & 500 & 5.42 &  3.38 &  3.57 &  3.53 &  5.64 &  0.225 &  1.94 &  3.15 \\
8 &~~~3.0 & 0.037 & 501 & 5.42 &  3.38 &  3.57 &  3.40 &  5.44 &  0.225 &  2.52 &  2.54 \\
8 &~~10.0 & 0.111 & 502 & 5.42 &  3.38 &  3.57 &  3.28 &  5.44 &  0.225 &  1.94 &  2.03 \\
8 &~~30.0 & 0.348 & 570 & 5.42 &  3.38 &  3.57 &  3.20 &  5.31 &  0.217 &  1.94 &  1.91 \\
8 & 100.0 & 0.908 & 650 & 5.42 &  3.38 &  3.57 &  2.99 &  5.16 &  0.217 &  1.60 &  1.82 \\
9 &~~~1.0 & 0.016 & 500 & 5.67 &  3.23 &  3.53 &  3.62 &  5.59 &  0.877 &  2.82 &  1.20 \\
9 &~~~3.0 & 0.038 & 501 & 5.36 &  3.78 &  3.40 &  3.36 &  5.63 &  1.059 &  4.28 &  1.24 \\
9 &~~10.0 & 0.107 & 510 & 5.36 &  3.78 &  3.40 &  3.26 &  5.40 &  4.624 &  9.84 &  1.36 \\
9 &~~30.0 & 0.328 & 575 & 5.36 &  3.78 &  3.40 &  3.14 &  5.34 &  9.462 & 26.24 &  1.93 \\
9 & 100.0 & 0.885 & 650 & 5.36 &  3.78 &  3.40 &  3.09 &  4.95 & 18.707 &102.53 &  3.48 \\
10&~~~1.0 & 0.011 & 500 & 5.09 &  3.55 &  3.67 &  3.38 &  5.63 &  0.072 &  1.28 &  6.40 \\
10&~~~3.0 & 0.031 & 500 & 5.09 &  3.55 &  3.67 &  3.26 &  5.41 &  0.072 &  1.23 &  4.52 \\
10&~~10.0 & 0.093 & 525 & 5.09 &  3.55 &  3.67 &  3.16 &  5.35 &  0.072 &  1.10 &  3.93 \\
10&~~30.0 & 0.286 & 545 & 5.09 &  3.55 &  3.67 &  3.13 &  5.01 &  0.072 &  1.06 &  3.75 \\
10& 100.0 & 0.813 & 600 & 5.09 &  3.55 &  3.67 &  2.91 &  5.10 &  0.072 &  1.02 &  3.63 \\
11&~~~1.0 & 0.008 & 500 & 5.21 &  3.50 &  3.76 &  3.32 &  5.62 &  0.056 &  1.33 & 10.44 \\
11&~~~3.0 & 0.024 & 500 & 5.21 &  3.50 &  3.76 &  3.19 &  5.38 &  0.054 &  1.33 &  7.17 \\
11&~~10.0 & 0.079 & 500 & 5.21 &  3.50 &  3.75 &  3.11 &  5.25 &  0.054 &  1.14 &  5.71 \\
11&~~30.0 & 0.227 & 550 & 5.21 &  3.50 &  3.75 &  3.10 &  4.81 &  0.054 &  1.10 &  5.44 \\
11& 100.0 & 0.705 & 590 & 5.21 &  3.50 &  3.75 &  2.90 &  4.98 &  0.054 &  1.10 &  5.27 \\
\enddata
\tablenotetext{a}{
For each model at 4.5~Gyr, the columns list $r_l$ in km, the ratio of the final mass to 
the initial mass, $r_m$ the radius of the largest object,
the slopes $q_1$, $q_2$, $q_3$, $q_4$, $q_5$ as defined in the main text,
$r_v$ the position of the first valley near the transition radius, 
$r_p$ the position of the first peak after the first valley, and
$A_{pv} = R(r_p) / R(r_v)$.
}
\end{deluxetable}

\subsection{Applications to TNOs}
\label{sec: casc-num-tnos}

To apply the results of collisional cascade calculations to the analysis of \citet{singer2019},
we focus on several basic features. In the \nh\ data, $R(r)$  is fairly flat at 1--20~km. From 
a well-defined peak at 1--2~km, the errors bars allow a shallow valley at 5--10~km. The last 
data point at $\sim$ 15~km places few constraints on any model. For this exercise, we assume 
that the size distribution remains fairly flat from 10~km to 50--100~km and then matches the 
steep power-law with $q \approx$ 5.0--5.5 for $r \gtrsim$ 100~km derived from ground-based 
observations.  Other options are possible, but this approach seems simplest. At the smallest
sizes probed by \nh, 0.1--1~km, the factor of 10--20 drop in $R(r)$ is well-defined. We make
no assumption about the behavior of $R(r)$ at $r \lesssim$ 0.1~km. 

Among the numerical calculations, several generate size distributions with the required 
features.  Models (5), (10), and (11) display a clear peak near 1~km and a valley at 0.05~km 
(see Fig.~\ref{fig: cc14}) that agrees with the peak and valley in the \nh\ data. From the peak 
to the valley in $R(r)$, the amplitude listed in Table~\ref{tab: sizedist-pars} grows from 
3--6 (\vc\ = 1~\kms) to 4--7 (\vc\ = 1.4~\kms) to 5--7 (\vc\ = 2~\kms). While smaller than
the amplitude in the \nh\ data, it seems plausible that a larger collision velocity might 
generate a larger amplitude.  Systems with $r_l$ = 1--10~km have a deep valley at 5--10~km
not observed in the \nh\ data. Calculations with $r_l$ = 30--100~km have a much shallower
valley with a peak at 30--100~km that lies outside the range covered by \nh. With power-law
slopes $q \approx$ 5 at $r \gtrsim$ 30--100~km, the models provide a reasonable match to the
ground-based data.

With a larger peak-to-valley amplitude, model (4) is also worth considering. For \vc\ = 1~\kms,
the size distributions of the model (4) calculations are somewhat offset from those required for
the \nh\ data, with peaks at 0.5~km and valleys at 0.025~km. While less well-placed than the 
peaks and valleys in models (5), (10), and (11), the amplitude of 7--9 is a better match to the
\nh\ data. At larger sizes, however, the model (4) size distributions with $r_l$ = 30--100~km
have a prominent valley at 3~km, a flat portion from 10--30~km, and a steep rise at 30--100~km 
(see Fig.~\ref{fig: cc12}).  These do not match the \nh\ data very well.

To learn whether calculations with parameters intermediate between model (4) and model (5)
provide a better match to the \nh\ data, we perform a limited set of calculations with the model 
(5) fragmentation parameters and substitute $Q_s = 10^6$~erg~g$^{-1}$~cm$^{0.4}$. Starting with 
$r_l = r_s$ = 100~km and $q_s$ = 0, these calculations seek to establish whether a somewhat
smaller bulk strength than in model (5) yields peaks and valleys close to those in model (5)
but with a somewhat larger wave amplitude. Calculations with \vc\ = 1~\kms, 2~\kms, and 
3~\kms\ allow a measure of the wave amplitude as a function of \vc\ at this $Q_s$.

\begin{figure}[t!]
\begin{center}
\includegraphics[width=4.5in]{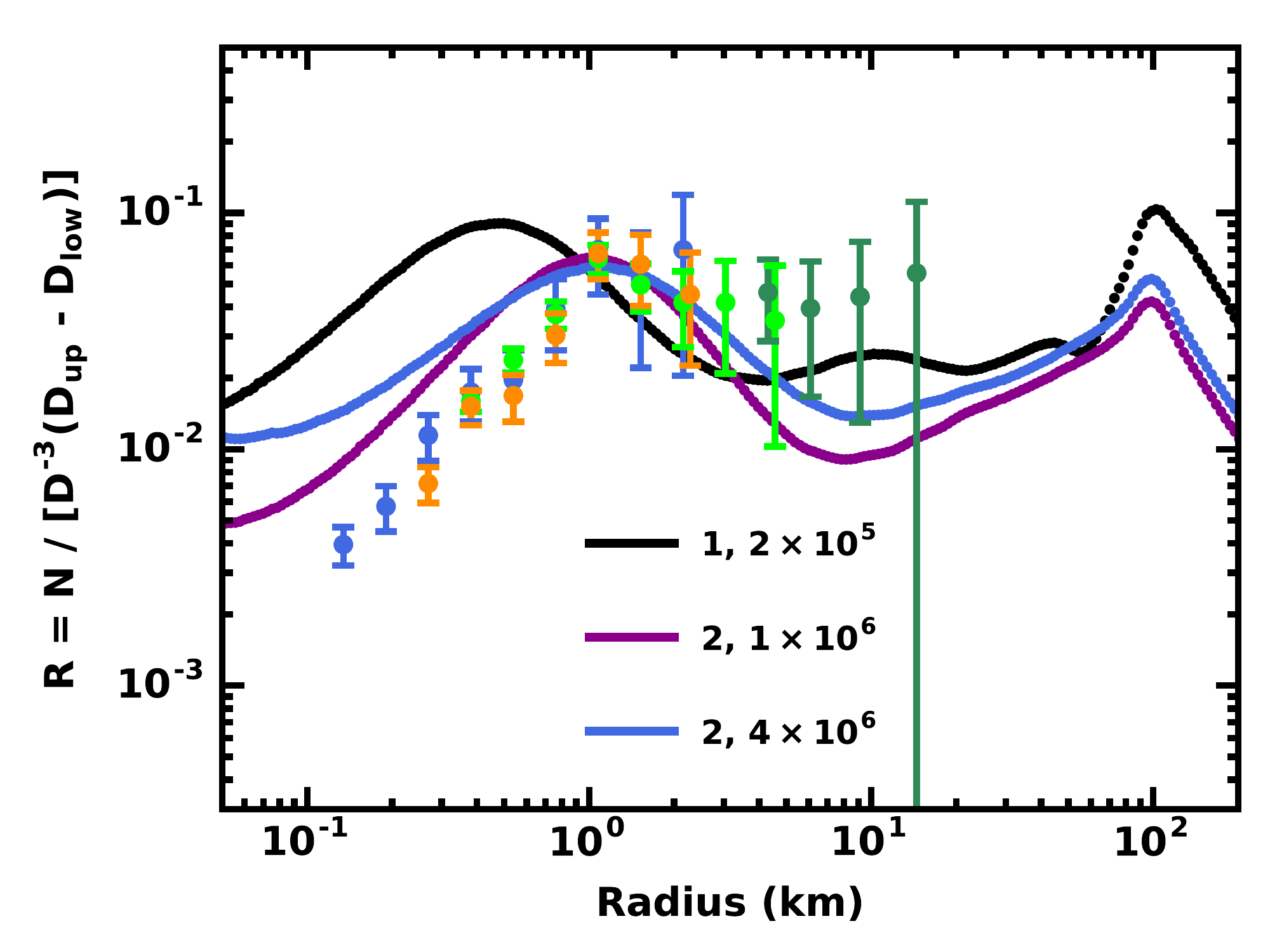}
\vskip -2ex
\caption{\label{fig: obs3}
Comparison of \nh\ data with $R(r)$ for three collisional cascades. All calculations use
$e_s = -0.4$ and the \citet{lein2012} parameters for the gravity component of \qdstar. 
The legend indicates values for \vc\ (in \kms) and $Q_s$. Models with \vc\ = 2~\kms\ and 
$Q_s = 10^6$~erg~g$^{-1}$~cm$^{0.4}$ provide a reasonable match to the \nh\ data.
}
\end{center}
\end{figure}

Fig.~\ref{fig: obs3} compares several model size distributions with the \nh\ data. 
With the model (4) parameters (Fig.~\ref{fig: obs3}, black curve), the model matches the
\nh\ data rather well at 1--20~km but fails at smaller sizes. Although the amplitude of
the wave for model (4) is large, the location of the peak and valley are displaced to
smaller sizes. Together with the normal ice fragmentation parameters, the larger collision 
velocity in model (11) matches the \nh\ data much better (Fig.~\ref{fig: obs3}, blue curve).
Although this size distribution has a deep valley at 5--10~km, the agreement with the
\nh\ data is reasonably good at 1--10~km. This model has a clear peak at 1~km, which agrees
with the \nh\ data, and a valley at 0.05~km, which is consistent with the \nh\ data. However,
the amplitude of the wave is a factor of 2--3 too small. 

For \vc\ = 2~\kms, models with $Q_s = 10^6$~erg~g$^{-1}$~cm$^{0.4}$ come close to matching 
the \nh\ data. At 1--100~km, this model has a peak at 1~km as in model (11), with a deeper 
valley at 5--10~km, and a gradual rise from 10~km to a sharp peak at 100~km. Solids with
$r \gtrsim$ 100~km have a steep size distribution with $q$ = 5. Overall, this size distribution
matches the \nh\ data well and has the steep slope at the largest sizes required by the
ground-based observations.  At smaller sizes, this model has a deep valley at 0.05~km. 
Unlike model (11), the depth of this valley almost matches the depth of the valley at 0.1~km
in the \nh\ data. 

In this model, with fragmentation parameters intermediate between those of the weak ice and
normal ice parameters, changing the collision velocity does not improve the match to the
observations. When \vc\ = 1~\kms, the deep valley at 5--10~km is smaller; however, the valley
at 0.05~km is also smaller. Thus, the lower collision velocity allows a better match to the
data at 1--10~km at the expense of a poorer match at 0.1--1~km. Increasing the collision
velocity to 3~\kms also fails to improve the match. At the higher collision velocities, the
deep valley at 0.05~km remains fixed. Although the amplitude of the wave is somewhat larger,
the peak is displaced to 2--3~km. In this size range, the match to the \nh\ data is somewhat
better. The cost for this better match is a deeper valley at 5--10~km. 

Compared to the analytical models, the $\chi^2$ per degree of freedom for the numerical models
is somewhat worse. The best analytical models have a $\chi^2$ per degree of freedom of 3--4. In
the numerical models, the $\chi^2$ grows to 5--10. In both types of calculations, poor models
have $\chi^2$ per degree of freedom exceeding 100. While not excellent fits, the best analytical
and numerical models are a significant improvement over other analytical and numerical models
where the slope of the size distribution at 0.1--100~km is much larger than 3 \citep[see the
discussion in][]{greenstreet2015}.

Overall, we prefer a model with \vc\ = 2~\kms\ and $Q_s = 10^6$~erg~g$^{-1}$~cm$^{0.4}$. 
Calculations with other initial values of $q_s$ ($q_s = -3$ and 3 instead of $q_s$ = 0) do 
not improve the match to the data. When $q_s = -3$, the valley at 5--10~km is deeper compared
to the model with $q_s$ = 0. At low collision velocities, \vc\ = 1~\kms, models with $q_s$ = 3
have much shallower valleys at 5--10~km than models woth smaller $q_s$. Increasing \vc\ to
2--3~\kms\ tends to eliminate this difference. Thus, systems with \vc\ = 2--3~\kms\ and
$q_s$ = 2--3 have size distributions similar to those with $q_s$ = 0.

\vskip 6ex
\section{DISCUSSION}
\label{sec: disc}

The analytic and numerical calculations described in \S\ref{sec: casc-an} and 
\S\ref{sec: casc-num} demonstrate that collisional cascades produce size distributions
with features similar to those observed in the \nh\ data for Charon impactors with 
$r \approx$ 0.1--10~km. In the
analytic models, reasonably good matches to the \nh\ data are possible when $Q_s$
is small ($\sim 10^4$~erg~g$^{-1}$~cm$^{0.4}$) or 
moderate ($\sim 4 \times 10^6$~erg~g$^{-1}$~cm$^{0.4}$). Values for $Q_s$ in between
these limits produce size distributions that disagree with the \nh\ data. Much larger
values for $Q_s$ also fail. The analytic models favor smaller values for the exponent
of the bulk strength component of \qdstar\ ($e_s \approx -0.2$ to $-0.4$). When $e_s$ is
larger, the waves in the size distribution are too small. Our analysis supports the
\citet{lein2012} parameters for the gravity component of \qdstar\ over the \citet{benz1999}
parameters. 

The numerical models strongly favor the normal ice fragmentation parameters. When $Q_s$
is small, the amplitudes of the waves in the size distribution are comparable to the
factor of 10--20 wave observed in the \nh\ size distribution. However, the peaks and valleys
in the numerical simulations are well off those in the data. When the calculations use
parameters for normal ice ($Q_s \sim 1-4 \times 10^6$~erg~g$^{-1}$~cm$^{0.4}$ and 
$e_s = -0.4$), the locations of peaks and valleys match those in the \nh\ data. For 
collision velocities \vc\ = 2~\kms, the amplitude of the wave falls about a factor of
two short of the amplitude observed in the \nh\ data. Changing the collision velocity 
does not change the amplitude of the wave significantly; larger \vc\ moves the peak away
from the peak in the \nh\ data. Although the match to the \nh\ data is not perfect, the
derived size distributions match much better than other results in the literature.

Comparison of Fig.~\ref{fig: obs2} and Fig.~\ref{fig: obs3} suggest the analytical models
match the \nh\ data better than the numerical models. Due to the cpu time required, the
numerical models sample a smaller set of fragmentation parameters than the analytical
models. Because the numerical models do not match the analytical models precisely (see
Figs.~\ref{fig: eqcc1}--\ref{fig: eqcc2}), it is not possible to anticipate the shape of 
the final size distribution in any numerical model. Choosing a fragmentation model
slightly different than `normal ice' might yield a better match to the \nh\ data. 
Within the numerical calculations, shot noise among large objects with infrequent collisions
and the redistribution of mass from cratering and catastrophic collisions appear to reduce 
the waviness of the size distribution somewhat relative to the analytical model. With the 
fragmentation parameters required to match the \nh\ data fairly well-constrained, it is 
possible to examine how the degree of waviness depends on other algorithms in the 
numerical model. We plan to consider this issue in a future study.

Despite the sensitivity to the fragmentation parameters, the shapes of the wavy size
distributions are remarkably insensitive to initial conditions.  After 100--200~Myr of 
collisional evolution at 45~au, there is little variation in $R(r)$ at 0.1--10~km among 
calculations with different $r_l$ and $q_s$. When $Q_s$ is fairly large, all $R(r)$ 
have a valley near the transition radius and a peak at 10--30 $r_t$; the position
of the peak and the valley agree well with analytic models \citep{obrien2003}.
Systems with smaller $Q_s$ have a valley at a size $r_v$ somewhat larger than $r_t$ 
and a peak $r_p \sim$ 10--30 $r_v$; these peaks/valleys lie off analytic predictions
due to interactions with waves generated by the small-size cutoff at 1~\mum.  
At 10--100~km, $R(r)$ is sensitive to $r_l$ and $q_s$. As the evolution proceeds 
past 100--200~Myr, these differences often disappear. 

Tests indicate that $R(r)$ is also insensitive to the fragmentation algorithm. 
Calculations with different values for $m_{l,0}$, $b_l$, and $b_d$ have similar size 
distributions at $t \gtrsim$ 100--200~Myr. Although we did not test alternative
approaches for catastrophic disruption or cratering \citep[e.g.,][and references
therein]{fraser2009b,benavidez2009,campo2012}, previous experience with other
fragmentation algorithms suggests that results will be similar to those described
above \citep[e.g.,][]{kb2002b,kb2004a,kb2004c,kb2008}. While calculations with other 
approaches may infer somewhat different `best' parameters for matches to the \nh\ data, 
the general conclusion that solids composed of `normal ice' provide a better match to 
the positions of peaks and valleys in the \nh\ data than `strong', `weak', or 
`very weak' ice should be independent of the fragmentation algorithm.

Another set of tests suggest that the evolution is independent of the algorithm for
velocity evolution. In all models described above, particles have initial $e$ and
$\imath$ that are fixed throughout the calculation. Allowing collisional damping, 
dynamical friction, and viscous stirring to modify particle velocities has little
impact. The cascade is too efficient at transporting material from large particles
to small particles. We did not investigate the possibility that large and small 
particles have different initial $e$ and $\imath$; however, we doubt that the results
would be much different from those described above.

The calculations that most closely match the \nh\ data also provide a good match to 
other observations of TNOs. By design, the calculations begin with a steep size 
distribution, $ q \approx 5.5$ for $r \gtrsim$ 100~km. After 4.5~Gyr of evolution, 
the final slope is somewhat shallower, $q \approx$ 5. Starting with a steeper 
slope, $q \approx 6$, instead of $q \approx$ 5.5 would result in a final slope 
close to the observed $q \approx$ 5.5 with little impact on the shape of the 
size distribution at smaller sizes. In simulations with $r_l \approx$ 30--100~km, 
the size distribution has a clear break in slope at $r_b \approx$ 50--100~km, 
close to the break inferred from ground-based and space-based imaging data. Some
of the calculations also generate a divot in the size distribution for objects 
with $r \lesssim r_b$. The magnitude of the divot in the calculations is 
comparable to that required in some surveys of TNOs \citep[e.g.,][]{shankman2013,
shankman2016,alexandersen2016,lawler2018}.  

Among the full suite of calculations, the power-law slope of the size distribution
at 10--100~km, $q \approx$ 1.5--3.5, includes the range derived from ground-based 
observations of TNOs, $q \approx$ 2.0--3.5 (see Fig.~\ref{fig: obs1}). Systems with
$r_l$ = 1--3~km generate the steepest slopes at $t \gtrsim$ 100--200~Myr, 
$q \approx$ 3.0--3.5, and have a break in the size distribution at smaller radii 
than observed, $r_b \approx$ 10--30~km. As we increase $r_l$, the slope at 10--100~km
decreases to $q \approx$ 2--2.5; the position of the break grows to 50--100~km as
observed. Some calculations with $r_l \gtrsim$ 10~km generate two breaks in $n(r)$,
one at 50--100~km and another at 10--30~km. In these systems, $q \approx$ 3.5 (3.0)
at $r \approx$ 10--30~km ($r \lesssim$ 10--20~km). 

Although results from the analytic model suggest that the placement of waves
at 0.1--10~km depends on \rmin, we fixed \rmin\ = 1~\mum\ in the suite of numerical
calculations.  For the normal and strong ice parameters, \rmin\ should have little 
impact on the shape of $R(r)$. However, it is plausible that a factor of 2--3 
smaller/larger \rmin\ could impact the positions of waves in calculations with the 
weak ice fragmentation parameters.  Shifting the positions of valleys and peaks in 
the size distributions of weak ice calculations to factor of 2--3 larger sizes would 
enable a better match to the \nh\ data. We plan to consider the impact of different 
choices for \rmin\ and \rmax\ in a separate publication.

In the calculations discussed here, expanding to a multi-annulus grid might produce
a somewhat different outcome. In a multi-annulus calculation, the size distribution
of solids in each annulus depends on the mix of collisions from material in other
annuli. In a system where we divide the 30--60~au annulus considered here into 8--16 
separate annuli with identical orbital $e$ and $\imath$, annuli closer to the Sun 
experience higher velocity collisions than more distant annuli. With no excitation
from nearby gas giants, we expect the average size distribution of the swarm to be
similar to the results of single annulus calculations. 

\subsection{Dynamical Evolution of the Solar System}
\label{sec: disc-migr}

In current models for the Solar System, the gas giants grow within a circumsolar
gaseous disk and reach their final masses before the disk dissipates $\sim$ 5--10~Myr 
after the Sun formed \citep[e.g.,][and references therein]{bk2011a,mordasini2015,
johan2017,chambers2018,bitsch2019}. 
To account for the orbital architecture of the trans-Neptunian region and other
features of the Solar System, dynamical calculations require Uranus and Neptune to
migrate outward through a remnant disk of solid material to reach their current orbits 
\citep[e.g.,][]{malhotra1993,malhotra1995,levison2003b,levison2008,nesv2015,nesv2016}. 
Viable migration models require proper timing of the migration and specific constraints
on the size distribution of solids within the remnant disk.  To avoid a collisional 
cascade removing significant material on short time scales, most of the mass in solids 
is in large objects with $r \gtrsim$ 100~km \citep[e.g.,][]{kbod2008,kb2010}. 
Achieving the observed mix of resonant and non-resonant objects requires migration
through a swarm of Plutos on time scales of 10--100~Myr \citep[e.g.,][]{nesv2015,
nesv2016}; otherwise, it is not possible to generate the correct mix of resonant and
non-resonant TNOs with the observed distribution of eccentricity and inclination.

To place the present calculations in the context of this evolution, we rely on previous 
results for the growth of solids at 15--150~au \citep[e.g.,][]{kb2008,kb2010,kb2012}.
Among any swarm of solids, the time scale for the size distribution to evolve scales
with the local surface density $\Sigma$ and orbital period $P$ as $ t \propto P / \Sigma$
\citep[see also][]{liss1987,gold2004,youdin2013}.  
Although the evolution of the collisional cascades described in this paper depends on 
$r_l$ and the fragmentation parameters, most calculations establish the main features of 
the final size distribution at 100--300~Myr. For any other combination of starting 
conditions, the time scale to achieve an approximate equilibrium size distribution is 
\begin{equation}
t_{eq} \approx 200 ~ {\rm Myr} 
\left ( \frac {a}{{\rm 45~au}} \right )^{3/2}
\left ( \frac {{\rm 0.15~g~cm^{-2}}}{\Sigma} \right ) ~ .
\label{eq: teq}
\end{equation}
Within the single annulus extending from 30~au to 60~au, $\Sigma$ = 0.15~\gcms\ yields 
a total mass in solids of 45~\mearth. For each calculation, the shape of the size 
distribution is nearly constant in time for $t \gtrsim t_{eq}$; the total mass in the 
swarm steadily declines with time.

For the initial conditions considered here, the time scale to generate the required 
features in the size distribution is a factor of 2--3 longer than the current estimate
for the migration time of Neptune, $\lesssim$ 100~Myr \citep{nesv2016}. However, several 
modifications to the starting conditions yield similar size distributions on shorter time 
scales. At 45~au, increasing the initial surface density by a factor of five shortens the 
time required to reach equilibrium by a similar factor. To avoid increasing the initial 
mass to an unpalatable 225~\mearth, it is necessary to reduce the size of the annulus 
from 30--60~au to 42--48~au. Shrinking the annulus further provides a way to maintain the 
same surface density (and collision time scale) while reducing the mass below 45~\mearth.

Although confining solids to a narrow annulus is an unconventional choice for models 
of the Solar System, recent high resolution observations of many protostellar disks
with ALMA reveal a variety of narrow rings at 25--100~au from the central star 
\citep[e.g.,][]{zhang2016,fedele2018,dullemond2018,huang2018,long2018,cieza2019,long2020}. 
Among pre-main sequence stars with large disks, rings are common. Although young stars
with compact disks are difficult to resolve with ALMA, \citet{long2020} show evidence for
structure on small scales, $a \sim$ 10--20~au in GQ~Lup~A. The physical properties of the
dusty rings in young stars are similar to those required to generate a suitably wavy size 
distribution of 0.1~km and larger TNOs in less than 100~Myr. Thus, this option is a 
plausible way to build a TNO size distribution that matches the \nh\ data on a time scale
consistent with Neptune migration.

Placing the solids at smaller $a$ is a viable alternative to a narrow ring at 45~au. 
Setting $a \approx$ 25~au and keeping the surface density fixed at $\Sigma \approx$ 
0.15~\gcms, the time scale to produce a wavy size distribution that closely matches 
the \nh\ data is $\sim$ 70~Myr. For the protosolar nebula model required to build the 
giant planets at 5--15~au, the surface density is $\Sigma \approx$ 0.5~\gcms\ at 25~au 
\citep[e.g.,][]{bk2011a,kb2012}. If this material extends from 20~au to 30~au,
the total mass is roughly 30~\mearth.  The evolution time is $\sim$ 25~Myr. While the
mass is somewhat larger than the $\sim$ 20~\mearth\ required in the \citet{nesv2016}
migration model, the time scale to generate a wavy size distribution is much smaller
than the 50--100~Myr migration time. In this scenario, a migrating Neptune places many
TNOs in resonant orbits. Our calculations suggest these TNOs would have a wavy size
distribution similar to the size distribution of Charon impactors.

In either model for the size distribution of TNOs, dynamical interactions with Neptune 
need to remove most of the mass \citep{levison2003b,kbod2008,nesv2016}. In our 
calculations, $R(r)$ achieves a characteristic shape with little total mass loss.
Once the solids have this shape, destructive collisions slowly reduce the total mass.
In the Solar System, where Neptune migrates outward from $\sim$ 15--20~au to 30~au, 
dynamical interactions with the planet eject mass from the vicinity of Neptune; 
subsequent interactions with Neptune and the other gas giants eventually eject this
mass from the Solar System.  If collisional processes at 20--30~au have time to 
generate the characteristic shape for the size distribution, dynamical processes can 
remove mass without changing the shape of the size distribution. Our analysis suggests 
that collisions at 20--30~au can generate the required shape. Testing this conclusion 
in more detail requires a collisional cascade calculation coupled to a migration simulation. 

Removing material in a narrow ring outside of Neptune's orbit is more challenging. Once
Neptune reaches its current orbit at $a \approx$ 30~au, it rapidly removes solids material
at 30--36~au and sculpts the orbits of solids at $a \gtrsim$ 36~au. The time scale to 
remove tens of Earth masses outside of 36~au depends on the position of the ring relative
to Neptune's orbit and the eccentricities and inclinations of solids within the ring.
It seems plausible that a collisional cascade can generate the necessary size distribution
of TNOs as Neptune migrates to its current orbit. Quantifying the ability of Neptune to
reduce the mass of solids to the level observed now requires coupling the collision
calculation to a dynamical calculation involving Neptune and the other gas giants.

\subsection{Long-term Evolution of Charon Impactors}
\label{sec: disc-imp}

As summarized in \citet{singer2019}, the craters on Charon represent the integrated 
history of impacts over the lifetime of the \pc\ system. After \pc\ forms 
\citep[e.g.,][]{canup2005,canup2011,kb2014b,desch2015,kb2019c}, impacts 
from TNOs and any material leftover from the formation of the small satellites 
\citep[e.g,][]{walsh2015,bk2020}
generate craters on the surface of each planet. Reconstructing the cratering 
history requires time-dependent models for the size distribution(s), orbital 
architecture(s), and impact rates of possible impactors. Combining these with 
a relation between the sizes of the impactor and its associated crater yields a 
size-frequency distribution of craters on the surface of Charon.

If the size distributions and the orbital architectures of TNOs are constant in 
time, analytic and semi-analytical prescriptions for the collision rates are 
sufficient to derive the impact history \citep[e.g.,][]{dellorro2013,
greenstreet2015,bierhaus2015}.
In this approach, the size distribution of the craters reflects the adopted size
disributions of TNOs. \citet{greenstreet2015} consider a broad range of size 
distributions derived from (i) power-law fits to ground-based TNO data (see 
\S\ref{sec: obs}) and (ii) numerical simulations \citep[e.g.,][]{schlicht2013}.
\citet{bierhaus2015} consider power-law size distributions with $q$ = 2 or 3.
In both studies, the predicted diameter of Charon's largest crater (equivalently,
the diameter of the largest impactor) is set by the total mass in TNOs and the 
fraction of TNO orbits that cross the orbit of \pc. Below this size, the $R(r)$
for impactors is identical to the adopted size distribution.

To derive the expected crater distribution on Charon from a collisional cascade,
we need to combine the calculations considered here with a dynamical model for
the evolution of (i) TNO orbits with time and (ii) the removal rate due to 
interactions with Neptune. Here, we consider several simple estimates for TNOs 
with $a \gtrsim$ 30~au to set the stage for a more detailed treatment in a future 
study. We assume that the TNOs generated at 20--30~au and placed onto larger $a$ 
orbits by Neptune have the space density and orbital architecture required to 
match current observations. Because the collisional cascade calculations generate 
a wavy size distibution on short time scales, all of the different dynamical classes
of TNOs have the same wavy size distribution. Of the calculations considered here,
those with the normal ice or weak ice fragmentation parameters have the best chance
of matching the size distribution of Charon impactors.

For the broad annulus at 30--60~au, only a fraction of the solids might collide 
with Pluto.  Neptune removes all solids inside 36~au. TNOs with $q \gtrsim$ 50~au 
never intersect Pluto's orbit; At aphelion, Pluto has a large height above the 
ecliptic plane. TNOs with $a \approx$ 50~au and lower $e$, lower inclination 
orbits also cannot collide with Pluto. Together, these constraints remove nearly
90\% of the solids in the calculation from further consideration. To estimate the
impact probability, we scale the number of TNOs with $H \le$ 9 ($r \gtrsim$ 40~km)
remaining in a calculation at 100--200~Myr to the results in Table 1 of 
\citet{greenstreet2015} and derive approximate rates for collisions with Pluto. 
Rates for Charon follow from the ratio of surface areas, with \rc/\rp\ = 0.26 
\citep{stern2015,nimmo2017}. 

Among the full suite of cascade calculations discussed here, the impact rates at
100--200~Myr range from a few to 100 times the rates quoted in \citet{greenstreet2015}.
Models with the smallest rate of impacts have most of the initial mass in 1--3~km 
planetesimals. Within 100--200~Myr, collisions convert 90\% to 95\% of the initial 
mass into debris with $r \lesssim$ 0.1~km. When most of the mass is initially in 
30--100~km objects, collisions take much longer to remove a significant amount of 
material from the swarm. These models retain most of their mass until 300--500~Myr.

For solids in a narrow annulus, the impact rate depends mainly on the location of 
the annulus. Because collision rates within the annulus are large, the mass declines
rapidly, on time scales of 20--100~Myr instead of 100--500~Myr for systems with
$r_l$ = 1--100~km. These loss rates limit the number of impacts onto \pc.  In an 
annulus at 38--42~au, most of the solids can interact with Pluto; however, Neptune 
dynamically removes objects not lost to the collisional cascade.  Although moving 
the annulus to 48--52~au limits removal by Neptune, the orbits of many fewer TNOs
cross the orbit of \pc. We suspect that it is possible to find a model where the 
size distribution at 20--30~Myr roughly matches the \nh\ observations and has the 
mass required to generate the observed cratering rate.  Compared to other options, 
this approach requires more fine tuning of the initial conditions.

Overall, it seems that impactors generated in a collisional cascade could explain
the \nh\ observations. Many calculations develop the required wavy size distributions 
in 20--30~Myr (for a narrow annulus at $\sim$ 45~au or a disk at 20--30~au) to
100--200~Myr (for a broad annulus at 30--60~au). Although the total mass in solids
continues to decline after these epochs, the shapes of these size distributions are 
relatively invariant over the rest of the age of the Solar System. Placing better
constraints on the ability of models to match the density of craters on Charon
requires calculations that track the dynamical interactions of TNOs with Neptune and 
the time-varying impact rate onto \pc. Although straightforward within the \orch\ code,
these calculations are time-consuming. Together with the relative insensitivity of
the calculations to the initial $q_s$, the success of ruling out solids with strong 
ice or very weak ice allows the next set of calculations to be more focused.

Aside from matching the density and the size distribution of craters on Charon, it is
important to match the current space density of the various dynamical classes of TNOs.
Of the options considered here, it is easiest to imagine that an initial disk of solids 
at 20--30~au can maintain a cascade that achieves all of the goals of a complete TNO
model. A well-placed narrow annulus outside the orbit of Neptune is the next most likely
option. Developing a successful model with a thick annulus at 30--60~au is a challenge.
More comprehensive numerical calculations can test these conclusions.

%Aside from the size distribution of impactors, \citet{singer2019} discuss several 
%reasons why the craters on Charon are old, $\sim$ 4~Gyr. If equilibrium size
%distributions require $\sim$ 4.5~Gyr of collisional evolution, then they are unlikely
%matches to the \nh\ data. For the initial conditions we consider, however, it takes
%only $\sim$ 500~Myr for collisional evolution at 45~au to develop a wavy $R(r)$ 
%that corresponds rather well to the \nh\ data. Starting from a more massive Kuiper belt 
%would shorten the evolution time. Thus, collisional evolution at 45~au can account for 
%the size distribution of impactors on Charon.

\subsection{Impactors on (486958) Arrokoth}
\label{sec: disc-arro}

As we completed the calculations for this study, \citet{spencer2020} reported 
an extensive analysis of the \nh\ data acquired during the 2019 January flyby of
the TNO (486958) Arrokoth. Aside from revealing a lack of small rings and small
satellites in the system, they describe statistics of the cratering record. For 
diameters 0.3--1~km, the frequency of craters follows a power-law with a slope, 
$q \approx -2$, that is indistinguishable from the frequency distribution of 
small craters with diameter 1--20~km on Charon \citep{singer2019}.  The density 
of 0.8--1~km craters on Arrokoth is roughly an order of magnitude larger than 
the density of 1--2~km craters on Charon, more than the factor of $\sim$ 2 
expected from detailed calculations of the impact frequency  prior to the flyby 
\citep{greenstreet2019}.

To compare the collisional casade calculations with these \nh\ data, we ignore 
the difference in absolute crater frequencies between Arrokoth and Charon and
focus on the shape of the frequency distribution.  We also gloss over the
impact of the trajectories and velocities of impactors and the porosity and 
other physical characteristics of ices on the relation between impactor size 
and crater diameter \citep[e.g.,][]{bierhaus2015,singer2019}.  For simplicity, 
we adopt the same scaling relation between impactor size and crater diamater, 
$D = D_c / 6.25$, used in \S\ref{sec: obs}.  Other options yield similar 
results \citep[e.g,][]{greenstreet2019}.

For the adopted scaling relation, it is challenging to find a cascade model 
capable of matching the combined set of Arrokoth--Charon cratering data. 
Within the \nh\ data, the combined power-law slope extends over nearly 
two orders of magnitude in crater diameter. If both sets of craters have 
the same relation between crater diameter and impactor size, a cascade model
needs to generate a power-law with slope $q \approx -2$ from 1~km to 
0.04--0.05~km. For the normal ice parameters, $R(r)$ has a maximum at 1~km 
and a minimum at 0.05~km; however, the amplitude from peak to trough is 
much smaller than observed in the combined \nh\ data set (see Figs.~\ref{fig: cc12},
\ref{fig: cc13},
and \ref{fig: obs3}. With the weak ice parameters, the amplitude is larger
but the maxima and minima are displaced to smaller sizes compared to the data.
All calculations with very weak ice have wave amplitudes large enough to 
match the combined \nh\ data. However, these cascades produce minima at 1~km
and maxima at 0.05~km.

Assuming the adopted scaling between crater and impactor sizes is correct, 
cascade calculations with very weak ice and a different \rmin\ might match
the combined \nh\ data. In the analytical models, shifting the minimum size
in the calculation from \rmin\ = 0.3~\mum\ to \rmin\ = 3~\mum\ resulted in
a significant shift in the positions of peaks and valleys in $R(r)$ at 0.1--1~km
for the very weak ice fragmentation parameters (see Fig.~\ref{fig: obs2}). 
If the numerical calculations with different \rmin\ generate similar shifts 
in the positions of peaks and valleys, then it should be possible to match
the \nh\ observations with the very weak ice fragmentation parameters.

Matching the \nh\ data with the very weak ice parameters would have interesting
implications for the long-term evolution of the TNO space density and cratering 
rate. When the cascade begins with a significant mass in small objects, 
$r \lesssim$ 1--10~km, swarms of solids with the very weak ice fragmentation 
parameters lose mass more rapidly than those with the stronger ice parameters. 
Thus, it is easier for swarms with the very weak ice parameters to lose enough 
mass to match the current space density of TNOs at 40--50~au. More rapid mass 
loss also lowers the rate of TNO impacts on other TNOs, enabling the cascade 
models to provide a better match to the impact rates on Pluto, Charon, Arrokoth,
and (eventually) other TNOs.

\subsection{Other Issues}
\label{sec: disc-issues}

In \S\ref{sec: obs}, we noted the limited constraints on the frequency of $\sim$
1~km TNOs from space-based occultation data \citep[e.g.,][]{schlicht2009,bianco2010,
schlicht2012,zhang2013,liu2015} and the need to convert small TNOs into Jupiter
family comets \citep[e.g.,][and references therein]{levison1997,emelyan2004,volk2008,
brasser2015}. In both examples, the apparent number of 1~km TNOs implies a very
steep power-law size distribution at 1--100~km with $q \approx$ 3.50--3.75. The 
\nh\ data for craters on Charon have $q \approx$ 3 at 1--10~km and $q \approx$ 2
at 0.1--1~km.  There are few good options to reconcile these differences. 
Normalizing the \nh\ counts at 1~km to level required for Jupiter family comets 
and the occultation data yields too many 10~km TNOs.

Our calculations do not address this issue.  It seems plausible that the source of 
the impactors on Charon differs from the source of Jupiter family comets, especially 
if TNOs are formed in different locations from $\sim$ 20~au to $\sim$ 50~au. 
Developing this idea in more detail requires a multi-annulus coagulation--dynamical 
calculation to follow the evolution of the dynamics and the size distribution of
TNOs at multiple points in the Solar System.

More complicated analyses of the space-based occultation data and ground-based TNO
surveys might also address the offset in number density relative to the \nh\ data.
For both data sets, analyses adopt power-law size distributions for TNOs with 
$r \lesssim$ 100~km. Repeating these analyses with wavy size distributions might 
provide some insight. Given the good general agreement between the analytical and
numerical calculations, it should be straightforward to incorporate our method for
generating an equilibrium size distribution into these analyses. In principle, these
data sets might place constraints on the fragmentation parameters needed to derive
an equilibrium size distribution. Evaluating these constraints together with the 
limits from \nh\ data might yield additional information on the bulk properties of
TNOs.

Finally, the size distribution of Charon impactors has features in common with the size
distributions of Jupiter's Trojan satellites and the main belt asteroids 
\citep[e.g.,][]{bottke2005,yoshida2005,yoshida2008,bottke2015,yoshida2017,singer2019,
yoshida2019}. In these other systems, $R(r)$ has a deep valley at $r_v \lesssim$ 0.1~km 
and a peak at $r_p \approx$ 5~km. Although we do not attempt to match $R(r)$ for 
asteroids or Jupiter's Trojans, collisional cascades with $r_l \approx$ 5~km and a
range of fragmentation parameters can produce a peak in $R(r)$ at 5~km.  Matching the 
deep valley probably requires solids composed of weak or normal ice rather than strong 
ice.

\vskip 6ex
\section{SUMMARY}
\label{sec: summary}

The analytic and numerical calculations described here demonstrate that collisional
cascades of 1--500~km icy solids generate wavy size distributions with features similar
to those inferred from counts of craters on Charon. For models with a standard prescription 
for the binding energy \qdstar\ (eq.~\ref{eq: qdstar}) and an energy-scaling algorithm
for collision outcomes, calculations that provide the best match to the \nh\ data have the
properties of normal ice. In swarms of strong ice solids, features in the size distribution 
do not match the amplitudes or positions of peaks and valleys in the \nh\ data. Calculations
of weak ice solids have the correct amplitudes for waves but not the correct positions of
peaks and valleys.

Aside from matching the \nh\ data, cascades can also maintain a steep size distribution
for solids with $r \gtrsim$ 100~km and a break in the size distribution at $r_b \approx$
50--100~km.  Some calculations also produce a divot in the size distribution of solids 
with $r \lesssim r_b$. Additional work is needed to learn whether the size distributions
that match the \nh\ data can account for the frequency of Jupiter family comets and recent
detections of small TNOs with occultations.

Adding \nh\ data on the cratering record of (486958) Arrokoth complicates this analysis. 
For the combined set of Arrokoth and Charon impactors, the normal ice calculations have a 
wave maximum that matches the change in slope of $R(r)$ at 10--20~km and; the wave minimum
is close to the smallest crater detected on Arrokoth. However, the wave amplitude from peak 
to trough is a factor of 3--4 smaller in the model than in the data. Although calculations
with very weak ice produce waves with roughly the required amplitude at 0.05--1~km, the 
models have a wave maximum (minimum) at the trough (peak) of the data. We speculate that
calculations with a different minimum radius for solids might shift wave position without
modifying the wave amplitude. We plan to test this possibility in a future study.

Although we perform calculations for one fixed annulus at 30--60~au from the Sun, scaling
relations developed from previous studies allow us to consider whether the cascade models
are consistent with the dynamical evolution of the gas giant planets and the cratering 
record on \pc. In rings of solids at 20--30~au where Neptune migrates to its current orbit
on time scales of 50--100~Myr \citep[e.g.,][]{nesv2016}, collisional evolution can produce
a wavy size distribution with features similar to the size distribution on Charon in 
20--30~Myr. This model may also yield a reasonable cratering rate on Charon. Outside 
Neptune's orbit, evolution of a swarm of solids at 30--60~au seems unlikely to produce
a small enough cratering rate. However, cascades in a narrow annulus at 42--48~au are a
viable alternative.

Making progress on a complete model for the size distribution of TNOs requires a calculation
that combines a collision model with a dynamical model to derive the time evolution of the
space density and orbital parameters of TNOs as functions of time. Aside from providing better
comparisons with data for the wavy size distribution of craters on Charon and Arrokoth, 
this model can also derive cratering rates for comparisons with \nh\ data.
With similarly wavy size distributions observed in main belt asteroids and Jupiter's Trojan 
satellites \citep{bierhaus2015,bottke2015,yoshida2019}, it seems plausible that collisional 
cascades among solids left over from the formation of the terrestrial planets and the gas 
giants play an important role in establishing the current properties of small objects 
throughout the Solar System.

\vskip 6ex

We acknowledge generous allotments of computer time on the NASA `discover' cluster.
Advice and comments from W. Fraser, M. Geller, and two reviewers greatly improved 
our presentation.  
Portions of this project were supported by the {\it NASA } {\it Outer Planets} and 
{\it Emerging Worlds} programs through grants NNX11AM37G and NNX17AE24G.
This research has made use of data and/or services provided by the International 
Astronomical Union's Minor Planet Center. 

%\bibliography{sfpl}
\bibliography{ms.bbl}

\end{document}